\documentclass[twocolumn,letterpaper,aps,prd,longbibliography,superscriptaddress,showpacs,nofootinbib,floatfix]{revtex4-1}

\usepackage{graphicx}	
\usepackage{amsmath}
\usepackage{xspace}	

\newcommand{\pt}{\mbox{$p_T$}\xspace}

\newcommand{\sqs}{\mbox{$\sqrt{s}$}\xspace}

\newcommand{\pp}{\mbox{$p$$+$$p$}\xspace}

\newcommand{\pion}{\pi^0}
\newcommand{\decpty}{Y_{\rm decay}^{\rm iso}}

\newcommand{\Rprime}{R_\gamma^{\rm iso}}
\newcommand{\mapfxn}{P(p_T^{\pi^0},p_T^\gamma)}
\newcommand{\pout}{p_{\rm out}}
\newcommand{\dphi}{\Delta\phi}
\newcommand{\ptassoc}{p_{T}^{\rm assoc}}
\newcommand{\pttrig}{p_T^{\rm trig}}

\newcommand{\rmspout}{\sqrt{\langle p_{\rm out}^2\rangle}}
\newcommand{\zt}{\langle z_T^{\pi^{0}}\rangle}
\newcommand{\jt}{\sqrt{\langle j_T^2\rangle}}

\begin{document}

\title{Nonperturbative-transverse-momentum effects and evolution in
dihadron and direct photon-hadron angular correlations in $p$$+$$p$
collisions at $\sqrt{s}$=510 GeV}

\newcommand{\abilene}{Abilene Christian University, Abilene, Texas 79699, USA}
\newcommand{\augie}{Department of Physics, Augustana University, Sioux Falls, South Dakota 57197, USA}
\newcommand{\banaras}{Department of Physics, Banaras Hindu University, Varanasi 221005, India}
\newcommand{\barc}{Bhabha Atomic Research Centre, Bombay 400 085, India}
\newcommand{\baruch}{Baruch College, City University of New York, New York, New York, 10010 USA}
\newcommand{\bnlcoll}{Collider-Accelerator Department, Brookhaven National Laboratory, Upton, New York 11973-5000, USA}
\newcommand{\bnlphys}{Physics Department, Brookhaven National Laboratory, Upton, New York 11973-5000, USA}
\newcommand{\caucr}{University of California-Riverside, Riverside, California 92521, USA}
\newcommand{\charlesczech}{Charles University, Ovocn\'{y} trh 5, Praha 1, 116 36, Prague, Czech Republic}
\newcommand{\chonbuk}{Chonbuk National University, Jeonju, 561-756, Korea}
\newcommand{\ciae}{Science and Technology on Nuclear Data Laboratory, China Institute of Atomic Energy, Beijing 102413, People's Republic of China}
\newcommand{\cns}{Center for Nuclear Study, Graduate School of Science, University of Tokyo, 7-3-1 Hongo, Bunkyo, Tokyo 113-0033, Japan}
\newcommand{\colorado}{University of Colorado, Boulder, Colorado 80309, USA}
\newcommand{\columbia}{Columbia University, New York, New York 10027 and Nevis Laboratories, Irvington, New York 10533, USA}
\newcommand{\czechtech}{Czech Technical University, Zikova 4, 166 36 Prague 6, Czech Republic}
\newcommand{\debrecen}{Debrecen University, H-4010 Debrecen, Egyetem t{\'e}r 1, Hungary}
\newcommand{\elte}{ELTE, E{\"o}tv{\"o}s Lor{\'a}nd University, H-1117 Budapest, P{\'a}zm{\'a}ny P.~s.~1/A, Hungary}
\newcommand{\eszterhazy}{Eszterh\'azy K\'aroly University, K\'aroly R\'obert Campus, H-3200 Gy\"ngy\"os, M\'atrai \'ut 36, Hungary}
\newcommand{\ewha}{Ewha Womans University, Seoul 120-750, Korea}
\newcommand{\fsu}{Florida State University, Tallahassee, Florida 32306, USA}
\newcommand{\gsu}{Georgia State University, Atlanta, Georgia 30303, USA}
\newcommand{\hanyang}{Hanyang University, Seoul 133-792, Korea}
\newcommand{\hiroshima}{Hiroshima University, Kagamiyama, Higashi-Hiroshima 739-8526, Japan}
\newcommand{\howard}{Department of Physics and Astronomy, Howard University, Washington, DC 20059, USA}
\newcommand{\ihepprot}{IHEP Protvino, State Research Center of Russian Federation, Institute for High Energy Physics, Protvino, 142281, Russia}
\newcommand{\illuiuc}{University of Illinois at Urbana-Champaign, Urbana, Illinois 61801, USA}
\newcommand{\inrras}{Institute for Nuclear Research of the Russian Academy of Sciences, prospekt 60-letiya Oktyabrya 7a, Moscow 117312, Russia}
\newcommand{\instpasczech}{Institute of Physics, Academy of Sciences of the Czech Republic, Na Slovance 2, 182 21 Prague 8, Czech Republic}
\newcommand{\isu}{Iowa State University, Ames, Iowa 50011, USA}
\newcommand{\jaea}{Advanced Science Research Center, Japan Atomic Energy Agency, 2-4 Shirakata Shirane, Tokai-mura, Naka-gun, Ibaraki-ken 319-1195, Japan}
\newcommand{\jyvaskyla}{Helsinki Institute of Physics and University of Jyv{\"a}skyl{\"a}, P.O.Box 35, FI-40014 Jyv{\"a}skyl{\"a}, Finland}
\newcommand{\kek}{KEK, High Energy Accelerator Research Organization, Tsukuba, Ibaraki 305-0801, Japan}
\newcommand{\korea}{Korea University, Seoul, 136-701, Korea}
\newcommand{\kurchatov}{National Research Center ``Kurchatov Institute", Moscow, 123098 Russia}
\newcommand{\kyoto}{Kyoto University, Kyoto 606-8502, Japan}
\newcommand{\labllr}{Laboratoire Leprince-Ringuet, Ecole Polytechnique, CNRS-IN2P3, Route de Saclay, F-91128, Palaiseau, France}
\newcommand{\lahorelums}{Physics Department, Lahore University of Management Sciences, Lahore 54792, Pakistan}
\newcommand{\lawllnl}{Lawrence Livermore National Laboratory, Livermore, California 94550, USA}
\newcommand{\losalamos}{Los Alamos National Laboratory, Los Alamos, New Mexico 87545, USA}
\newcommand{\lund}{Department of Physics, Lund University, Box 118, SE-221 00 Lund, Sweden}
\newcommand{\maryland}{University of Maryland, College Park, Maryland 20742, USA}
\newcommand{\mass}{Department of Physics, University of Massachusetts, Amherst, Massachusetts 01003-9337, USA}
\newcommand{\michigan}{Department of Physics, University of Michigan, Ann Arbor, Michigan 48109-1040, USA}
\newcommand{\muhlenberg}{Muhlenberg College, Allentown, Pennsylvania 18104-5586, USA}
\newcommand{\myongji}{Myongji University, Yongin, Kyonggido 449-728, Korea}
\newcommand{\nagasaki}{Nagasaki Institute of Applied Science, Nagasaki-shi, Nagasaki 851-0193, Japan}
\newcommand{\nara}{Nara Women's University, Kita-uoya Nishi-machi Nara 630-8506, Japan}
\newcommand{\natmephi}{National Research Nuclear University, MEPhI, Moscow Engineering Physics Institute, Moscow, 115409, Russia}
\newcommand{\newmex}{University of New Mexico, Albuquerque, New Mexico 87131, USA}
\newcommand{\nmsu}{New Mexico State University, Las Cruces, New Mexico 88003, USA}
\newcommand{\ohio}{Department of Physics and Astronomy, Ohio University, Athens, Ohio 45701, USA}
\newcommand{\ornl}{Oak Ridge National Laboratory, Oak Ridge, Tennessee 37831, USA}
\newcommand{\orsay}{IPN-Orsay, Univ.~Paris-Sud, CNRS/IN2P3, Universit\'e Paris-Saclay, BP1, F-91406, Orsay, France}
\newcommand{\peking}{Peking University, Beijing 100871, People's Republic of China}
\newcommand{\pnpi}{PNPI, Petersburg Nuclear Physics Institute, Gatchina, Leningrad region, 188300, Russia}
\newcommand{\riken}{RIKEN Nishina Center for Accelerator-Based Science, Wako, Saitama 351-0198, Japan}
\newcommand{\rikjrbrc}{RIKEN BNL Research Center, Brookhaven National Laboratory, Upton, New York 11973-5000, USA}
\newcommand{\rikkyo}{Physics Department, Rikkyo University, 3-34-1 Nishi-Ikebukuro, Toshima, Tokyo 171-8501, Japan}
\newcommand{\saispbstu}{Saint Petersburg State Polytechnic University, St.~Petersburg, 195251 Russia}
\newcommand{\seoulnat}{Department of Physics and Astronomy, Seoul National University, Seoul 151-742, Korea}
\newcommand{\stonybrkc}{Chemistry Department, Stony Brook University, SUNY, Stony Brook, New York 11794-3400, USA}
\newcommand{\stonycrkp}{Department of Physics and Astronomy, Stony Brook University, SUNY, Stony Brook, New York 11794-3800, USA}
\newcommand{\sungskku}{Sungkyunkwan University, Suwon, 440-746, Korea}
\newcommand{\tenn}{University of Tennessee, Knoxville, Tennessee 37996, USA}
\newcommand{\titech}{Department of Physics, Tokyo Institute of Technology, Oh-okayama, Meguro, Tokyo 152-8551, Japan}
\newcommand{\tsukuba}{Center for Integrated Research in Fundamental Science and Engineering, University of Tsukuba, Tsukuba, Ibaraki 305, Japan}
\newcommand{\vandy}{Vanderbilt University, Nashville, Tennessee 37235, USA}
\newcommand{\weizmann}{Weizmann Institute, Rehovot 76100, Israel}
\newcommand{\wigner}{Institute for Particle and Nuclear Physics, Wigner Research Centre for Physics, Hungarian Academy of Sciences (Wigner RCP, RMKI) H-1525 Budapest 114, POBox 49, Budapest, Hungary}
\newcommand{\yonsei}{Yonsei University, IPAP, Seoul 120-749, Korea}
\newcommand{\zagreb}{University of Zagreb, Faculty of Science, Department of Physics, Bijeni\v{c}ka 32, HR-10002 Zagreb, Croatia}
\affiliation{\abilene}
\affiliation{\augie}
\affiliation{\banaras}
\affiliation{\barc}
\affiliation{\baruch}
\affiliation{\bnlcoll}
\affiliation{\bnlphys}
\affiliation{\caucr}
\affiliation{\charlesczech}
\affiliation{\chonbuk}
\affiliation{\ciae}
\affiliation{\cns}
\affiliation{\colorado}
\affiliation{\columbia}
\affiliation{\czechtech}
\affiliation{\debrecen}
\affiliation{\elte}
\affiliation{\eszterhazy}
\affiliation{\ewha}
\affiliation{\fsu}
\affiliation{\gsu}
\affiliation{\hanyang}
\affiliation{\hiroshima}
\affiliation{\howard}
\affiliation{\ihepprot}
\affiliation{\illuiuc}
\affiliation{\inrras}
\affiliation{\instpasczech}
\affiliation{\isu}
\affiliation{\jaea}
\affiliation{\jyvaskyla}
\affiliation{\kek}
\affiliation{\korea}
\affiliation{\kurchatov}
\affiliation{\kyoto}
\affiliation{\labllr}
\affiliation{\lahorelums}
\affiliation{\lawllnl}
\affiliation{\losalamos}
\affiliation{\lund}
\affiliation{\maryland}
\affiliation{\mass}
\affiliation{\michigan}
\affiliation{\muhlenberg}
\affiliation{\myongji}
\affiliation{\nagasaki}
\affiliation{\nara}
\affiliation{\natmephi}
\affiliation{\newmex}
\affiliation{\nmsu}
\affiliation{\ohio}
\affiliation{\ornl}
\affiliation{\orsay}
\affiliation{\peking}
\affiliation{\pnpi}
\affiliation{\riken}
\affiliation{\rikjrbrc}
\affiliation{\rikkyo}
\affiliation{\saispbstu}
\affiliation{\seoulnat}
\affiliation{\stonybrkc}
\affiliation{\stonycrkp}
\affiliation{\sungskku}
\affiliation{\tenn}
\affiliation{\titech}
\affiliation{\tsukuba}
\affiliation{\vandy}
\affiliation{\weizmann}
\affiliation{\wigner}
\affiliation{\yonsei}
\affiliation{\zagreb}
\author{A.~Adare} \affiliation{\colorado} 
\author{C.~Aidala} \affiliation{\losalamos} \affiliation{\michigan} 
\author{N.N.~Ajitanand} \affiliation{\stonybrkc} 
\author{Y.~Akiba} \email[PHENIX Spokesperson: ]{akiba@rcf.rhic.bnl.gov} \affiliation{\riken} \affiliation{\rikjrbrc} 
\author{R.~Akimoto} \affiliation{\cns} 
\author{J.~Alexander} \affiliation{\stonybrkc} 
\author{M.~Alfred} \affiliation{\howard} 
\author{V.~Andrieux} \affiliation{\michigan} 
\author{K.~Aoki} \affiliation{\kek} \affiliation{\riken} 
\author{N.~Apadula} \affiliation{\isu} \affiliation{\stonycrkp} 
% ...
\collaboration{PHENIX Collaboration} \noaffiliation

\date{\today}


\begin{abstract}


Dihadron and isolated direct photon-hadron angular correlations are 
measured in $p$$+$$p$ collisions at $\sqrt{s}=510$ GeV. Correlations of 
charged hadrons of $0.7<p_T<10$ GeV/$c$ with $\pi^0$ mesons of $4<p_T<15$ 
GeV/$c$ or isolated direct photons of $7<p_T<15$ GeV/$c$ are used to study 
nonperturbative effects generated by initial-state partonic transverse 
momentum and final-state transverse momentum from fragmentation. The 
nonperturbative behavior is characterized by measuring the out-of-plane 
transverse momentum component $p_{\rm out}$ perpendicular to the axis of 
the trigger particle, which is the high-$p_T$ direct photon or $\pi^0$. 
Nonperturbative evolution effects are extracted from Gaussian fits to the 
away-side inclusive-charged-hadron yields for different trigger-particle 
transverse momenta ($p_T^{\rm trig}$). The Gaussian widths and root mean 
square of $p_{\rm out}$ are reported as a function of the interaction hard 
scale $p_T^{\rm trig}$ to investigate possible 
transverse-momentum-dependent evolution differences between the 
$\pi^0$-h$^\pm$ and direct photon-h$^\pm$ correlations and factorization 
breaking effects. The widths are found to decrease with $p_T^{\rm trig}$, 
which indicates that the Collins-Soper-Sterman soft factor is not driving 
the evolution with the hard scale in nearly back-to-back dihadron and 
direct photon-hadron production in $p$$+$$p$ collisions. This behavior is 
in contrast to Drell-Yan and semi-inclusive deep-inelastic scattering 
measurements.

\end{abstract}

	

\maketitle


\section{\label{introduction}Introduction}

In the last two decades, the study of quantum chromodynamics (QCD) bound 
states has evolved from static, one-dimensional snapshots of quarks and 
gluons to focus on multidimensional structure and the dynamics of partons.  
The theoretical framework that has been developed to describe parton 
dynamics in hadrons involves transverse-momentum-dependent (TMD) parton 
distribution functions (PDFs) and fragmentation functions (FFs). In 
traditional collinear PDFs and FFs, any momentum of the partons transverse 
to the hadron boost axis is integrated over.  In TMD PDFs or FFs, the 
transverse momentum of the partons is not integrated out and instead 
remains explicit in the PDF or FF, offering a means of describing the 
transverse momentum distribution of unpolarized partons within an 
unpolarized hadron, as well as a variety of spin-momentum correlations 
when polarized hadrons and/or partons are considered.

Early theoretical work in TMD PDFs took place in the 1980s by Collins, 
Soper, and Sterman~\cite{CS:1981, CS:1982,css_evolution}, with extensive 
further development in the 1990s (see e.g.~\cite{Sivers:1990, Sivers:1991, 
Mulders-Tangerman:1995}). However, some theoretical details regarding the 
definition of TMD PDFs within a perturbative QCD (pQCD) framework have 
only been clarified in the last five years~\cite{Collins_Book}.  We note 
that due to confinement, the behavior of partons within hadrons is 
nonperturbative in that it cannot be calculated theoretically within pQCD.  
Collinear or TMD PDFs are nonperturbative functions that can be 
constrained by and/or used to predict high-energy scattering processes 
within a pQCD framework.  In such a framework, the nonperturbative 
functions such as PDFs as well as FFs factorize from the perturbatively 
calculable partonic hard scattering cross section and from each other.  
Lattice QCD offers an alternative, complementary approach to pQCD, 
performing numerical nonperturbative calculations directly.  In the past 
lattice QCD could only calculate moments of PDFs, integrated over parton 
collinear momentum fraction $x$ as well as parton transverse momentum. 
However, recent developments have demonstrated the potential to go beyond 
these limitations. These efforts are still in very early 
stages~\cite{Musch:lQCD,Musch:2011,Ji:lQCD}.

There is already experimental evidence from semi-inclusive deep-inelastic 
scattering (SIDIS) and Drell-Yan (DY) measurements that several TMD PDFs 
describing spin-momentum correlations are nonzero~\cite{hermes_tmds, 
compass_tmds,E866_DY,NA10_DY,E615_DY,STAR_W_An,JLAB_Collins_Sivers, 
JLAB_clas_asym}.  In addition, there is empirical evidence for nonzero 
spin-momentum correlations in the process of hadronization from 
electron-positron annihilation as well as 
SIDIS~\cite{BELLE_asymm,BABAR_asymm,compass_tmds,HERMES_collins, 
JLAB_Collins_Sivers}. Furthermore, transverse single-spin asymmetries up 
to $\sim$40\% have been measured in inclusive hadron production in 
hadronic collisions, indicating large nonperturbative spin-momentum 
correlations in these processes (see 
e.g.~\cite{eta_TSSA,STAR_pi0_TSSA,BRAHMs_TSSA}). However, these 
measurements cannot probe TMD functions directly because there is no 
simultaneous observation of perturbative and nonperturbative momentum 
scales.

The recent focus on multidimensional structure and parton dynamics has not 
only offered richer information on the behavior of partons confined within 
hadrons, but has moreover brought to light fundamental predictions 
regarding QCD as a nonAbelian gauge-invariant quantum field theory.  In 
particular, the role of color interactions due to soft gluon exchanges 
with the remnants of the hard scattering have become clearer. Because TMD 
functions preserve more nonperturbative information compared to collinear 
functions, TMD functions can differ from collinear ones with regards to 
universality and factorization. For example, the Sivers TMD 
PDF~\cite{Sivers:1990}, a correlation between the proton spin and quark 
transverse momentum, was shown to possibly be nonzero due to phase 
interference effects from soft gluon exchanges in 
SIDIS~\cite{BelitskyYuan,Brodsky_SIDIS}. Shortly afterward, 
Ref.~\cite{collins_sivers_prediction} showed that, due to the gauge 
invariant nature of QCD and the parity and time (PT) odd nature of the 
Sivers TMD PDF, the function should be the same magnitude but opposite in 
sign when measured in Drell-Yan vs. SIDIS processes because of the 
different color flows possible in the initial state vs. final state. 
Twist-2 TMD PDFs that involve one polarization vector are odd under PT 
transformations, leading to this predicted effect. The nonvanishing nature 
of the Sivers function has already been measured in polarized 
SIDIS~\cite{hermes_tmds}; there is not yet a measurement of this function 
in polarized Drell-Yan. A first indication from the Drell-Yan like W boson 
production exists~\cite{STAR_W_An}. The results favor a sign-change if TMD 
evolution effects are small, but at this stage the error bars are still 
large enough that a definitive statement can not be drawn from this single 
measurement. It is only for the TMD PDFs odd under PT transformations, 
where such sign-change behavior is expected, that gluon exchanges cannot 
be completely eliminated via a gauge transformation. 

In the more complicated QCD process \pp to hadrons, soft gluon exchanges 
in both the initial and final state are possible, leading to new predicted 
effects for observables sensitive to a small transverse momentum scale. In 
such processes, factorization breaking has been 
predicted~\cite{trogers_factbreaking,Mulders:2006,Collins:2007,Collins:2007preprint} 
in both polarized and unpolarized interactions. Here the nonperturbative 
objects in the cross section become correlated with one another and cannot 
be factorized into a convolution of TMD PDFs or TMD FFs. However, there 
are no theoretical claims that the perturbative partonic cross section 
does not factorize from the nonperturbative physics. Similarly to the case 
of the TMD PDFs that are odd under PT transformations, gluon exchanges 
that lead to the predicted factorization breaking cannot be eliminated via 
a gauge transformation. It is important to recognize that the ideas behind 
the predicted sign change of certain TMD PDFs and factorization breaking 
represent a major qualitative departure from previous purely perturbative 
approaches that do not account for soft gluon exchanges with remnants of 
the hard scattering. Possibly related effects known as ``color coherence" 
have been studied and observed in multijet states in hadronic 
collisions~\cite{CDF_cc,D0_cc,CMS_cc}, but these types of effects have not 
been rigorously treated in a TMD framework. 

In calculations of TMD processes where factorization is predicted to hold, 
the evolution with the hard scale of the interaction is known to be 
governed by the Collins-Soper (CS) evolution 
equation~\cite{CS:1981,CS:1982}. Note that the CS evolution equation comes 
directly out of the derivation of TMD 
factorization~\cite{collins_CSfact_preprint}. In contrast to the DGLAP 
collinear evolution equations~\cite{DGLAP_equations,DGLAP_2,DGLAP_3}, 
which are purely perturbative, the kernel for the CS evolution equation 
for TMD processes involves the Collins-Soper-Sterman (CSS) soft 
factor~\cite{css_evolution}, which generally contains nonperturbative 
contributions. The soft factor is understood to be strongly universal, the 
same for unpolarized and polarized processes, PDFs and FFs, with the only 
difference being between quarks and 
gluons~\cite{collinsrogers_tmdevolutionkernel}. Because lattice calculations 
of the soft factor are currently not possible, the soft factor must be 
extracted from parameterizations of experimental measurements within a 
pQCD framework. For a discussion of the CSS soft factor and TMD evolution 
phenomenology, see Ref.~\cite{collinsrogers_tmdevolutionkernel}. 

The theoretical expectation from CSS evolution is that any momentum width 
sensitive to nonperturbative $k_T$ would grow as the hard scale increases. 
This can be understood intuitively as a broadening of the phase space for 
gluon radiation with increasing hard scale. In addition this has been 
studied and observed in multiple phenomenological analyses of Drell-Yan 
and Z boson data~(see 
e.g.~\cite{Tevatron_Zboson_resum,Nadolsky_DYZ_globfit,Metz_intrin_kt}), as 
well as phenomenological analyses of SIDIS data, where factorization is 
also predicted to hold~(see e.g.~\cite{aidala_rogers_2014, 
SIDIS_evolution,Metz_intrin_kt}). As mentioned above, because the CS 
evolution equation comes directly out of the derivation of TMD 
factorization, it then follows that a promising avenue to investigate 
factorization breaking effects is by looking for qualitative differences 
from CSS evolution in processes where factorization breaking is expected, 
such as nearly back-to-back dihadron correlations produced in \pp 
collisions. 

\begin{figure}[thb]
 \includegraphics[width=1.0\linewidth]{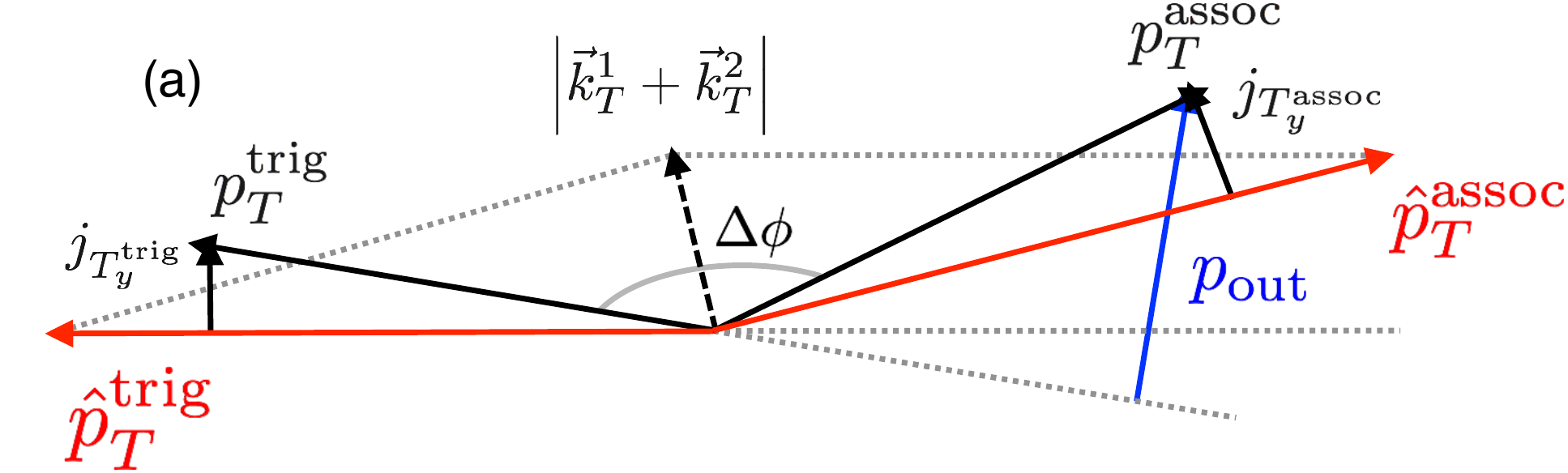} \\
 \includegraphics[width=1.0\linewidth]{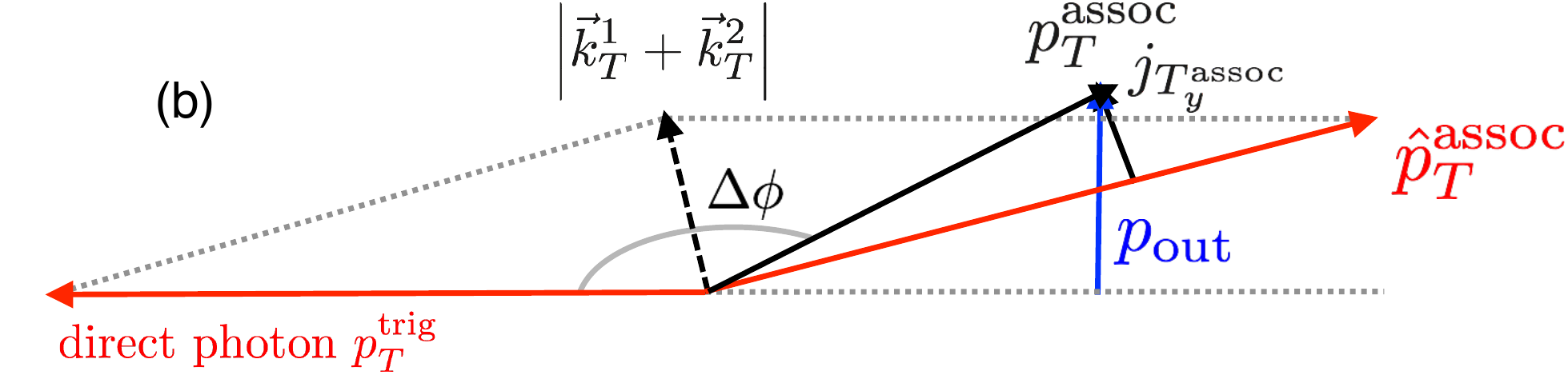}
 \caption{\label{fig:kTkinematics}
A diagram showing the hard-scattering kinematics of (a) dihadron and (b) 
direct photon-hadron event in the transverse plane. Two hard-scattered 
partons with transverse momenta $\hat{p}_T^{\rm trig}$ and $\hat{p}_T^{\rm 
assoc}$ [red lines] are acoplanar due to the initial-state $\vec{k}_T^1$ 
and $\vec{k}_T^2$ from each parton. These result in a trigger and 
associated jet fragment $\pttrig$ and $\ptassoc$ with a transverse 
momentum component perpendicular to the jet axis $j_{T^{\rm trig}_y}$ and 
$j_{T^{\rm assoc}_y}$ in the transverse plane, which are assumed to be 
Gaussian such that $\jt=\sqrt{2\langle j^{2}_{T^{\rm trig}_y}\rangle} 
=\sqrt{2\langle j^{2}_{T_y^{\rm assoc}}\rangle}$.  For direct photons (b) 
only one jet fragment $\ptassoc$ is produced because the direct photon is 
produced from the hard scattering. The quantity $\pout$ [blue] is the 
transverse momentum component of the away-side hadron perpendicular to the 
trigger particle axis.
}
 \end{figure}

To have sensitivity to possible factorization breaking and modified TMD 
evolution effects, a particular observable must be sensitive to a small 
scale on the order of $\Lambda_{QCD}$ and measured over a range of hard 
scales. Nearly back-to-back dihadron production has long been used as a 
proxy for measuring initial-state partonic transverse momentum 
$k_T$~\cite{ppg029,ppg089,e706_kt,CCOR_kt}, which is defined in 
Fig.~\ref{fig:kTkinematics}.  First used in predictions by Ref.~\cite{FFF} 
as a method for understanding large differences in hard scattering cross 
sections between theory and data, nearly back-to-back two-particle and 
dijet angular correlations have since been used to measure $k_T$ over a 
large range of center of mass 
energies~\cite{ppg029,ppg095,e706_kt,ALICE_dijet_kt}. Direct photon-hadron 
correlations are of particular interest because the photon comes directly 
from the partonic hard scattering, and thus carries initial-state 
information without any final-state fragmentation effects. The direct 
photon approximates the away-side jet energy at leading order (LO) while 
still being directly sensitive to the partonic transverse momentum scale. 
Direct photons also give an interesting comparison to dihadron production 
because they do not carry color charge, thus, assuming factorization 
holds, only two TMD PDFs and one TMD FF are necessary in the cross section 
calculation compared to two TMD PDFs and two TMD FFs in dihadron 
production. Therefore, there should be more avenues for gluon exchange in 
nearly back-to-back dihadron events when compared to direct photon-hadron 
events.

Figure~\ref{fig:kTkinematics} shows the hard scattering kinematics of a 
nearly back-to-back dihadron event in the transverse plane. The effect of 
initial-state $k_T$ and final-state $j_T$, the transverse momentum of the 
hadron with respect to the jet axis, can be probed in hadronic collisions 
by measuring the out-of-plane momentum component $\pout$ with respect to 
the near-side hadron or direct photon, collectively referred to as the 
trigger particle. $\pout$ thus quantifies the acoplanarity of the 
two-particle pair, with $\pout=0$ signifying exactly back-to-back particle 
production. Using the trigger particle as a proxy for the jet, the 
1-dimensional quantity $\pout$ is transverse to the \pt of the trigger 
particle, $\pttrig$, and has a magnitude of:

\begin{equation}\label{eq:pout}
	\pout = \ptassoc\sin\dphi
\end{equation}

\noindent where $\ptassoc$ is the \pt of the associated hadron and 
$\dphi$ is the azimuthal angular separation between the trigger and 
associated particle as shown in Fig.~\ref{fig:kTkinematics}. 
Reference~\cite{ppg029} has shown that the root mean square of $\pout$ and 
$k_T$ are related by

\begin{equation} \label{eq:kteqn}
\frac{\langle z_T\rangle\sqrt{\langle k_T^2\rangle}}{\hat{x}_h} = \frac{1}{x_h}\sqrt{\langle\pout^2\rangle-\langle j_{T_y}^2\rangle(1+x_h^2)}
\end{equation}

\noindent where $\langle z_T\rangle=\pttrig/\hat{p}_T^{\rm trig}$ and 
$x_h=\langle \ptassoc\rangle/\langle\pttrig\rangle$, and quantities with a 
hat indicate partonic-level quantities. Note that in the determination of Eq.~\ref{eq:kteqn}, it was assumed in Ref.~\cite{ppg029} that the component $j_{T_y}$ for both the trigger and associated jet axes was sampled from the same Gaussian distribution of $\jt$. All the quantities on the left 
side of Eq.~\ref{eq:kteqn} are partonic, while those on the right side can 
be measured via the correlated away-side hadron. Equation~\ref{eq:kteqn} 
gives a clear definition for how to relate the root mean square 
initial-state $k_T$ and final-state $j_T$ to the observable $\pout$.

The Relativistic Heavy Ion Collider (RHIC) is an ideal facility to study 
nonperturbative factorization breaking effects because they are only 
predicted in hadronic collisions where at least one final-state hadron is 
measured, and the measurement has sensitivity to a small initial- and 
final-state transverse momentum scale. Observables of interest are final 
states where at least one particle has a large \pt, defining a hard scale, 
at least one final-state hadron is measured, and the observable is also 
sensitive to initial- and final-state $k_T$ and $j_T$. At RHIC energies, 
the $p_T$ reach for direct photons and pions is sufficiently large to have 
separation from the nonperturbative momentum scale. Direct photon-hadron 
and $\pion$-hadron correlations were chosen specifically because of 
experimental capabilities and because of the differing number of 
final-state hadrons in the event; the $\pion$-hadron correlations probe an 
extra nonperturbative function, assuming factorization, and thus one more 
Gaussian $j_T$ convolution than the direct photon-hadron correlations.


\section{\label{experiment_details}Experiment Details}

In 2012 and 2013 the PHENIX experiment collected data from $\pp$ 
collisions at $\sqs=510$ GeV. After data quality and vertex cuts, 
integrated luminosities of approximately 30 pb$^{-1}$ in 2012 and 152 
pb$^{-1}$ in 2013 were used for the analysis of dihadron and direct 
photon-hadron correlated pairs. The measured $\pout$ distributions 
presented here are at a higher center of mass energy and have 
significantly reduced statistical uncertainties compared to~\cite{ppg095}. 
The higher center of mass energy also allows the probing of smaller $x$ 
values of the TMD PDFs. Additionally, because the focus of this work is 
identifying possible nonperturbative factorization breaking effects, one 
of the observables presented here specifically isolates effects from 
nonperturbative $k_T$ and $j_T$, extending previous measurements which 
only observed effects sensitive to both perturbative and nonperturbative 
contributions.

\begin{figure}[thb]
 \includegraphics[width=0.98\linewidth]{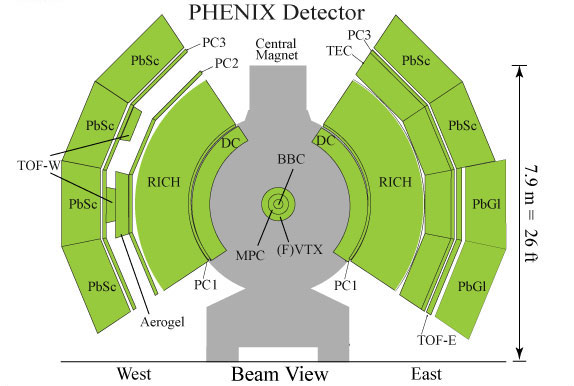}  
 \caption{\label{fig:phenix}
Cross section view along the beam line of the PHENIX detector, showing the 
detectors composing the central arms in 2012 and 2013. The relevant 
subsystems for this analysis are described in the text.
 }
 \end{figure}

The PHENIX detector can measure two-particle correlations between photons 
and hadrons with its electromagnetic calorimeter (EMCal) and drift chamber 
(DC) plus pad chamber (PC) tracking system located in two central arms. 
The central arms are nearly back-to-back in azimuth, with each arm 
covering approximately $\pi/2$ radians in azimuthal angle and 0.7 units of 
pseudorapidity about midrapidity~\cite{phenix_central_arms}. A schematic 
showing the two central arms is shown in Fig.~\ref{fig:phenix}. 

The EMCal~\cite{EMCal} is located at a radial distance of approximately 5 
meters from the beam pipe and is composed of 8 sectors, 4 in each arm. Six 
sectors are lead-scintillator (PbSc) sampling calorimeters, and the other 
two are lead glass (PbGl) $\check{\rm C}$erenkov calorimeters. The PbSc 
and PbGl calorimeters measure electromagnetic showers with intrinsic 
resolution $\sigma_E/E = 2.1\%\oplus8.1\%/\sqrt{E}$ and 
$0.8\%\oplus5.9\%/\sqrt{E}$, respectively.  High energy photons are 
identified with a cluster shower shape cut and charged particle veto. The 
shower shape cut also removes most high energy photons that overlap too 
closely with another photon, which helps eliminate $\pion$ merging effects 
at energies greater than $\sim$12 GeV in the PbSc and $\sim$ 17 GeV in the 
PbGl. The granularity of the EMCal is 
$\Delta\eta\times\Delta\phi\sim0.011\times0.011$ for PbSc and 
$0.008\times0.008$ for PbGl, where $\Delta\eta$ and $\Delta\phi$ refer to 
the pseudorapidity and azimuthal angular segmentation, respectively. The 
high granularity of the EMCal along with the shower shape cut allows for 
$\pion$ and $\eta$ reconstruction via the diphoton channel up to $\pt\sim$ 
17 GeV.  Previous direct-photon, $\eta$, and $\pion$ cross sections 
measured in the PHENIX central arm can be found 
in~\cite{directphoton_crosssection,ppg186,ppg107}.

The $\pion$ and $\eta$ mesons are tagged in the EMCal via their two-photon 
decay for the purposes of removing decay photon background to identify 
direct photons and constructing the $\pion$-h$^\pm$ correlated pairs. To 
reduce the combinatorial background, only photons with energy greater than 
$1$ GeV are considered.  The invariant mass windows were 120--160 
MeV/$c^2$ for $\pion$ and 500--600 MeV/$c^2$ for $\eta$ mesons.

The PHENIX tracking system~\cite{DC} allows charged hadron detection via a 
drift chamber (DC) in each central arm along with two pad chambers (PC) 
directly behind the drift chambers. The momentum resolution was determined 
to be $\delta p/p = 0.7\%\oplus1.0\%p$ with $p$ in GeV/$c$. Tracks are 
identified via the DC, covering a radial distance of $2.02<r<2.49$ meters 
from the beam pipe. Secondary tracks from decays or conversions are 
reduced by a condition that matches tracks in the DC to hits in the 
outermost PC3, located at a radial distance of 4.98 meters from the beam 
pipe. The charged particle veto suppresses hadronic showers in the EMCal 
by matching tracks from the full tracking system to clusters in the EMCal.


\section{\label{statistical_subtraction}Correlation Functions}

The correlation functions are constructed following the methods of 
Refs.~\cite{ppg029,ppg089,ppg095}.  The number of correlated hadrons per 
trigger particle is referred to as the per trigger yield, and is collected 
for the different types of trigger particle-associated hadron pairs. To 
quantify the inefficiencies of the PHENIX detector, the hadron yields are 
corrected by a charged hadron efficiency determined from a Monte Carlo 
single-particle generator and a {\sc geant}-based simulation of the detector. 
Additionally, due to the limited acceptance of the PHENIX detector, the 
per-trigger yields are divided by a mixed event distribution. Mixed event 
distributions are collected on a run-by-run basis to quantify any changing 
inefficiencies with time in the acceptance. The collected trigger particle 
is mixed with charged hadrons from different events and a mixed event 
correlation function is constructed to correct for the acceptance of the 
detector. In total the full correlation function is determined by the 
following equation

\begin{equation}\label{eq:corrfunceq}
\frac{1}{N_{\rm trig}}\frac{dN}{d\dphi} = \frac{1}{N_{\rm trig}}\frac{dN/d\dphi_{\rm raw}}{dN/d\dphi_{\rm mixed}\epsilon(\pt)}
\end{equation}

\noindent where $\epsilon(\pt)$ is the hadron efficiency described above. 
Note that this definition is general for any observable that could be 
constructed in a two-particle correlation, so it applies to the 
determination of the $\pout$ distributions also. For a complete 
description of two-particle correlation analyses in the PHENIX central 
arms, see Refs.~\cite{ppg029,ppg089,ppg090,ppg095}.

\subsection{Statistical Subtraction of Decay Photons}

To identify direct photons, Ref.~\cite{ppg090} used a method that is based 
upon identifying a total sample of inclusive per-trigger yield 
correlations, then subtracting the decay component. From 
Ref.~\cite{ppg090}, the yield of charged hadrons per direct photon was 
determined with the following equation

\begin{equation}\label{eq:statsubrgamma}
Y_{\rm direct} = \frac{1}{R_\gamma-1}(R_\gamma Y_{\rm inclusive}-Y_{\rm decay})
\end{equation}

\noindent Here $Y$ is the per-trigger yield where the trigger particle for 
each per-trigger yield is indicated as direct, inclusive, or decay, and 
$R_\gamma$ is the relative contribution of direct photons to decay photons 
such that $R_\gamma=N_{\rm inclusive}/N_{\rm decay}$. The total yield of 
photons, the inclusive photons, comes from adding all of the decay and 
direct photons, $N_{\rm inclusive} = N_{\rm direct}+N_{\rm decay}$. In 
Ref.~\cite{ppg090} direct photons are defined as any photon not from a 
decay process, which includes next-to-leading order (NLO) photons that 
emerge from parton-to-photon fragmentation.  

To eliminate the presence of NLO fragmentation photons, Ref.~\cite{ppg095} 
implemented isolation and tagging cuts; thus Eq.~\ref{eq:statsubrgamma} 
was modified to include these cuts. To determine the per-trigger yield of 
isolated direct photons, the number of isolated decay photons was 
subtracted from the isolated inclusive photon sample, where 
$N^{\rm iso}_{\rm inclusive}=N_{\rm decay}^{\rm iso}
+N_{\rm direct}^{\rm iso}$. The subtraction procedure results in the 
following equation for per-trigger yields of isolated photon 
quantities~\cite{ppg095}

\begin{equation}\label{eq:statsubeqn} 
Y_{\rm direct}^{\rm iso}=\frac{1}{\Rprime-1}\left(\Rprime 
Y_{\rm inclusive}^{\rm iso}-Y_{\rm decay}^{\rm iso}\right) 
\end{equation} 

\noindent where the trigger particles are noted as direct, inclusive, or 
decay for a given per-trigger yield $Y$ and ``iso" refers to ``isolated". 
$\Rprime$ is the relative contribution of isolated direct and decay 
photons, where $\Rprime=N_{\rm inclusive}^{\rm iso}/N_{\rm decay}^{\rm 
iso}$ and indicates isolated direct photon production for $\Rprime>1$. The 
subtraction procedure eliminates remaining background due to isolated 
decay photons that appear direct, which are due most often to asymmetric 
$\pion\rightarrow\gamma\gamma$ decays where the low \pt photon is not 
detected.

To suppress sources of background photons, tagging and isolation cuts are 
implemented at the event-by-event level. To reduce the contribution from 
decay photons, candidate inclusive photons are tagged and removed if a 
partner photon of $\pt>1$ GeV is found such that the invariant mass of the 
pair falls within the regions of 118--162 or 500--600 MeV/$c^2$. The tagging 
cuts use a larger $\pion$ invariant mass range than for identifying 
$\pion$ for dihadron correlations to err on the side of removing more 
decay photons. An isolation cut further suppresses decay photons as well 
as NLO fragmentation photons by requiring that the sum of the EMCal energy 
deposits and \pt of charged tracks within a radius of 0.4 radians around 
the candidate photon be less than 10\% of the photon's total energy. To 
reduce the impact of detector acceptance effects, photons that pass the 
isolation and tagging cuts are also required to be $\sim$0.1 radians from 
the edge of the detector in both $\eta$ and $\phi$ forcing a large portion 
of the isolation cone to fall inside the PHENIX acceptance. 

Because the number of isolated decay photons is not \textit{a priori} known, 
the decay photon per-trigger yield is determined with a probability 
density function. Isolated $\pion$-h$^\pm$ correlated pairs are weighted 
by a probability density function to map these per-trigger yields to the 
isolated decay photon hadron correlated per-trigger yields. This function, 
determined in Ref.~\cite{ppg095}, gives the probability of an isolated 
$\pion$ with $p_T^{\pion}$ to decay to a photon with $p_T^\gamma$ in the 
PHENIX acceptance where the photon was unable to be tagged as a decay 
photon. In the PHENIX central arms, the inability to tag a decay photon 
happens most often from asymmetric $\pion$ decays, where one photon misses 
the detector completely. A 4\% systematic uncertainty was assigned to the 
decay photon statistical subtraction method as a whole, which includes not 
considering backgrounds due to higher mass states such as the $\eta$, 
$\omega$, and $\rho$. To determine the per-trigger yield of isolated decay 
pairs, the number of isolated $\pion$s is mapped via the probability 
function to the number of isolated decay pairs in a given \pt bin. The 
per-trigger yield of isolated decay pairs is then

\begin{equation}
\decpty = \frac{\sum_{N_{\pi}^{\rm iso}}\mapfxn N_{\pi-h}^{\rm iso}}{\sum_{N_{\pi}^{\rm iso}}\mapfxn N_{\pi}^{\rm iso}}
\end{equation}

\noindent where $\mapfxn$ is the probability density function described 
above and contains all of the dependence and efficiencies of the detector 
on $p_T^{\pion}$ and $p_T^{\gamma}$.  The $N_{\pi}^{\rm iso}$ and 
$N_{\pi-h}^{\rm iso}$ are simply the number of isolated $\pion$-trigger 
particles measured and the number of isolated $\pion$-h$^\pm$ pairs 
measured, respectively.

The $\Rprime$ is determined by measuring $R_\gamma$ and correcting 
$R_\gamma$ with tagging and isolation efficiencies. Because the quantity is 
the ratio of the inclusive photons to decay photons after tagging and 
isolation cuts, it can be written as

\begin{eqnarray}
\Rprime &=& \frac{N_{\rm inclusive}^{\rm iso}}{N_{\rm decay}^{\rm iso}} \\ \nonumber
&=& \frac{N_{\rm inclusive}-N_{\rm decay}^{\rm tag}
-N_{\rm inclusive}^{\rm niso}}{N_{\rm decay}-N^{\rm tag}_{\rm decay}
-N^{\rm niso}_{\rm decay}} \\ \nonumber
&=& \frac{R_\gamma}{(1-\epsilon_{\rm decay}^{\rm tag})
(1-\epsilon_{\rm decay}^{\rm niso})}\frac{N_{\rm inclusive}
-N_{\rm decay}^{\rm tag}-N_{\rm inclusive}^{\rm niso}}{N_{\rm inclusive}}
\label{eq:rprimeequation}
\end{eqnarray}

\noindent where ``niso'' refers to ``not isolated.'' Because the tagging cuts 
are applied before the isolation cut, $N_{\rm decay}^{\rm tag}$ is the number 
of photons tagged as decay photons regardless of the isolation cut, 
while $N_{\rm inclusive}^{\rm niso}$ is the number of not isolated photons 
that were not able to be tagged. 
$\Rprime$ is now written in terms of values that can be measured. 
$R_\gamma$ and the tagging efficiency $\epsilon_{\rm decay}^{\rm tag} = 
N_{\rm dec}^{\rm tag}R_\gamma/N_{\rm inclusive}$ can be determined without 
the probability function because these quantities do not depend on possible 
isolated decay photons. The right-most fraction in Eq.~7 is simply the 
number of photons that pass the isolation and tagging cuts divided by the 
total number of inclusive photons and can be determined by counting the 
number of photons that pass the described cuts. The efficiency with which 
the isolation cut removes decay photons $\epsilon_{\rm decay}^{\rm niso}$ 
is determined by applying the probability function at the level of the 
isolated parent meson and mapping the effect to the daughter photon

\begin{equation}
\epsilon_{\rm decay}^{\rm niso} =\left(1+\frac{\sum_\pi 
P(p_T^\pi,p_T^\gamma)\cdot N^{\rm iso}_\pi}{\sum_\pi 
P(p_T^\pi,p_T^\gamma)\cdot N^{\rm niso}_\pi}\right)^{-1}
\label{eq:epsilonisoeq}
\end{equation}

\begin{figure}[tbh]
 	\includegraphics[width=1.0\linewidth]{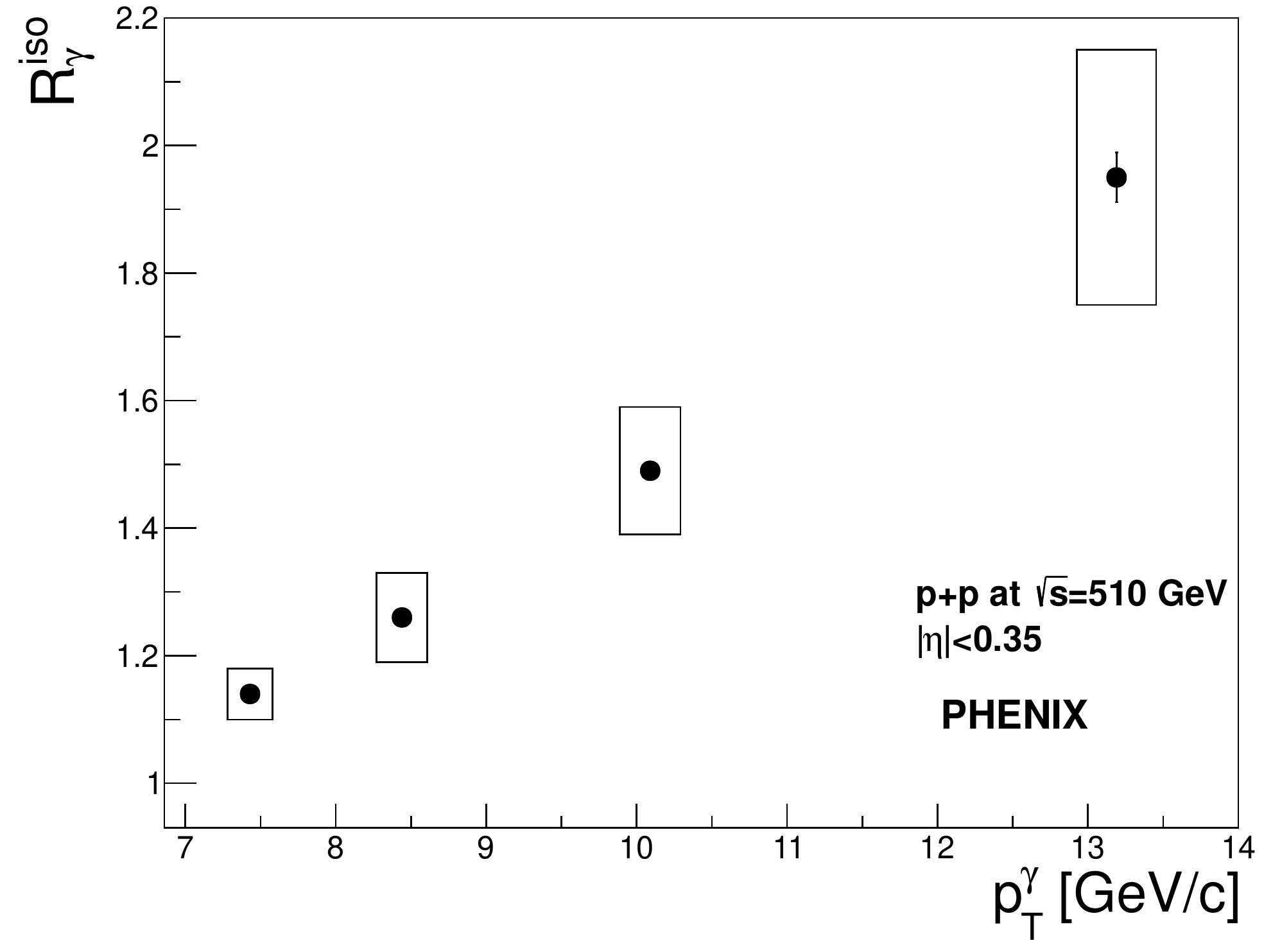}  
 	\caption{\label{fig:rgammaprime}
Measured $\Rprime$ for use in the statistical subtraction, 
Eq.~\ref{eq:statsubeqn}. The boxes quantify the systematic uncertainty.
 	}
 \end{figure}

Each of the quantities for determining $\Rprime$ is found by counting the 
number of photons that pass the various cuts except for the isolation 
efficiency $\epsilon^{\rm niso}_{\rm decay}$, which is found by measuring 
the number of isolated and not isolated $\pion$ that pass the cuts and 
weighting by the probability function as in Eq.~\ref{eq:epsilonisoeq}. 
$R_\gamma$ was found by dividing the number of inclusive photons $N_{\rm 
inclusive}$ by the number of decay photons $N_{\rm decay}$; $N_{\rm 
decay}$ was determined by counting the number of photons tagged from 
$\pion$ decays and correcting for higher mass states and the PHENIX single 
and diphoton detection efficiencies derived from a {\sc geant}-based 
simulation. $\epsilon_{\rm decay}^{\rm tag}$ can then be calculated with 
$R_\gamma$ and the number of tagged decay photons $N_{\rm dec}^{\rm tag}$ 
and inclusive photons $N_{\rm inclusive}$. As a cross check, systematic 
uncertainties on $R_\gamma$ were evaluated using the direct photon and 
$\pion$ pQCD cross sections with the CT10 PDFs~\cite{CT10_PDFs} and DSS14 
FFs~\cite{DSS14_FFs}.  The tagging efficiency is 0.36--0.43 and the 
isolation efficiency is 0.61--0.69 from the lowest to highest $\pttrig$ 
bins. Note that each quantity in Eq.~7 is dependent only on $\pttrig$. 
Figure~\ref{fig:rgammaprime} shows the values of $\Rprime$ as a function 
of $p_T^\gamma$; for which values greater than unity indicate isolated 
direct-photon production.

\section{\label{results}Results}

\subsection{\label{dphicorrelations}Azimuthal Correlations}

\begin{figure*}[tbh]
 	
\includegraphics[width=0.99\linewidth]{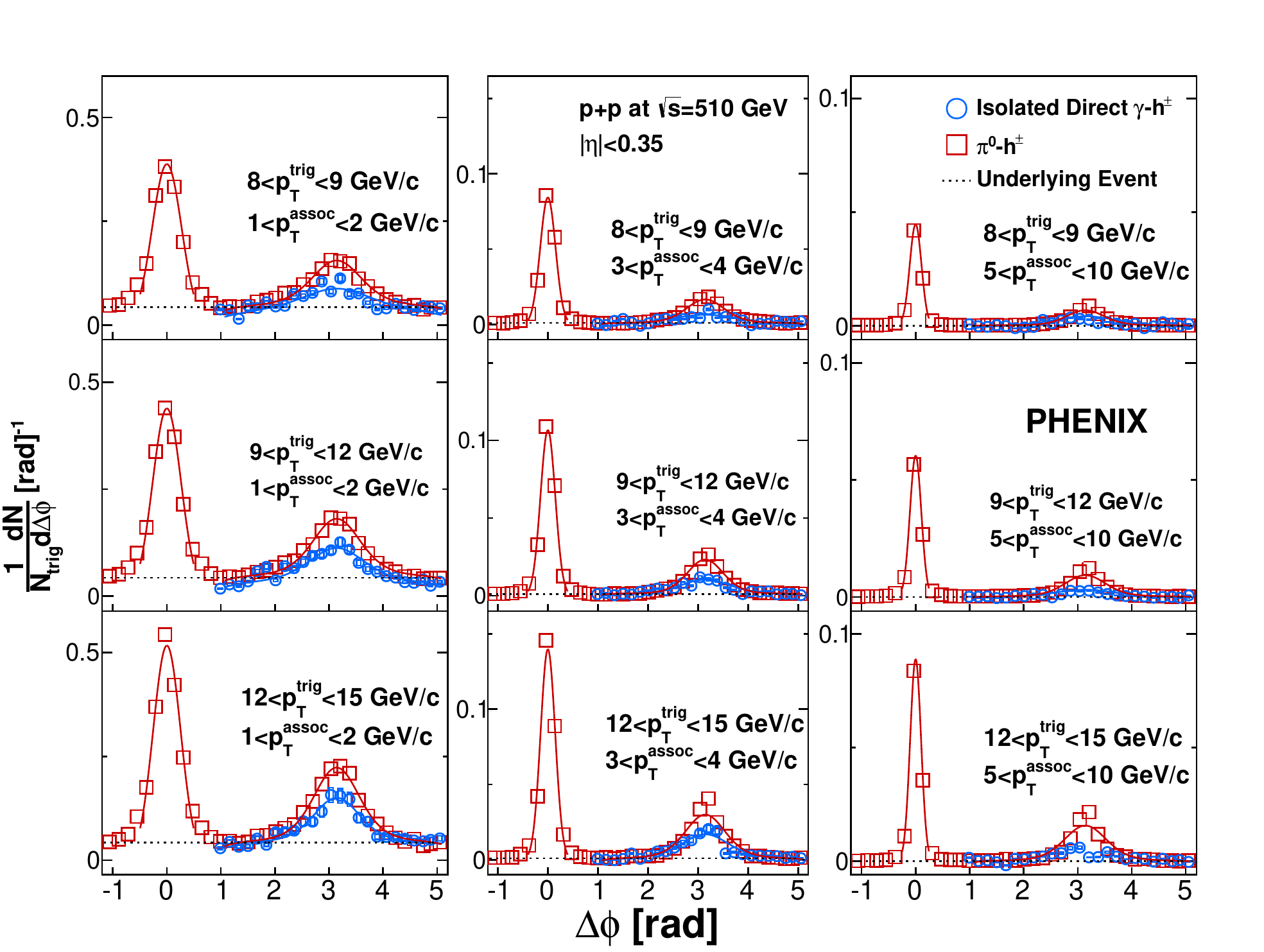}  
 	\caption{\label{fig:dphicorrelations}
Per-trigger yield of charged hadrons shown as a function of the 
azimuthal angle between the $\pion$ or direct photon trigger particle 
and associated charged hadron. The dashed [black] lines show an estimate 
of the underlying event yield and are drawn to guide the eye in 
distinguishing the underlying event from the away-side jet. The solid 
lines through the open squares [red] and open circles [blue] are fits to 
extract the widths of the near and away sides.  A 9\% overall 
normalization uncertainty on the charged hadron yields is not shown in 
the figure.
 	}
 \end{figure*}

Figure~\ref{fig:dphicorrelations} shows a few examples of the per-trigger 
yields of associated charged hadrons for both $\pion$ and direct photon 
triggers as a function of $\dphi$ in bins of $\pttrig$ and $\ptassoc$. The 
azimuthal correlations as a function of $\dphi$ show standard jet 
structure characteristics. The $\pion$ yields have clear peaks at both 
$\dphi=0$ and $\dphi=\pi$, indicating nearly back-to-back jet production. 
The near-side yields of the isolated direct photons are not plotted, 
similarly to Ref.~\cite{cms_gammajet}, because the yields within the 
isolation cone are physically uninterpretable. In addition, the effect of 
$k_T$ smearing is characterized with the away-side peaks. The near-side 
$\pion$ peaks are larger than the away sides due to the effect of so 
called ``trigger bias," discussed in Ref.~\cite{ppg029}. The away-side 
yields of the direct photons are smaller than those from the $\pion$ 
triggers due to the smaller jet energy sampled by direct photons; the 
$\pion$ has some fractional energy $z_{\pion} E_{\rm jet}$ where 
$z_{\pion}$ refers to the momentum fraction of the $\pion$ from the 
scattered parton and $E_{\rm jet}$ is the energy of the jet, whereas the 
direct photon approximates the away-side jet energy at LO. The underlying 
event levels for $\pion$ and direct photon triggers are similar, which 
would be expected if the underlying event structure is completely 
uncorrelated from the partonic hard scattering. A 9\% normalization 
uncertainty on the charged hadron yields is not shown on this figure or 
any of the following per-trigger yields. This uncertainty is of similar 
magnitude to~\cite{ppg095} and is largely due to the uncertainty when 
matching tracks from the DC to the PC. All of the per-trigger yields as a 
function of $\dphi$ can be found in the Supplemental 
Material~\cite{supp_matt}.


\subsection{\label{jt}$\jt$ Determination}

\begin{figure}[tbh]
 	\includegraphics[width=1.0\linewidth]{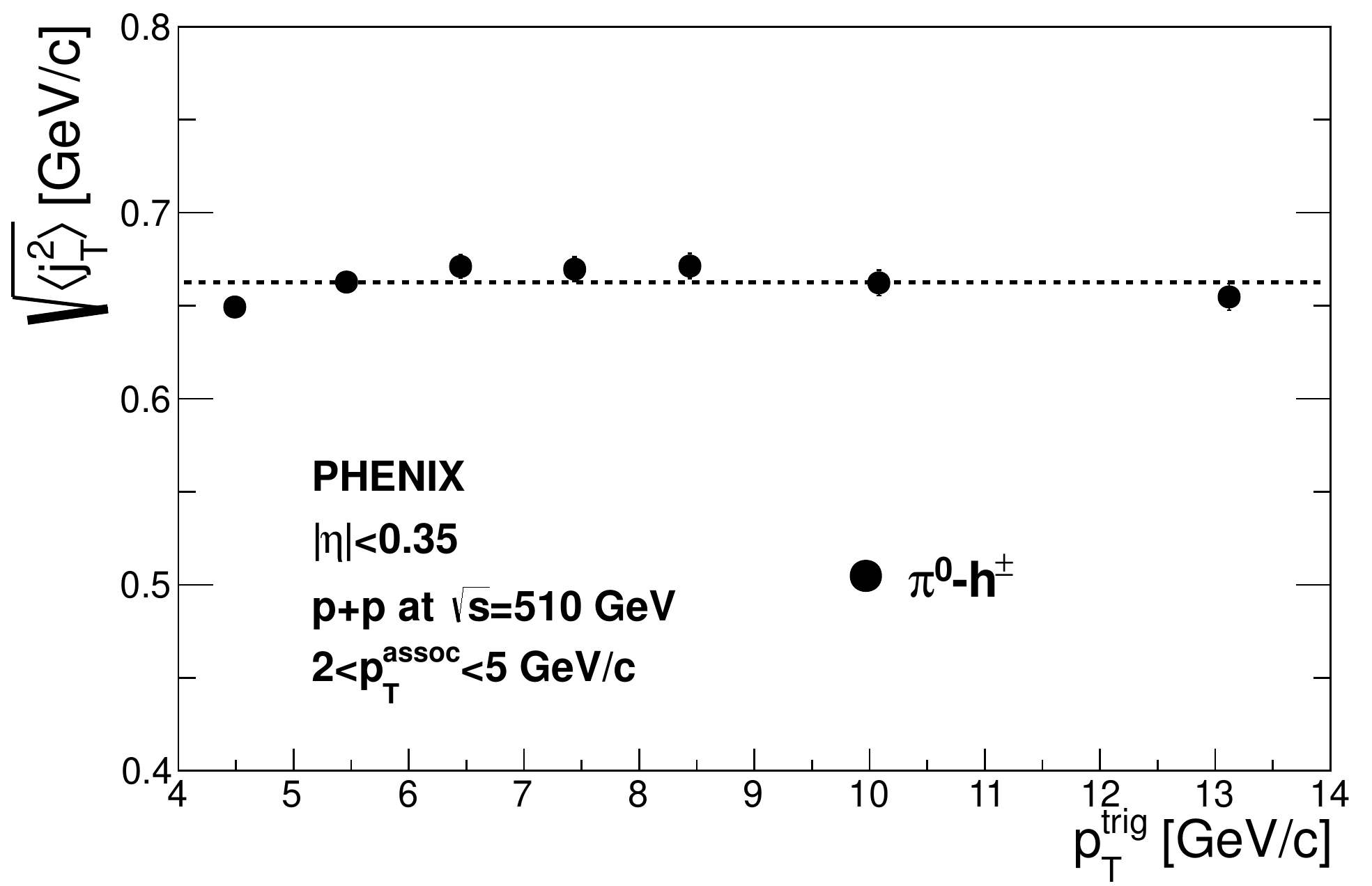}  
 	\caption{\label{fig:jts}
The $\jt$ shown as a function of $\pttrig$ is determined with 
Eq.~\ref{eq:jteqn} and has been shown to be approximately constant with 
$\sqs$ and \pt in the limited \pt range examined here. The line shows 
a constant fit to average over $\pttrig$.
 	}
 \end{figure}

The value $\jt$ was determined by the widths of Gaussian fits to the 
near-side of the $\pion$ correlation functions, similarly to 
Ref.~\cite{ppg029}. Examples of the fits are shown on the near-side 
$\pion$ peaks in Fig.~\ref{fig:dphicorrelations}. Values of $\jt$ were 
calculated with the following equation
\begin{equation} 
\jt =\sqrt{2\langle j_{T_y}^2\rangle}\simeq 
\sqrt{2}\frac{\pttrig\ptassoc}{\sqrt{p_{T}^{\rm trig^{2}}
+p_T^{\rm assoc^{2}}}}\sigma_N
\label{eq:jteqn}
\end{equation}

\noindent where $\sigma_N$ is the Gaussian width. Previous measurements 
have shown $\jt$ to be approximately constant with $\sqs$ and $\pttrig$ in 
a similar $\pttrig$ range to that examined 
here~\cite{ppg029,ppg089,CCOR_kt, jt_cern}. Only bins $\ptassoc>2$ GeV/$c$ 
were used to satisfy the assumption $\ptassoc\gg\sqrt{2}~j_{T}$ which was 
made to determine Eq.~\ref{eq:jteqn}. Each $\pttrig$ bin was fit to a 
constant and averaged $\jt$ over $\ptassoc$. Figure~\ref{fig:jts} shows 
the results, which were then fit with a constant to average over 
$\pttrig$, which is shown as a dotted line in Fig.~\ref{fig:jts}.  After 
averaging, $\jt$ was determined to be

\begin{equation}
\jt = 0.662\pm0.003\text({\rm stat})\pm0.012({\rm sys}) {\rm GeV}/c
\end{equation}

\noindent where the systematic uncertainty is due to the momentum 
resolution of the detector as well as approximations made to determine 
Eq.~\ref{eq:jteqn} in Ref.~\cite{ppg029}. Recent ATLAS results show a 
similar fragmentation variable over a significantly larger range of 
hundreds of GeV/$c$ in \pt, and show that the average transverse 
momentum with respect to the jet axis rises slowly with $p_T^{\rm jet}$ 
over this significantly larger \pt range~\cite{atlas_fragfunc}.

\subsection{\label{rmspoutresults}$\rmspout$ Determination}

The quantity $\rmspout$ was extracted from the $\dphi$ correlations as was 
done in previous measurements\footnote{We note that the fit function used 
here has the $\sqrt{2}$ in the error function in the denominator, not the 
numerator as was done in the previous references. In order for the 
normalization of the function to be unity across the range 
$[\pi/2,3\pi/2]$ this $\sqrt{2}$ should be in the denominator of the error 
function. We have studied the effect of this change and find that it does 
not change the value of $\rmspout$ extracted. This is because, as Fig. 31 
of Ref.~\cite{ppg029} shows, the quantity $\rmspout$ is determined from 
the exponential component of the fit function. The yield parameter 
extracted in Ref.~\cite{ppg029} changes slightly, but within the quoted 
systematic uncertainties.}~\cite{ppg029,ppg089,ppg095}. The value of 
$\rmspout$ quantifies the width of the away-side jet. The correlation 
functions are fit in bins of $\pttrig$ and $\ptassoc$ with the following 
function in the range $\pi/3<\dphi<5\pi/3$:

\begin{equation}\label{eq:rmspouteq}
\frac{dN}{d\dphi} = C_0+C_1\cdot\frac{dN_{\rm far}}{d\dphi}
\end{equation}
with

\begin{equation}
\frac{dN_{\rm far}}{d\dphi} =
 \begin{cases} 
 	\hfil 
 0 & |\dphi-\pi|>\frac{\pi}{2}\\
         \\
	 \frac{-\ptassoc\cos\dphi}{\sqrt{2\pi\langle p_{\rm out}^2\rangle}\textrm{Erf}(\ptassoc/\sqrt{2\langle p_{\rm out}^2})} \nonumber \\
	 \times \textrm{exp}\left(-\frac{|\ptassoc|^2\sin^2\dphi}{2\langle\pout^2\rangle}\right )
	 
	  & |\dphi-\pi|\leq\frac{\pi}{2}, \\
   \end{cases}
\end{equation}

\noindent where the parameters $C_0$, $C_1$, and $\rmspout$ are left as 
free parameters, with $C_0$ quantifying the underlying event, $C_1$ a 
normalization constant, and $\rmspout$ the parameter of interest. The fit 
extends to $\pi/3$ and $5\pi/3$ in order to accurately quantify the 
underlying event. Example fits are drawn on the correlation functions in 
Fig.~\ref{fig:dphicorrelations}. Systematic uncertainties were evaluated 
by altering the fit region by $\pm$0.2 radians and taking the absolute 
value of the difference of the resulting $\rmspout$. The systematic 
uncertainties for the direct photons are larger due to the increased 
fluctuations in the underlying event due to the statistical subtraction 
technique. 

 \begin{figure}[tbh]
 	\includegraphics[width=1.0\linewidth]{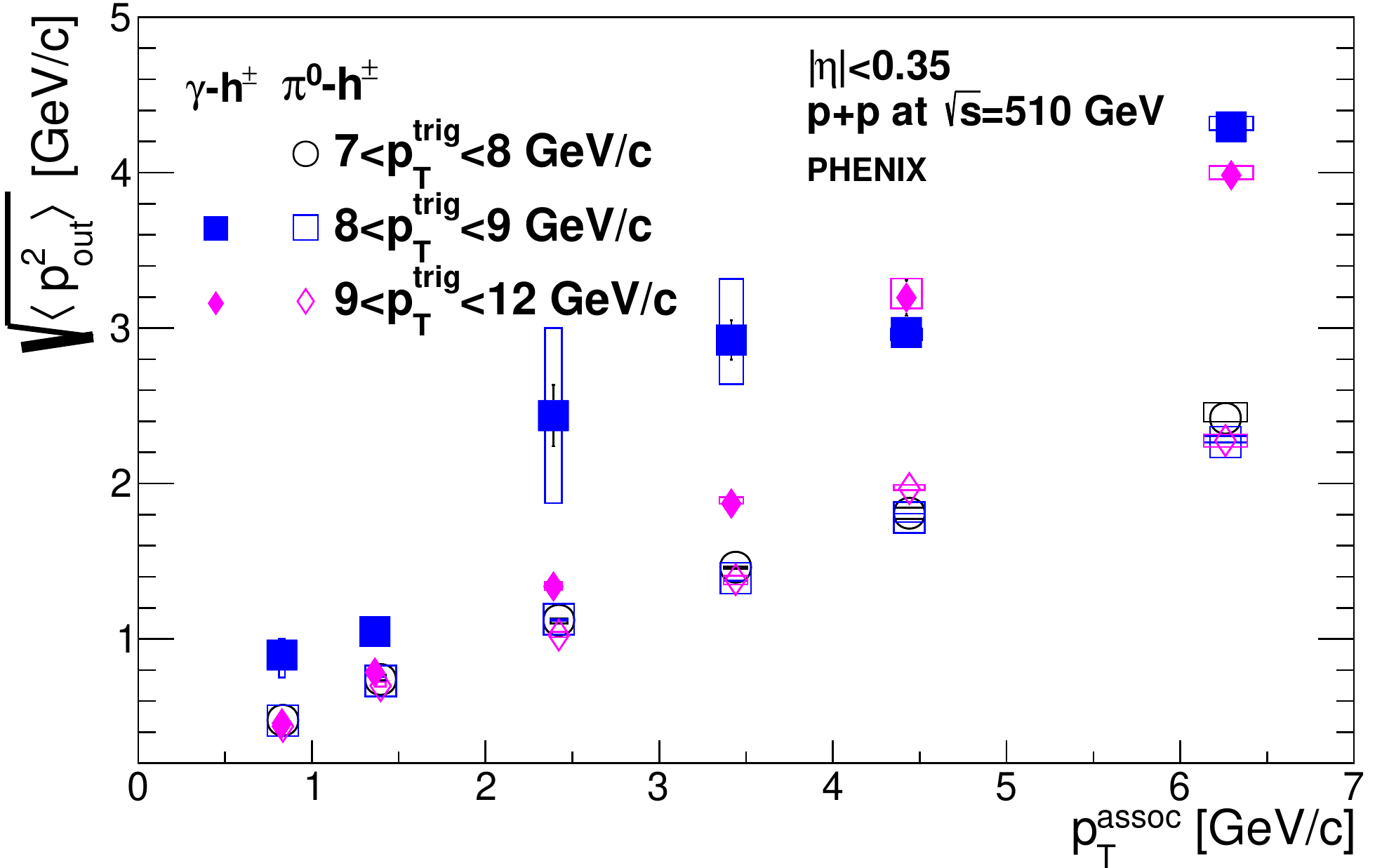}  
 	\caption{\label{fig:rmspoutvsptassoc}
$\rmspout$ for $\pion$ and direct photon triggers as a function of 
$\ptassoc$ in several $\pttrig$ bins.
 	}
 \end{figure}
 
Figure~\ref{fig:rmspoutvsptassoc} shows $\rmspout$ as a function of 
$\ptassoc$ in several $\pttrig$ bins for both $\pion$ and direct photon 
triggers.  All of the $\rmspout$ values can be found in Tables I and II of 
the Supplemental Material~\cite{supp_matt}. In the following figures 
showing measured quantities, filled points are for isolated direct photons 
and open points are for $\pion$ triggers. Both $\rmspout$ distributions 
for $\pion$ and direct photon triggers show a clear dependence on 
$\ptassoc$, with the direct photons having a stronger dependence. The 
direct photon $\rmspout$ quantities are larger due to the smaller jet 
energy being sampled compared to the $\pion$ jet energies. The strong 
dependence of $\rmspout$ on $\ptassoc$ is anticipated as could be 
ascertained from the definition of $\pout=\ptassoc\sin\dphi$. In the same 
$\dphi$ region, as $\ptassoc$ gets larger, $\pout$ will also get larger.

\begin{figure}[tbh]
 	\includegraphics[width=1.0\linewidth]{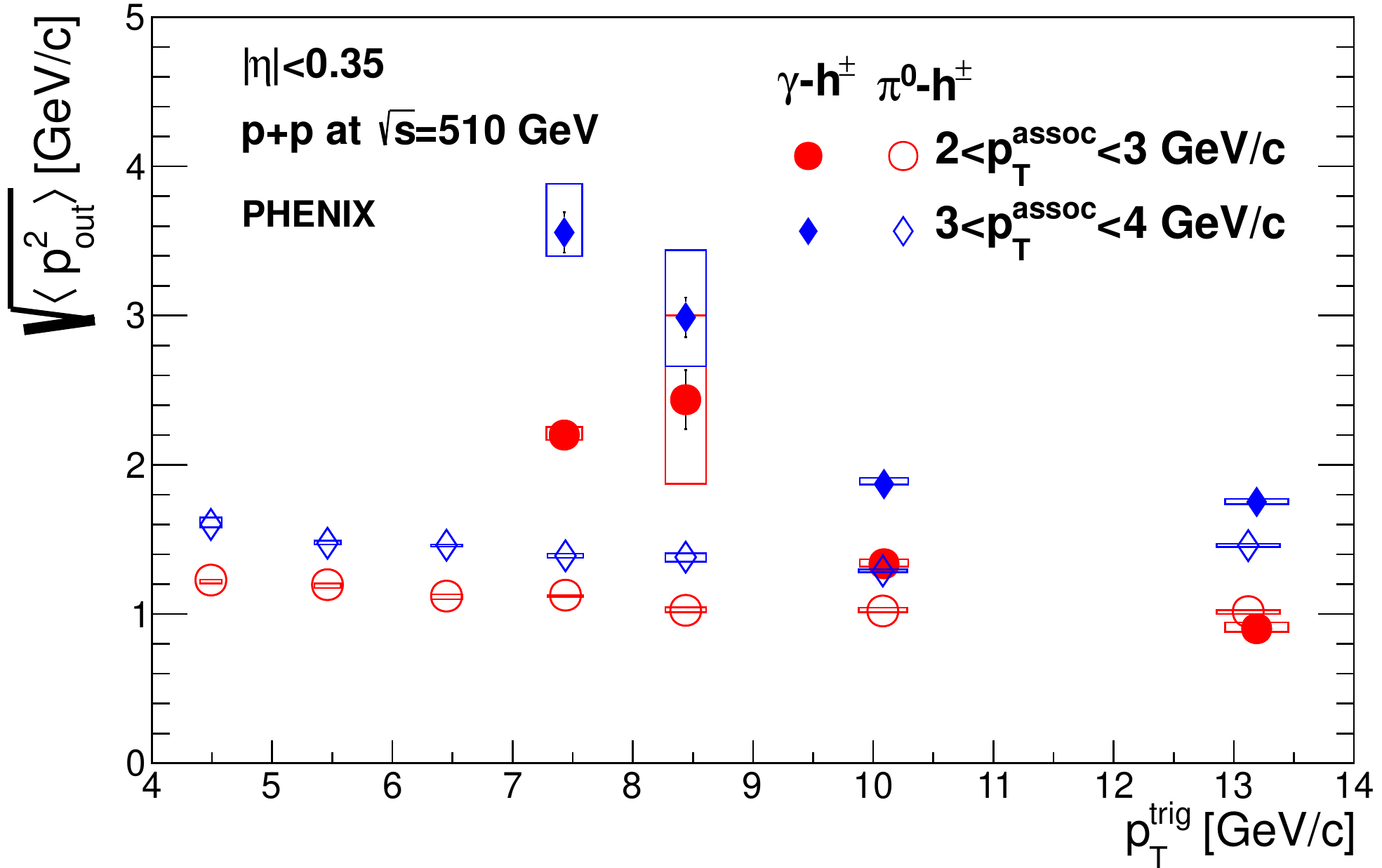}  
 	\caption{\label{fig:rmspoutvspttrig}
$\rmspout$ for $\pion$ and direct photon triggers as a function of 
$\pttrig$ in two different $\ptassoc$ bins.
 	}
 \end{figure}

\begin{figure}[tbh]
 	\includegraphics[width=1.0\linewidth]{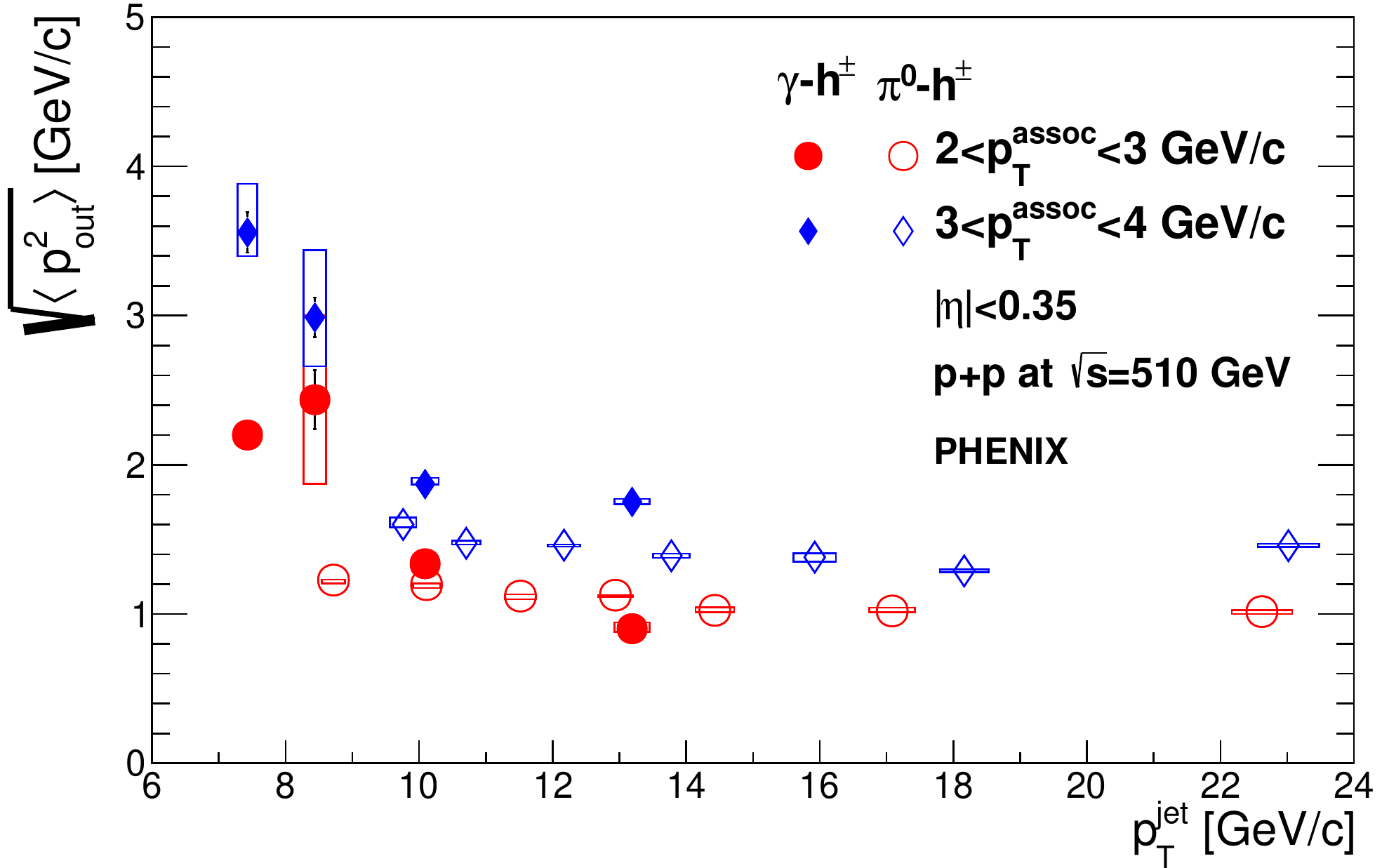}  
 	\caption{\label{fig:rmspoutvsptjet}
$\rmspout$ for $\pion$ and direct photon triggers as a function of 
$p_T^{\rm jet}$ in two different $\ptassoc$ bins. For $\pion$ triggers, 
$p_T^{\rm jet}=\pttrig/\zt$ where $\zt$ was determined from {\sc pythia}.
 	}
 \end{figure}

Figure~\ref{fig:rmspoutvspttrig} shows a subset of the $\rmspout$ results 
for both direct photon and $\pion$ triggers as a function of $\pttrig$ in 
the $\ptassoc$ range 2--4 GeV/$c$. The $\pion$-triggered $\rmspout$ 
decreases with $\pttrig$, although this dependence is small. The direct 
photons clearly have a strong dependence on $\pttrig$ relative to the 
$\pion$ triggers. The $\pion$ data shown in Fig.~\ref{fig:rmspoutvspttrig} 
contain a dependence on the fragmentation function not present in the 
direct photon data because the direct photons emerge directly from the 
hard scattering. To explore this dependence, {\sc pythia} 
6.4~\cite{pythia} hard-scattered QCD events were analyzed to determine the 
average $z_T=\pttrig/\hat{p}_T^{\rm trig}$ of a $\pion$ where the hat 
quantity refers to the hard scattered parton. $\zt$ was determined in the 
same bins used in the data to correct the $\pion$ $\pttrig$ to an 
estimated jet \pt in order to make a better comparison between the direct 
photons and $\pion$. Figure~\ref{fig:rmspoutvsptjet} shows the same 
$\rmspout$ plotted as a function of $p_T^{\rm jet}$, where $p_T^{\rm jet}$ 
refers to $\pttrig$ for the direct photons and to $\pttrig/\zt$ for the 
$\pion$. The $\zt$ correction ranges from 0.45--0.63 as a function of 
$\pttrig$. After the correction, the $\pion$ and direct photon $\rmspout$ 
do not appear to form a single continuous function; rather the $\pion$ 
$\rmspout$ continue approximately linearly to lower $p_T^{\rm jet}$. It is 
possible that the the stronger dependence of $\rmspout$ on $\pttrig$ for 
the direct photons is due to the smaller jet energy being probed. This 
effect may also be seen for the low $\pttrig$ dihadron correlations in 
Table II of Ref.~\cite{ppg089}, where $\rmspout$ has been observed to show 
a stronger dependence at small $\pttrig$ than for larger $\pttrig$.

\subsection{\label{poutdistributions} $\pout$ Distributions}

 \begin{figure*}[thb]
 \includegraphics[width=0.99\linewidth]{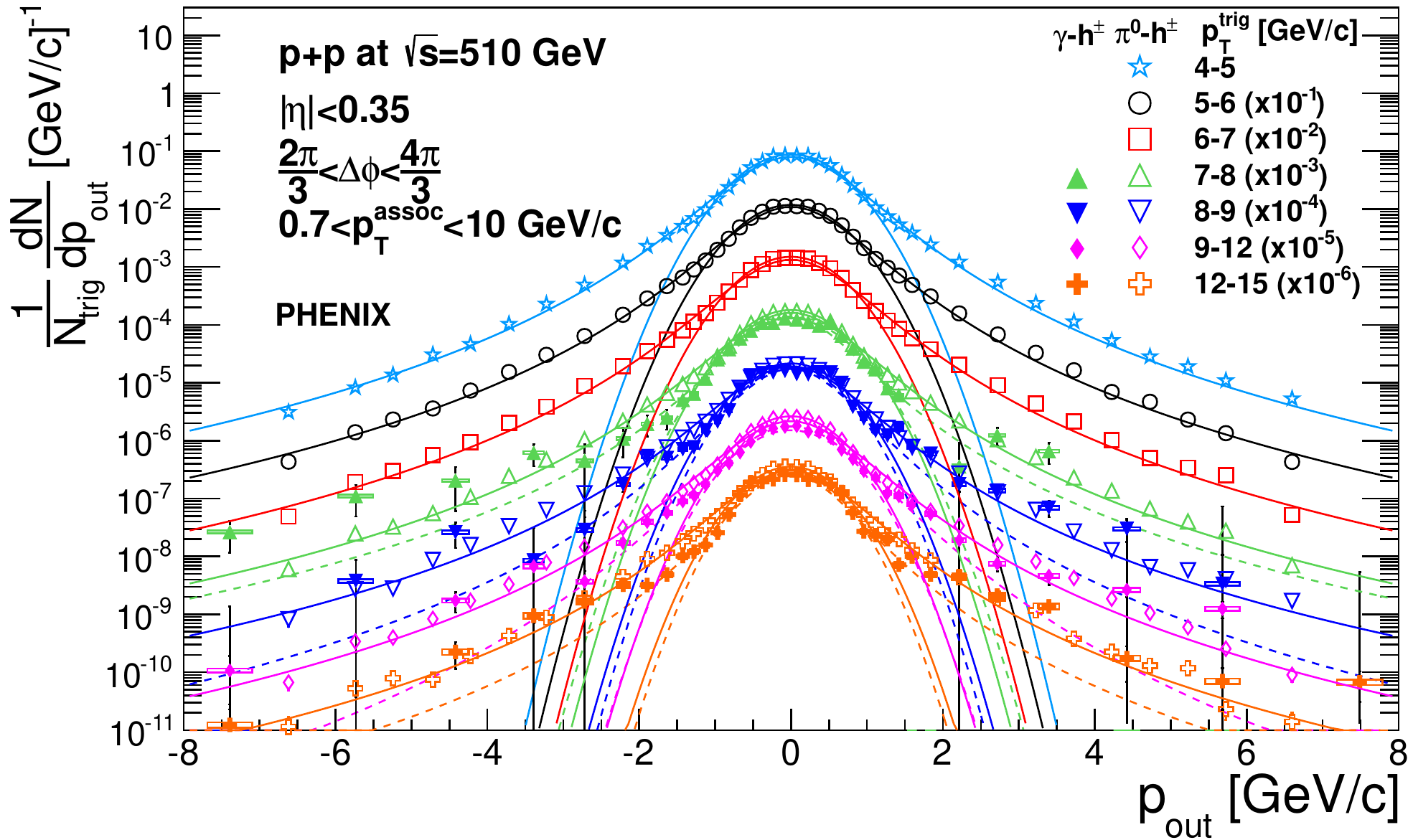}  
 \caption{\label{fig:poutdistributions}
Per trigger yields of charged hadrons as a function of $\pout$. The 
$\pion$ and direct photon distributions are fit with Gaussian functions at 
small $\pout$ and Kaplan functions over the whole range, showing the 
transition from nonperturbative behavior generated by initial-state $k_T$ 
to perturbative behavior generated by hard gluon radiation. A 9\% overall 
normalization uncertainty on the charged hadron yields is not shown in the 
figure.
 }
 \end{figure*}

Figure~\ref{fig:poutdistributions} shows the per-trigger yields of the 
$\pout$ distributions for $\pion$ and direct photon triggers. Only 
away-side hadrons were used in making the distributions, with the 
requirement that the correlated hadron satisfy $2\pi/3<\dphi<4\pi/3$. The 
underlying event was statistically subtracted out from the $\pout$ 
per-trigger yields using the parameters from the fits to the $\dphi$ 
correlations with Eq.~\ref{eq:rmspouteq} in order to identify only charged 
hadron yield associated with the hard scattering. The underlying event 
yield for a given bin was statistically subtracted by applying a factor 
$N_{\rm UE} = 1-f(\dphi)$ where $f(\dphi)$ is the correction function 
determined by $C_0$ divided by the fits to the $\dphi$ correlations. For 
the smaller $\pttrig$ bins, this is an important subtraction because in 
the signal region $\dphi\sim\pi$ the underlying event contributes roughly 
50\% of the away-side yield, as could be ascertained from 
Fig~\ref{fig:dphicorrelations}. The yield corrected by the underlying 
event factor $N_{\rm UE}$ is then subjected to the usual construction of 
the correlation function outlined in Sec.~\ref{statistical_subtraction}. 
Systematic uncertainties on the underlying event background subtraction 
were evaluated by performing the subtraction after changing the underlying 
event parameter $C_0$ to $C_0\pm 1\sigma$, where $\sigma$ is the error on 
$C_0$ from the fit. These uncertainties were found to be on the order of 
tenths of a percent in the $\pout\approx0$ region. The values of the 
$\pout$ distributions can be found in Tables III-XIII of the Supplemental 
Material~\cite{supp_matt}. Note that a 4\% systematic uncertainty is 
assigned to $\pout$ due to the detector resolution on $\ptassoc$ and 
$\dphi$.

The distributions are fit with a Gaussian at small $\pout$ in the region 
[-1.1,1.1] GeV/$c$ as well as a Kaplan function over the whole range, with 
the Kaplan function parameterized by $a(1+\frac{\pout^2}{b})^{-c}$ where 
$a$, $b$, and $c$ are free parameters. In Fig.~\ref{fig:poutdistributions} 
the solid lines are fits to the $\pion$ distributions and the dashed lines 
are fits to the isolated direct photon distributions. The Gaussian 
functions clearly fail past $\sim$1.3 GeV, showing a transition to power 
law behavior which the Kaplan functions accurately describe. The power law 
behavior is generated from hard gluon radiation in the initial state or 
final state, whereas the Gaussian behavior is generated from the soft 
$k_T$ and $j_T$ and is demonstrated in the nearly back-to-back hadrons 
that are produced around $\pout\approx0$.

\begin{figure}[thb]
\includegraphics[width=1.0\linewidth]{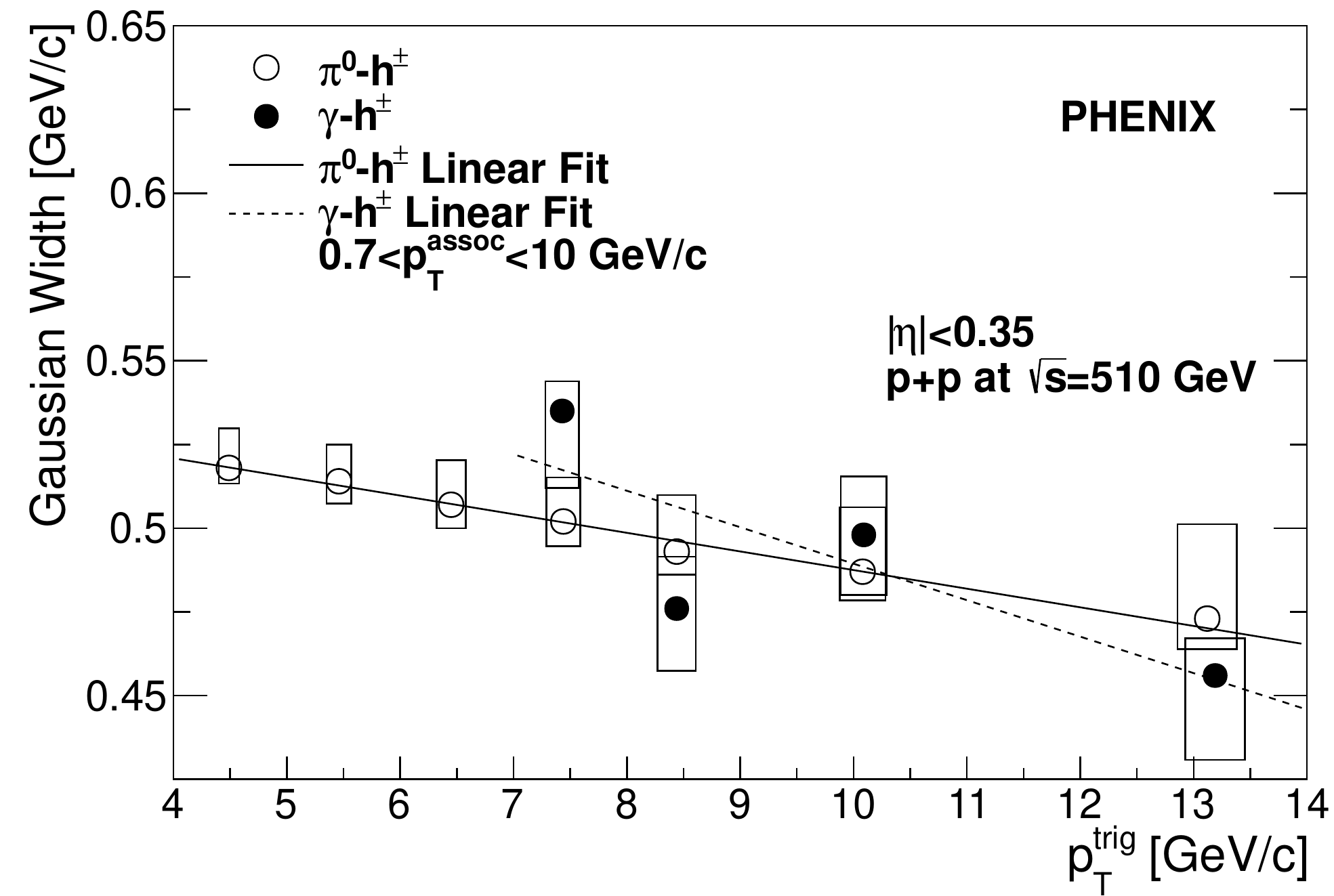}
\caption{\label{fig:gausswidths}
Gaussian widths of $\pout$ for $\pion$ and direct photon triggers as a 
function of $\pttrig$.
	}
\end{figure}

\begin{figure}[thb]
\includegraphics[width=1.0\linewidth]{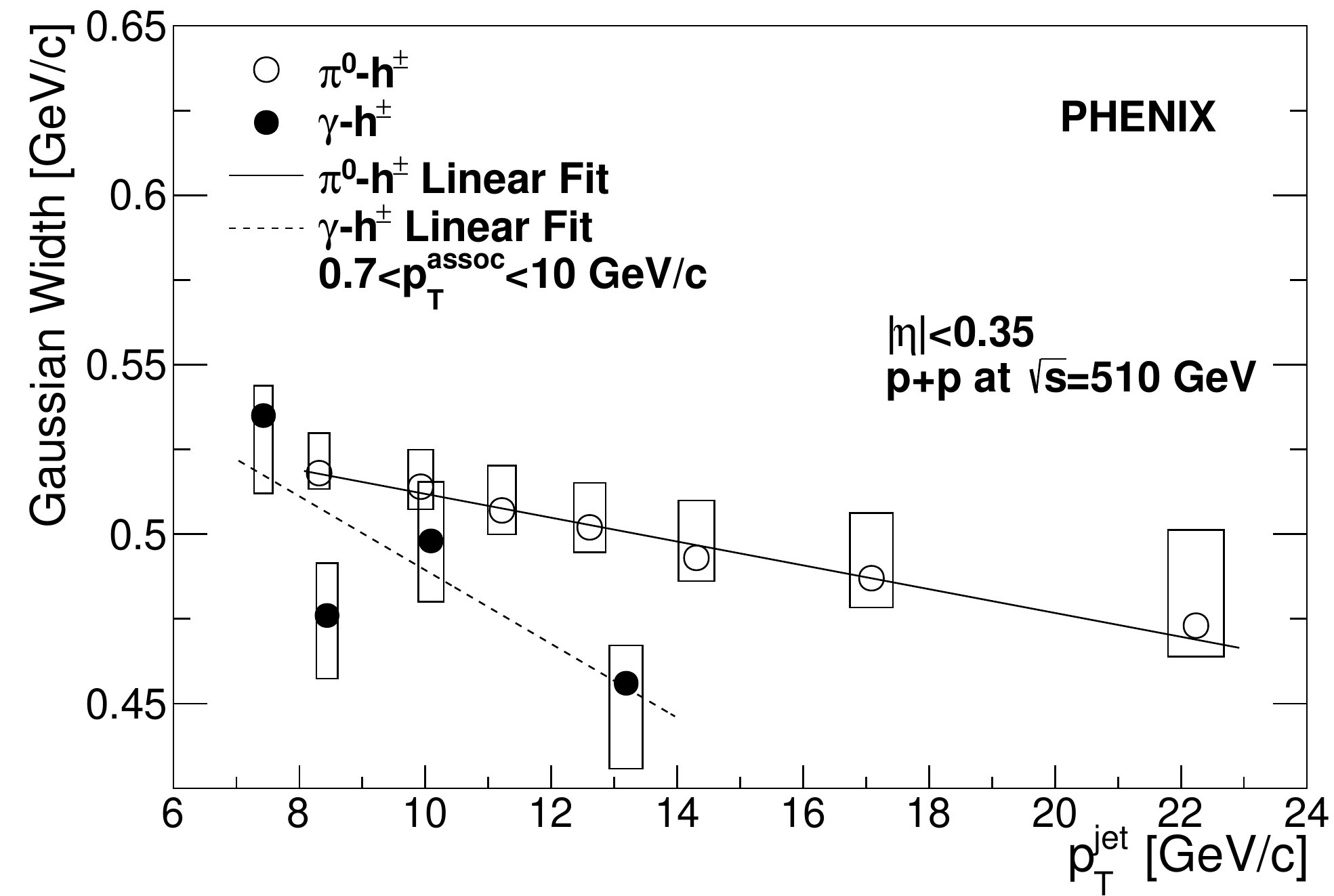}  
\caption{\label{fig:gausswidthscorrected}
Gaussian widths of $\pout$ for $\pion$ and direct photon triggers as a 
function of $p_T^{\rm jet}$, where $p_T^{\rm jet}$ is $\pttrig$ for the 
direct photons and $\pttrig/\zt$ for the $\pion$ triggers. $\zt$ was 
determined from a {\sc pythia} simulation.
 }
 \end{figure}

The evolution of $\pout$ as a function of $\pttrig$ was characterized by 
the Gaussian widths at small $\pout$. Figure~\ref{fig:gausswidths} shows 
the widths from Gaussian fits to both $\pion$ and direct photon triggers 
as a function of $\pttrig$. Systematic uncertainties were evaluated by 
altering the Gaussian fit region by $\pm$ 0.15 GeV/$c$ and taking the 
absolute value of the difference of the resulting widths. As the 
systematic uncertainties dominate the uncertainties of the widths, the 
error bars shown in Fig.~\ref{fig:gausswidths} are the statistical and 
systematic uncertainties combined in quadrature.  Similarly to 
$\rmspout$, the direct photons and $\pion$ both show decreasing widths 
with $\pttrig$. Linear fits to the two sets of widths give slopes of 
$-0.0055\pm0.0018{\rm (stat)}\pm0.0010{\rm (syst)}$ for $\pion$ mesons 
and $-0.0109\pm0.0039{\rm (stat)}\pm0.0016{\rm (syst)}$ for direct 
photons.  Systematic uncertainties on the slopes were conservatively 
estimated by evaluating the fit when the points were placed at the 
limits given by the systematic uncertainties, and then taking the 
difference of the slopes. Similarly to $\rmspout$ the $\pion$ triggers 
were corrected by the same $\zt$ corrections from {\sc pythia}. The 
result is shown in Fig.~\ref{fig:gausswidthscorrected}; again the $\zt$ 
correction amounts to a scale factor of approximately two for the 
$\pttrig$ of the $\pion$ triggers. When plotted against $\pttrig/\zt$ 
the magnitude of the slope for the $\pion$ triggers is 
$-0.0035\pm0.0012{\rm (stat)}\pm0.0006{\rm (syst)}$. It should be noted 
that the slope of the widths changes if the minimum $\ptassoc$ cut is 
increased, but that the slope always remains negative. Integrating over 
the full range of $0.7<\ptassoc<10$ GeV/$c$ allowed by the PHENIX 
detector gives the smallest magnitude slope, thus it is the most 
conservative measurement for comparing to CSS evolution. For example, 
the slope of the Gaussian widths of $\pout$ for $1.2<\ptassoc<10$ 
GeV/$c$ was determined to be $-0.012\pm0.003{\rm (stat)}\pm0.001{\rm 
(syst)}$ for $\pion$-meson and $-0.023\pm0.007{\rm (stat)}\pm0.003{\rm 
(syst)}$ for direct-photon triggers.  The same behavior can be seen in 
the values of $\rmspout$ in Fig.~\ref{fig:rmspoutvspttrig} and in the 
Supplemental Material~\cite{supp_matt}.

\begingroup \squeezetable
\begin{table}[ht]
\caption{Gaussian widths from fits to the $\pout$ 
distributions.}\label{t:gausswidths}
\begin{ruledtabular}  \begin{tabular}{ccc}
Trigger Type & $\langle\pttrig\rangle$ [GeV/$c$] & Gaussian Width 
[GeV/$c$] \\
	\hline

$\pion$ & 4.49 & 0.518  $^{+0.012}_{-0.005}$ \\ 
		& 5.46 & 0.514  $^{+0.011}_{-0.007}$ \\ 
		& 6.45 & 0.507  $^{+0.013}_{-0.007}$ \\ 
	 	& 7.44 & 0.502  $^{+0.013}_{-0.007}$ \\ 
		& 8.44 & 0.493  $^{+0.017}_{-0.007}$ \\ 
		& 10.1 & 0.487  $^{+0.019}_{-0.009}$ \\ 
		& 13.1 & 0.473  $^{+0.028}_{-0.009}$ \\
\\
Direct photon & 7.43 & 0.535  $^{+0.009}_{-0.023}$ \\ 
				& 8.44 & 0.476  $^{+0.015}_{-0.019}$ \\ 
				& 10.1 & 0.498  $^{+0.017}_{-0.018}$ \\ 
				& 13.2 & 0.456  $^{+0.011}_{-0.025}$ \\ 
\end{tabular} \end{ruledtabular}
\end{table}
\endgroup


\section{\label{discussion}Discussion}

\subsection{\label{pionphotonconsiderations}Measured Results}

Figures~\ref{fig:rmspoutvspttrig} and~\ref{fig:gausswidths} show that, 
consistent with previous RHIC measurements, $\rmspout$ and the Gaussian 
widths of $\pout$ sensitive to initial-state and final-state $k_T$ and 
$j_T$ decrease with the hard scale. Interpretation of $\rmspout$ is 
slightly different than that of the Gaussian widths from the $\pout$ 
distributions, because the Gaussian widths are extracted from fits to the 
nearly back-to-back region, which is generated only by nonperturbative 
$k_T$ and $j_T$. The $\rmspout$ values are extracted from fits to the 
entire away-side jet region in the $\dphi$ correlations; therefore, these 
quantities inherently include the charged hadrons in the perturbatively 
generated tail away from $\dphi\sim\pi$ whereas the Gaussian widths 
measured from the $\pout$ distributions \emph{only} have contributions 
from $\dphi\sim\pi$. Nonetheless the values of $\rmspout$ are dominated by 
the nearly back-to-back region as this is where most of the away-side 
charged hadrons are, but this subtle difference between the two 
observables should be noted. The widths quantified by $\rmspout$ have the 
benefit that they can be extracted from the finely binned 
$\pttrig\otimes\ptassoc$ $\dphi$ angular correlations. Throughout this 
discussion, we will use the term ``width" to refer to both the $\rmspout$ 
and Gaussian widths extracted from $\pout$. 

\begin{figure}[thb]
\includegraphics[width=1.0\linewidth]{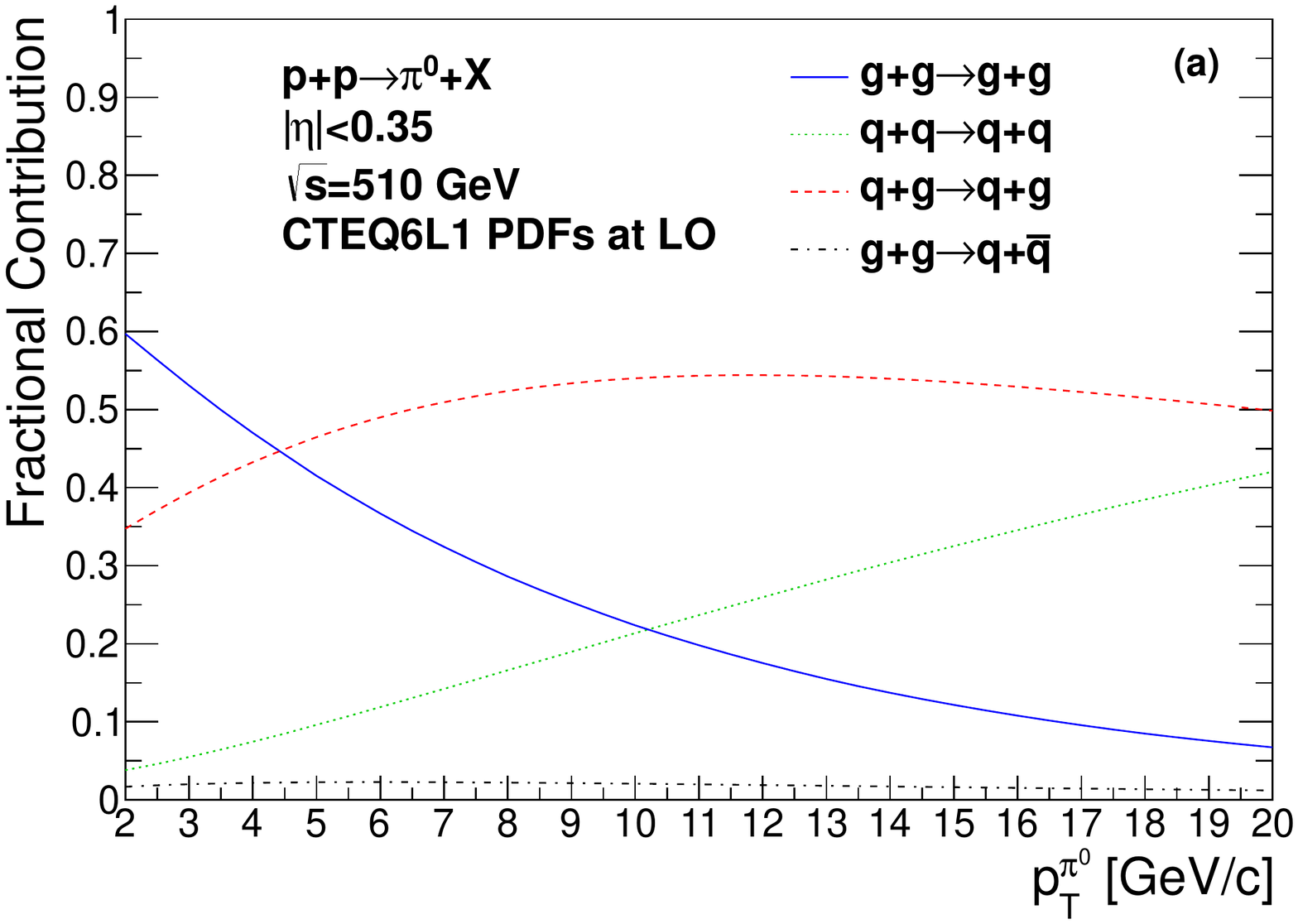}  
\includegraphics[width=1.0\linewidth]{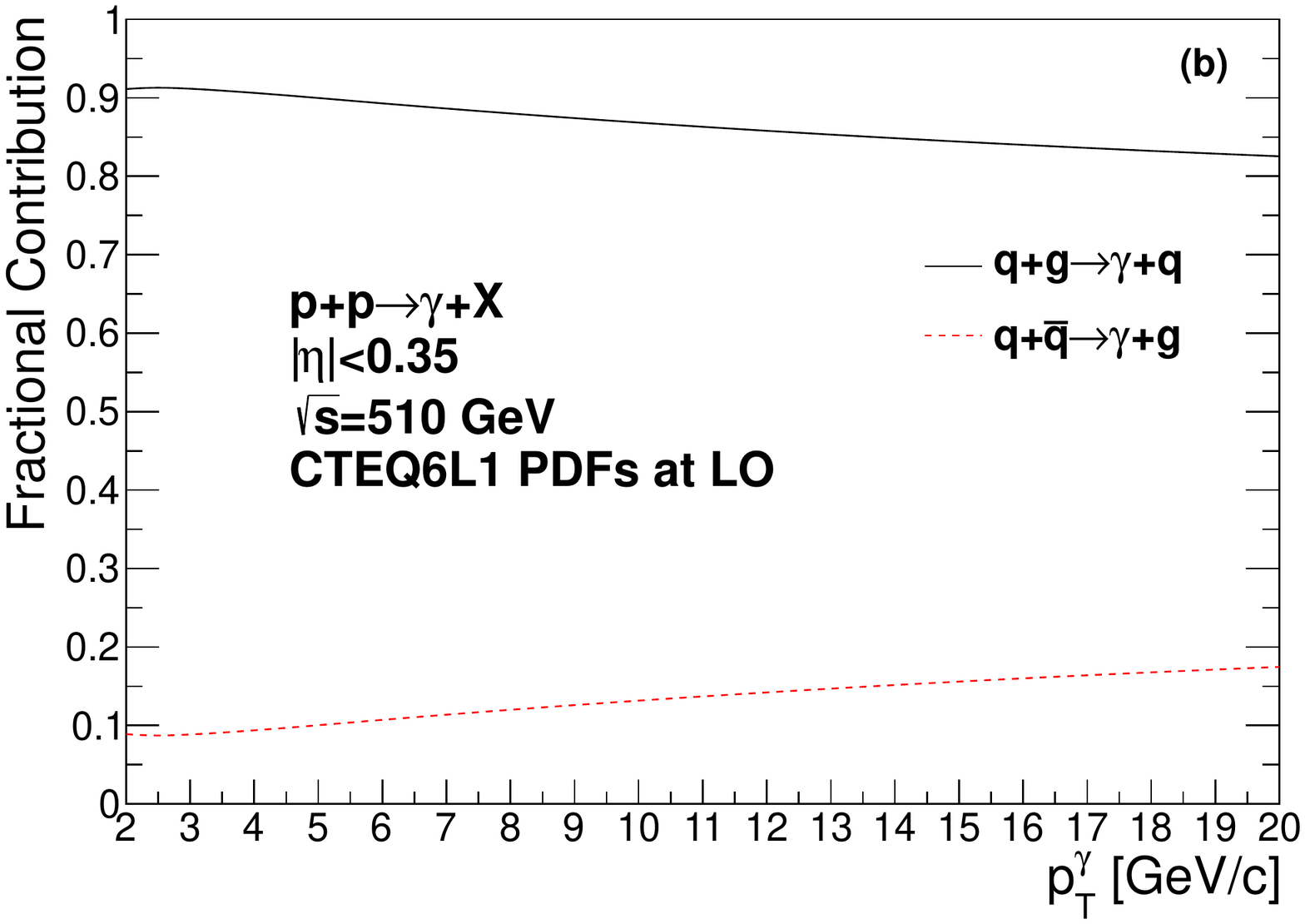}
\caption{\label{fig:pi0_partonic}
The fractional contribution of partonic scattering processes to (a) 
inclusive $\pion$ and (b) direct-photon production at LO in \pp 
collisions at $\sqs=510$ GeV in the PHENIX pseudorapidity region. Note 
that the process $q\bar{q}\rightarrow gg$ is not drawn in panel (a) 
because its contribution is less than one percent in this $p_T$ range.
 }
 \end{figure}

There is a difference in the mix of scattered away-side partons probed by 
inclusive-$\pion$ and direct-photon triggers.  
Figure~\ref{fig:pi0_partonic} shows the fractional contribution to the 
total cross section calculated in pQCD for the LO diagrams for (a) 
inclusive-$\pion$ and (b) direct-photon production. The 
CTEQ6L1~\cite{CTEQ6L1} PDFs were used for the calculations in addition to 
the DSS14 FFs~\cite{DSS14_FFs} for the $\pion$. At LO, quark-gluon Compton 
scattering accounts for approximately 85\% of direct photons produced at 
midrapidity, while~$\pion$ triggers are instead generated by a significant 
contribution of $qg$, $gg$, and $qq$ scatterings. Therefore, any 
comparison between direct photon and $\pion$ triggers could be affected by 
the fact that the away-side charged hadrons are produced by quark 
jets~$\sim$85\% of the time for direct photons and a mix of gluon and 
quark jets for the $\pion$. For the direct photon partonic fractions, NLO 
corrections do not make a significant difference in the dominance of the 
quark-gluon Compton scattering process in the central rapidity region 
studied here~\cite{NLO_dirphots}.

Many correlation measurements similar to the one presented here have been 
made at RHIC~\cite{ppg029,ppg089,ppg095,STARcorr:2014PRL}, as discussed in 
the Introduction. Although the same conclusion regarding the evolution of 
the widths found here can be drawn from these measurements, they were made 
with different physics goals; examples include understanding partonic 
energy loss in a nuclear medium or characterizing fragmentation functions. 
Earlier correlation measurements were largely motivated by the heavy ion 
or hard scattering high-energy-physics community, and it was not 
until recently that the nucleon-structure community began to understand 
how to look for possible factorization breaking effects in these types of 
measurements~\cite{AybatRogers:2011}. This came as a result of the recent 
interest in understanding TMD evolution, especially understanding the 
nonperturbative contributions to TMD evolution. 
Reference~\cite{collinsrogers_tmdevolutionkernel} gives a comprehensive 
discussion of phenomenology including TMD evolution and how this 
phenomenology came to the forefront in 2011. 

\subsection{\label{evolutiondiscussion}Expectations from CSS Evolution}

Consistent with previous measurements, the data presented here clearly 
show that momentum widths sensitive to nonperturbative $k_T$ and $j_T$ 
decrease with the hard scale in $\pion$- and direct photon-charged hadron 
correlations. As was mentioned in the Introduction, the expectation from 
CSS evolution is that momentum widths sensitive to nonperturbative 
transverse momentum scales should increase with the hard scale. To compare 
to what is predicted by CSS evolution, the slopes were compared to a slope 
of zero as this quantifies the boundary between narrowing and increasing 
widths with $\pttrig$. The confidence interval excludes a slope of zero at 
the 2.6$\sigma$ level for both the $\pion$ and direct photon triggered 
correlations. The likelihood ratio from a slope of zero was calculated to 
be 0.03 for both the $\pion$ and direct photon triggered correlations, 
which implies that the data is not consistent with a flat line. 

Because $k_T$ and $j_T$ have been measured to be approximately constant in 
the $\pttrig$ region probed here~\cite{ppg029,ppg095}, kinematically it 
would be expected that the acoplanarity decrease with $\pttrig$. However, 
this same argument would apply for both DY and SIDIS, showing that the 
effect of decreasing widths seen in $\pion-h^\pm$ and direct 
$\gamma-h^\pm$ correlations cannot be a kinematic or fragmentation effect 
alone. It is also interesting that Ref.~\cite{ALICE_dijet_kt} shows that 
in dijet correlations at very high \pt and \sqs, momentum widths sensitive 
to initial-state $k_T$ increase with the \pt of the jet. These 
measurements are at large \pt and sensitive to large $k_T$ at the higher 
\sqs of the Large Hadron Collider, and thus follow the leading-log 
approximation which is also purely perturbative and predicts increasing 
widths with the hard scale (see e.g.~\cite{CDF_kt_MLLA} and references 
within).

The CSS evolution framework was motivated by understanding perturbative 
QCD dynamics. At this time, QCD was still in the early stages of 
development, and nonperturbative dynamics were not the focus within the 
framework of pQCD. As QCD became well established as the theory of the 
strong force, measurements performed at high enough energies for 
perturbative techniques to be applicable began to be used routinely to 
constrain nonperturbative physics in the form of collinear PDFs and FFs.  
It is only in the last two decades that there has been increasing focus on 
using perturbative techniques to understand nonperturbative parton 
dynamics.  The study of such nonperturbative dynamics provides information 
on parton behavior within bound states and the process of hadronization by 
defining and constraining TMD PDFs and FFs. Importantly, it is 
additionally offering new insights on fundamental aspects of QCD as a 
nonAbelian gauge-invariant quantum field theory, for example through the 
predicted relative sign difference of the Sivers TMD PDF when probed via 
SIDIS versus DY~\cite{collins_sivers_prediction}, and through TMD 
factorization breaking in certain processes~\cite{trogers_factbreaking}. 
Factorization breaking results from basic QCD principles. Namely, 
nonAbelian phase interferences from the exchange of gluons between 
colored objects cannot, in general, be disentangled. Similarly, phase 
interferences from gluon exchange play a role in the Sivers effect where 
it implies a sign change for the Sivers function between DY and SIDIS 
interactions. The reason that gluon exchange in DY and SIDIS does not lead 
to factorization breaking is because DY and SIDIS are both quantum 
electrodynamic processes at LO, so there are limited paths for gluon 
exchange. Only initial-state exchange in DY and final-state exchange in 
SIDIS are possible, whereas in hadronic collisions with a final-state 
hadron measured both initial- and final-state exchanges are possible. 

Observing differences in the evolution of momentum widths as a function of 
the hard scale is a powerful observable due to the qualitative conclusion 
that can be drawn from the data when comparing to the expectation of CSS 
evolution. Before the recent interest in understanding TMD evolution, 
measured deviations from calculations at some given scale assuming 
factorization holds were the only obvious way to look for factorization 
breaking effects. Such calculations are not available. Simply looking for 
qualitative differences in the evolution of the observable gives a clear 
discrepancy with the expectation from CSS evolution, and this is 
significantly more powerful than trying to compare with a calculation that 
requires greater knowledge of the nonperturbative functions. It is 
furthermore interesting to point out that the inclusive hadron transverse 
single-spin asymmetries in hadronic collisions measured at forward 
rapidities also deviate from the expectation provided by standard 
perturbative evolution. In charged pion production, the asymmetry changes 
strikingly little from $\sqs=4.9$ GeV to $\sqs=62.4$ 
GeV~\cite{physrevmod}. Asymmetries have been measured to be nonzero at 
center of mass energies up to 200 GeV and appear to plateau at \pt up to 5 
GeV/$c$~\cite{ppg135}, while perturbative techniques give a clear 
prediction that the asymmetry should fall off as \pt 
increases~\cite{rogers:2013:extra_asymm}.

\subsection{\label{pythia_comparison}Comparison to {\sc pythia}}

Without any available theoretical calculations of our observable, the 
results found in the data as well as the expectation from CSS evolution 
were investigated with a {\sc pythia} simulation. Factorization breaking is not 
predicted in DY production because there are no final-state hadrons 
produced directly from the hard scattering. The same observable $\pout$ 
can be constructed in DY events with two nearly back-to-back leptons. For 
dileptons, it would be expected that the Gaussian width of $\pout$ would 
broaden as one increases the hard scale of the interaction, as DY is known 
to follow CSS evolution. The Perugia0 tune~\cite{perugia_tune} should be 
ideal to study this because it was tuned to the CDF Z$^0$ cross section data 
at low \pt~\cite{CDFzbosons}. Therefore the Perugia tunes should be 
reasonably adequate at reproducing DY events where the total \pt is small.

\begin{figure}[thb]
\includegraphics[width=1.0\linewidth]{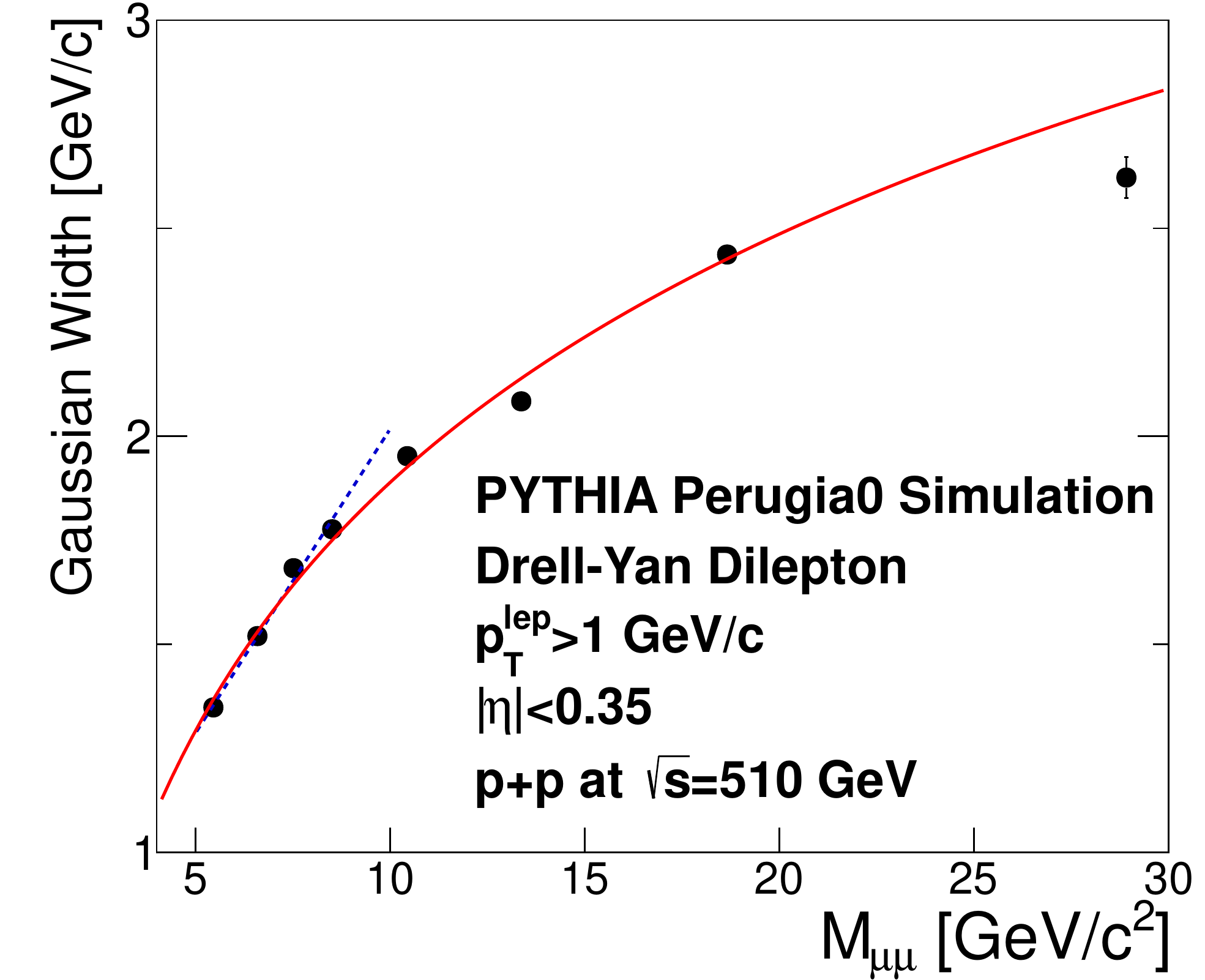}
\caption{\label{fig:DYgausswidths}
Gaussian widths extracted from {\sc pythia} Drell-Yan $\pout$ distributions. In 
Drell-Yan factorization breaking is not predicted. Here the widths show a 
positive slope with the invariant mass of the dilepton, as predicted by 
CSS evolution. The solid red line shows a log fit over the full invariant 
mass range and the blue dotted line shows a linear fit in the region 5--10 
GeV/$c^2$.
	}
\end{figure}

{\sc pythia} 6.4 DY events were generated and $\pout$ was determined for 
the correlated dileptons to confirm the expectation from CSS evolution for 
this observable. $\pout$ is defined similarly to Eq.~\ref{eq:pout}, 
$\pout=p_T^{\rm lep}\sin\dphi$ where the higher-\pt lepton is taken as the 
near-side trigger particle and the lower-\pt lepton is taken as the 
away-side associated particle used in the determination of $\pout$. The 
distributions were fit with Gaussian functions in the nonperturbative 
nearly back-to-back region, and the widths of the dilepton $\pout$ 
distributions are shown in Fig.~\ref{fig:DYgausswidths}. {\sc pythia} 
reproduces the expectation that the widths of the DY pairs increase with 
the $Q^2$ of the interaction when $\pout$ is only sensitive to 
initial-state $k_T$ and there are no final-state hadrons. The widths are 
quantitatively much larger than the dihadron or direct photon-hadron 
widths because the DY dileptons emerge from the virtual photon, which 
means that, in the PHENIX pseudorapidity region, their \pt is large. When 
measuring a final-state hadron, $\pout$ by definition must be smaller for 
the case of a measured final-state hadron vs.~a DY lepton because the \pt 
of the charged hadron must be smaller than or equal to the \pt of the 
scattered parton due to the fragmentation process. Any quantitative value 
of $\pout$ will naturally be dependent on the $\ptassoc$ measured; see 
Fig.~\ref{fig:rmspoutvsptassoc}. In DY, each lepton will have 
approximately half the momentum of the interaction hard scale, and the 
larger the momentum the larger $\pout$ can be while still being in the 
nearly back-to-back region $\dphi\sim\pi$, i.e. nonperturbatively 
generated. What is relevant is the evolution of this width with the hard 
scale of the interaction, not the quantitative value, as this is just 
indicative of what away-side $\ptassoc$ is observed. The DY widths were 
fit with a linear function shown as the blue dotted line in the region 
which was most linear, 5--10 GeV/$c^2$, and the slope of the line was 
determined to be 0.146$\pm$0.004. Additionally the red solid line shows a 
log fit over the full invariant mass range. The DY slope is the opposite 
sign from the direct photon-hadron and dihadron correlations and it is 
also approximately one order of magnitude larger making it significantly 
different from the dihadron and direct photon-hadron slopes.

\begin{figure*}[thb]
\includegraphics[width=0.99\linewidth]{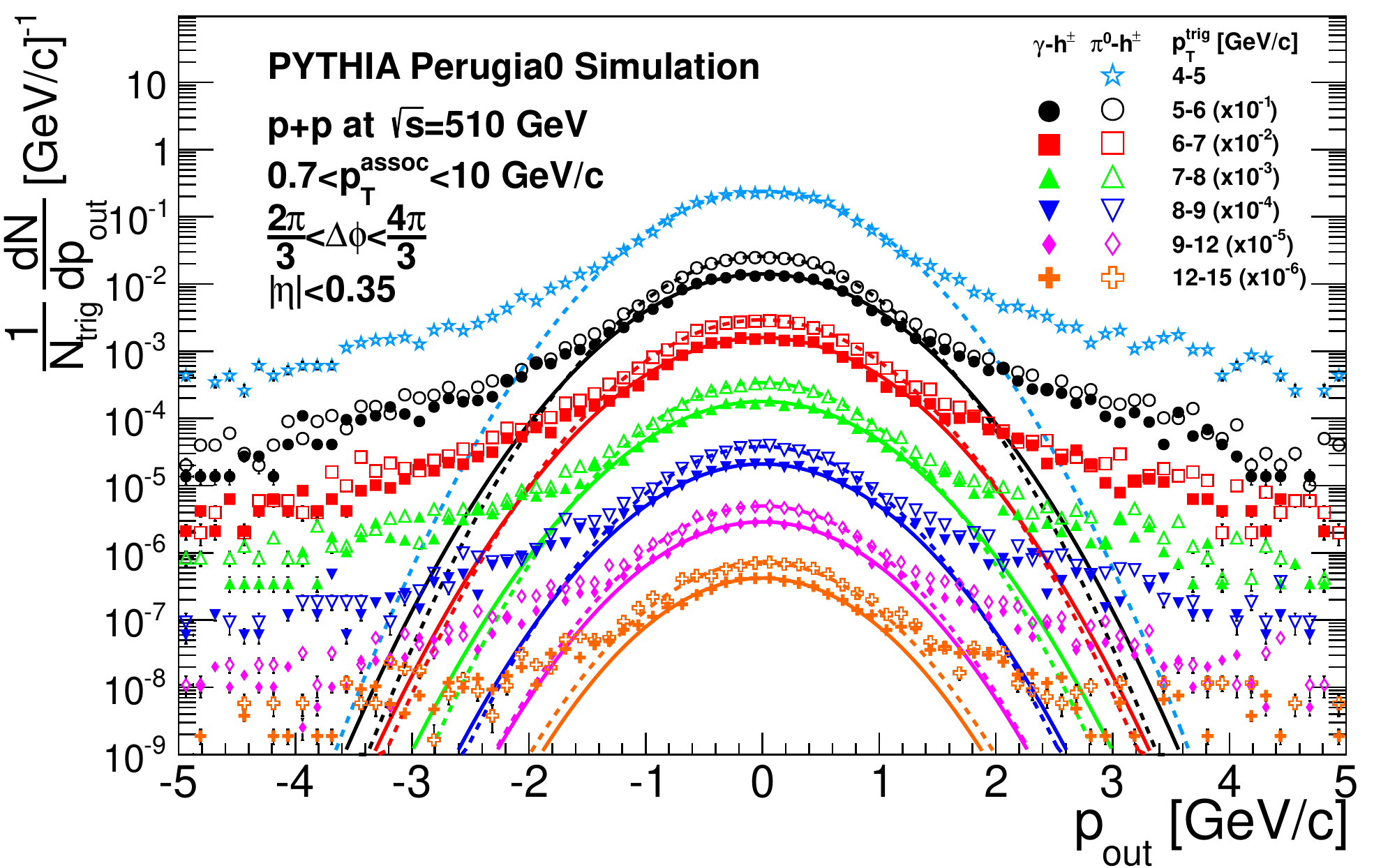}  
\caption{\label{fig:pout_pythia}
$\pion$-h$^\pm$ and direct photon-h$^\pm$ $\pout$ distributions from 
{\sc pythia} correlations. The nonperturbative region was fit with Gaussian 
functions similarly to what was done with the data. {\sc pythia} shows similar 
behavior to the data: a clear nonperturbative region transitioning to a 
perturbative power-law tail.
 }
 \end{figure*}

Similarly, {\sc pythia} direct photon and dijet events were generated at 
$\sqs=510$ GeV with the Perugia0 tune, changing the Gaussian intrinsic 
$k_T$ parameter PARP(91) setting to $3.2$ GeV/$c$ as should be expected at 
$\sqs=510$ GeV from Ref.~\cite{ppg029}. The Perugia0 tune was used again 
for the dihadron and direct photon-hadron correlations so that a direct 
comparison could be made to the DY Perugia0 tune Gaussian widths. The 
direct photons were required to be isolated similarly to what was done in 
data. Correlated pairs of $\pion$ or direct photon and $\pi^\pm$, $K^\pm$, 
and $p$,$\bar{p}$ were collected in the PHENIX pseudorapidity, and the 
observables $\dphi$ and $\pout$ were determined from the correlated pairs. 
Similarly to what was done in data, the background from the underlying 
event was statistically subtracted out to make the $\pout$ distributions. 
{\sc pythia} correlated pairs show the same features as the data do as can be 
seen in Fig.~\ref{fig:pout_pythia}; $\pout$ exhibits a Gaussian shape at 
small $\pout$ which transitions to a power law shape at large $\pout$.

\begin{figure}[thb]
\includegraphics[width=1.0\linewidth]{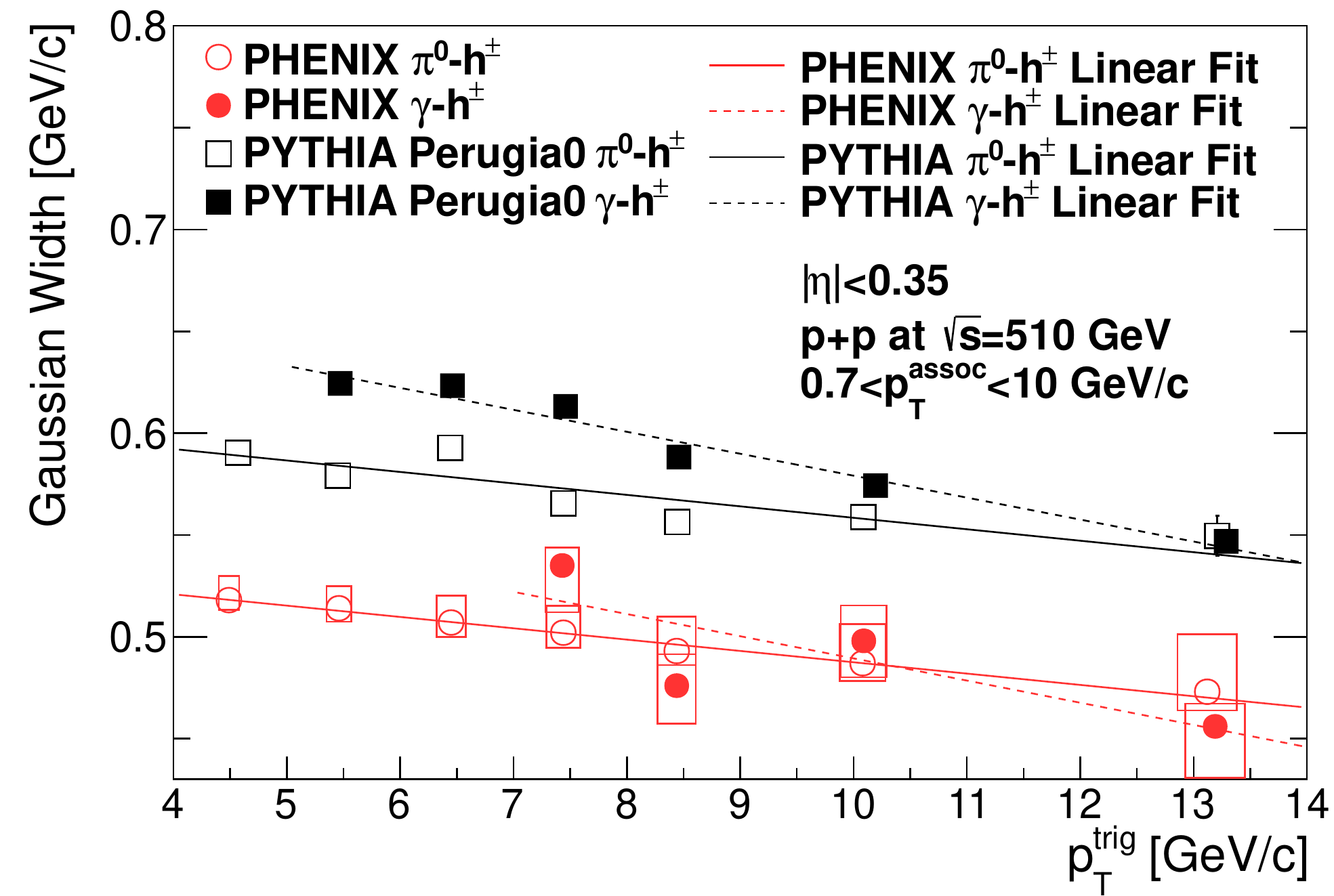}  
\caption{\label{fig:pythia_gauss}
Gaussian widths extracted from {\sc pythia} $\pout$ distributions are shown 
compared to the results measured at PHENIX. {\sc pythia} produces consistent 
evolution rates as what was determined in PHENIX, although there is a 
10\%--15\% difference in the quantitative values in each $\pttrig$ bin.
 }
\end{figure}

Gaussian widths were extracted from the {\sc pythia}-generated correlations in 
the same way that was done for the data. The widths from the {\sc pythia} 
$\pout$ distributions are shown with the measured widths in 
Fig.~\ref{fig:pythia_gauss}. Remarkably, the {\sc pythia} results reproduce the 
measured slopes in both sign and magnitude for both $\pion$ and direct 
photon triggers. The slope values from {\sc pythia} were 
$-0.0056\pm0.0007$ for $\pion$-meson
$-0.0107\pm0.0006$ for direct-photon triggers. The measured slopes are 
$-0.0055\pm0.0018{\rm (stat)}\pm0.0010{\rm (syst)}$ for $\pion$-meson 
and $-0.0109\pm0.0039{\rm (stat)}\pm0.0016{\rm (syst)}$ for direct-photon 
triggers.  The negative sign of the slope was found in both the 
quark-gluon Compton and quark-antiquark annihilation processes for 
isolated direct photon production, indicating that the effect in {\sc 
pythia} is not due to a difference in quark vs. gluon fragmentation. Additionally, as the minimum $\ptassoc$ cut is increased when constructing the $\pout$ distributions, the slope of the Gaussian widths increases, similarly to what is seen in data. One 
noticeable difference between {\sc pythia} and the data is the 
quantitative values of the widths. Here the results from {\sc pythia} 
differ by about $\sim$15\% for both the $\pion$ and direct photon 
triggers, depending on the $\pttrig$ bin.

The nonperturbative Gaussian behavior of $\pout$ is generated by the soft 
initial-state $k_T$ and final-state $j_T$ as indicated in 
Fig.~\ref{fig:kTkinematics}. In the nearly back-to-back region 
$\pout\lesssim1.3$ GeV/$c$, $\pout$ is small and thus can only be 
generated by soft gluon radiation because the two particles are nearly 
coplanar. It is unsurprising that {\sc pythia} does not replicate the 
quantitative values of the Gaussian widths well as there is little data 
that would offer constraints to this region. What is striking is that 
{\sc pythia} replicates the evolution rate for both $\pion$ and direct photon 
Gaussian widths. 

While {\sc pythia} certainly does not explicitly consider analytical 
factorization breaking effects as it assumes collinear factorization, in 
contrast to a collinear pQCD calculation it does include initial- and 
final-state interactions. After a parton interacts in {\sc pythia}, the 
remnants of the two protons are free to interact with other objects in the 
event, and every object in the interaction is forced to color neutralize. 
Factorization breaking effects are predicted in dihadron and direct 
photon-hadron correlations due to the possibility of gluon exchange in 
both the initial and final states. This includes gluon exchange with 
remnants of the interaction, because the remnants of the interacting 
protons can exchange gluons with partons in both the initial and final 
states. Sensitivity to these effects requires a small transverse momentum 
scale; in the transverse-momentum-integrated case observables no longer 
have this sensitivity. For this reason it is plausible that {\sc pythia} 
could be sensitive to these effects because of interactions between the 
proton remnants and partons involved in the hard scattering. Because {\sc 
pythia} allows initial- and final-state interactions via gluon exchanges, 
the necessary interactions to allow for factorization breaking effects are 
present within the {\sc pythia} framework. It should also be noted that 
{\sc pythia} replicates the color coherence effects in 
Refs.~\cite{CDF_cc,D0_cc,CMS_cc} as well.

The underlying mechanism that leads to the prediction of the sign change 
in the Sivers function or factorization breaking is gluon exchange between 
partons associated with the hard scattering and colored remnants. In cases 
where factorization breaking is predicted, it implies that the traditional 
organization of the nonperturbative objects into separate PDFs and FFs for 
each colliding proton and produced hadron no longer holds.  However, so 
far we have no knowledge of how to approach a reorganization of the 
nonperturbative objects, which would presumably include novel correlation 
functions describing partons correlated across the colliding protons. The 
fact that {\sc pythia} accurately describes both the qualitative and 
quantitative nature of the slopes of the widths as a function of hard 
scale offers a potential path forward to greater understanding and further 
advancing what can be calculated within the rigors of pQCD.

\section{\label{summary}Summary and Conclusions}

Dihadron and direct photon-hadron correlations sensitive to 
nonperturbative transverse momentum effects have been measured in the 
PHENIX experiment at RHIC in $\sqs=510$ GeV $\pp$ collisions, motivated by 
the prediction of factorization breaking in such 
processes~\cite{trogers_factbreaking,Mulders:2006,Collins:2007,Collins:2007preprint}. 
Correlations between $\pion$ or direct photons with charged hadrons were 
measured. The azimuthal angular separation $\dphi$ and out-of-plane 
transverse momentum component $\pout$ for the correlated pairs were 
measured. $\pout$ has sensitivity to nonperturbative transverse momentum 
in the initial state, as well as in the final state when a produced hadron 
is measured. The $\rmspout$ and Gaussian widths of the $\pout$ 
distributions were measured from the correlations, and both observables 
decrease with the hard scale of the interaction $\pttrig$. The direct 
photons exhibit a larger dependence than the $\pion$ triggers on $\pttrig$ 
for both $\rmspout$ and the Gaussian widths of $\pout$. The narrowing of 
the Gaussian widths as a function of $\pttrig$ indicates that the 
Collins-Soper-Sterman soft factor cannot be driving the evolution, in 
contrast with Drell-Yan dilepton production and SIDIS where factorization 
is predicted to hold and the widths are empirically known to increase with 
hard scale.  Study of the same observables via the {\sc pythia} event 
generator, which allows for gluon exchange between partons involved in the 
hard scattering and the proton remnants, reveals strikingly similar 
characteristics.  The similarity between {\sc pythia} and the experimental 
data offers a promising path forward to understand the mechanism in QCD 
driving the observed evolution in more detail.


\section*{ACKNOWLEDGMENTS}  

We thank the staff of the Collider-Accelerator and Physics
Departments at Brookhaven National Laboratory and the staff of
the other PHENIX participating institutions for their vital
contributions.   We also thank T.C.~Rogers, J.C.~Collins, and 
L.~L{\"o}nnblad for valuable discussions regarding the 
interpretation of these results, as well as T.~Kaufmann for 
providing several pQCD calculations.
We acknowledge support from the 
Office of Nuclear Physics in the
Office of Science of the Department of Energy,
the National Science Foundation, 
Abilene Christian University Research Council, 
Research Foundation of SUNY, and
Dean of the College of Arts and Sciences, Vanderbilt University 
(U.S.A),
Ministry of Education, Culture, Sports, Science, and Technology
and the Japan Society for the Promotion of Science (Japan),
Conselho Nacional de Desenvolvimento Cient\'{\i}fico e
Tecnol{\'o}gico and Funda\c c{\~a}o de Amparo {\`a} Pesquisa do
Estado de S{\~a}o Paulo (Brazil),
Natural Science Foundation of China (People's Republic of China),
Croatian Science Foundation and
Ministry of Science, Education, and Sports (Croatia),
Ministry of Education, Youth and Sports (Czech Republic),
Centre National de la Recherche Scientifique, Commissariat
{\`a} l'{\'E}nergie Atomique, and Institut National de Physique
Nucl{\'e}aire et de Physique des Particules (France),
Bundesministerium f\"ur Bildung und Forschung, Deutscher
Akademischer Austausch Dienst, and Alexander von Humboldt Stiftung (Germany),
National Science Fund, OTKA, K\'aroly R\'obert University College, 
and the Ch. Simonyi Fund (Hungary),
Department of Atomic Energy and Department of Science and Technology (India), 
Israel Science Foundation (Israel), 
Basic Science Research Program through NRF of the Ministry of Education (Korea),
Physics Department, Lahore University of Management Sciences (Pakistan),
Ministry of Education and Science, Russian Academy of Sciences,
Federal Agency of Atomic Energy (Russia),
VR and Wallenberg Foundation (Sweden), 
the U.S. Civilian Research and Development Foundation for the
Independent States of the Former Soviet Union, 
the Hungarian American Enterprise Scholarship Fund,
and the US-Israel Binational Science Foundation.



\begin{thebibliography}{75}%
\makeatletter
\providecommand \@ifxundefined [1]{%
 \@ifx{#1\undefined}
}%
\providecommand \@ifnum [1]{%
 \ifnum #1\expandafter \@firstoftwo
 \else \expandafter \@secondoftwo
 \fi
}%
\providecommand \@ifx [1]{%
 \ifx #1\expandafter \@firstoftwo
 \else \expandafter \@secondoftwo
 \fi
}%
\providecommand \natexlab [1]{#1}%
\providecommand \enquote  [1]{``#1''}%
\providecommand \bibnamefont  [1]{#1}%
\providecommand \bibfnamefont [1]{#1}%
\providecommand \citenamefont [1]{#1}%
\providecommand \href@noop [0]{\@secondoftwo}%
\providecommand \href [0]{\begingroup \@sanitize@url \@href}%
\providecommand \@href[1]{\@@startlink{#1}\@@href}%
\providecommand \@@href[1]{\endgroup#1\@@endlink}%
\providecommand \@sanitize@url [0]{\catcode `\\12\catcode `\$12\catcode
  `\&12\catcode `\#12\catcode `\^12\catcode `\_12\catcode `\%12\relax}%
\providecommand \@@startlink[1]{}%
\providecommand \@@endlink[0]{}%
\providecommand \url  [0]{\begingroup\@sanitize@url \@url }%
\providecommand \@url [1]{\endgroup\@href {#1}{\urlprefix }}%
\providecommand \urlprefix  [0]{URL }%
\providecommand \Eprint [0]{\href }%
\providecommand \doibase [0]{http://dx.doi.org/}%
\providecommand \selectlanguage [0]{\@gobble}%
\providecommand \bibinfo  [0]{\@secondoftwo}%
\providecommand \bibfield  [0]{\@secondoftwo}%
\providecommand \translation [1]{[#1]}%
\providecommand \BibitemOpen [0]{}%
\providecommand \bibitemStop [0]{}%
\providecommand \bibitemNoStop [0]{.\EOS\space}%
\providecommand \EOS [0]{\spacefactor3000\relax}%
\providecommand \BibitemShut  [1]{\csname bibitem#1\endcsname}%
\let\auto@bib@innerbib\@empty
\bibitem [{\citenamefont {Collins}\ and\ \citenamefont
  {Soper}(1981)}]{CS:1981}%
  \BibitemOpen
  \bibfield  {author} {\bibinfo {author} {\bibfnamefont {J.~C.}\ \bibnamefont
  {Collins}}\ and\ \bibinfo {author} {\bibfnamefont {D.~E.}\ \bibnamefont
  {Soper}},\ }\bibfield  {title} {\enquote {\bibinfo {title} {{Back-To-Back
  Jets in QCD}},}\ }\href {\doibase 10.1016/0550-3213(81)90339-4} {\bibfield
  {journal} {\bibinfo  {journal} {Nucl. Phys. B}\ }\textbf {\bibinfo {volume}
  {193}},\ \bibinfo {pages} {381} (\bibinfo {year} {1981})},\ \bibinfo {note}
  {[Erratum: Nucl. Phys.B213,545(1983)]}\BibitemShut {NoStop}%
\bibitem [{\citenamefont {Collins}\ and\ \citenamefont
  {Soper}(1982)}]{CS:1982}%
  \BibitemOpen
  \bibfield  {author} {\bibinfo {author} {\bibfnamefont {J.~C.}\ \bibnamefont
  {Collins}}\ and\ \bibinfo {author} {\bibfnamefont {D.~E.}\ \bibnamefont
  {Soper}},\ }\bibfield  {title} {\enquote {\bibinfo {title} {{Parton
  Distribution and Decay Functions}},}\ }\href {\doibase
  10.1016/0550-3213(82)90021-9} {\bibfield  {journal} {\bibinfo  {journal}
  {Nucl. Phys. B}\ }\textbf {\bibinfo {volume} {194}},\ \bibinfo {pages} {445}
  (\bibinfo {year} {1982})}\BibitemShut {NoStop}%
\bibitem [{\citenamefont {Collins}\ \emph {et~al.}(1985)\citenamefont
  {Collins}, \citenamefont {Soper},\ and\ \citenamefont
  {Sterman}}]{css_evolution}%
  \BibitemOpen
  \bibfield  {author} {\bibinfo {author} {\bibfnamefont {J.~C.}\ \bibnamefont
  {Collins}}, \bibinfo {author} {\bibfnamefont {D.~E.}\ \bibnamefont {Soper}},
  \ and\ \bibinfo {author} {\bibfnamefont {G.~F.}\ \bibnamefont {Sterman}},\
  }\bibfield  {title} {\enquote {\bibinfo {title} {{Transverse Momentum
  Distribution in Drell-Yan Pair and $W$ and $Z$ Boson Production}},}\ }\href
  {\doibase 10.1016/0550-3213(85)90479-1} {\bibfield  {journal} {\bibinfo
  {journal} {Nucl. Phys. B}\ }\textbf {\bibinfo {volume} {250}},\ \bibinfo
  {pages} {199} (\bibinfo {year} {1985})}\BibitemShut {NoStop}%
\bibitem [{\citenamefont {Sivers}(1990)}]{Sivers:1990}%
  \BibitemOpen
  \bibfield  {author} {\bibinfo {author} {\bibfnamefont {D.~W.}\ \bibnamefont
  {Sivers}},\ }\bibfield  {title} {\enquote {\bibinfo {title} {{Single Spin
  Production Asymmetries from the Hard Scattering of Point-Like
  Constituents}},}\ }\href {\doibase 10.1103/PhysRevD.41.83} {\bibfield
  {journal} {\bibinfo  {journal} {Phys. Rev. D}\ }\textbf {\bibinfo {volume}
  {41}},\ \bibinfo {pages} {83} (\bibinfo {year} {1990})}\BibitemShut {NoStop}%
\bibitem [{\citenamefont {Sivers}(1991)}]{Sivers:1991}%
  \BibitemOpen
  \bibfield  {author} {\bibinfo {author} {\bibfnamefont {D.~W.}\ \bibnamefont
  {Sivers}},\ }\bibfield  {title} {\enquote {\bibinfo {title} {{Hard scattering
  scaling laws for single spin production asymmetries}},}\ }\href {\doibase
  10.1103/PhysRevD.43.261} {\bibfield  {journal} {\bibinfo  {journal} {Phys.
  Rev. D}\ }\textbf {\bibinfo {volume} {43}},\ \bibinfo {pages} {261} (\bibinfo
  {year} {1991})}\BibitemShut {NoStop}%
\bibitem [{\citenamefont {Mulders}\ and\ \citenamefont
  {Tangerman}(1996)}]{Mulders-Tangerman:1995}%
  \BibitemOpen
  \bibfield  {author} {\bibinfo {author} {\bibfnamefont {P.~J.}\ \bibnamefont
  {Mulders}}\ and\ \bibinfo {author} {\bibfnamefont {R.~D.}\ \bibnamefont
  {Tangerman}},\ }\bibfield  {title} {\enquote {\bibinfo {title} {{The Complete
  tree level result up to order 1/Q for polarized deep inelastic
  leptoproduction}},}\ }\href {\doibase doi:10.1016/0550-3213(95)00632-X}
  {\bibfield  {journal} {\bibinfo  {journal} {Nucl. Phys. B}\ }\textbf
  {\bibinfo {volume} {461}},\ \bibinfo {pages} {197} (\bibinfo {year}
  {1996})},\ \bibinfo {note} {[Erratum: Nucl. Phys.B484,538(1997)]}\BibitemShut
  {NoStop}%
\bibitem [{\citenamefont {Collins}(2013)}]{Collins_Book}%
  \BibitemOpen
  \bibfield  {author} {\bibinfo {author} {\bibfnamefont {J.~C.}\ \bibnamefont
  {Collins}},\ }\href {http://www.cambridge.org/de/knowledge/isbn/item5756723}
  {\emph {\bibinfo {title} {{Foundations of perturbative QCD}}}}\ (\bibinfo
  {publisher} {Cambridge University Press},\ \bibinfo {year}
  {2013})\BibitemShut {NoStop}%
\bibitem [{\citenamefont {Musch}\ \emph {et~al.}(2012)\citenamefont {Musch},
  \citenamefont {Hagler}, \citenamefont {Engelhardt}, \citenamefont {Negele},\
  and\ \citenamefont {Schafer}}]{Musch:lQCD}%
  \BibitemOpen
  \bibfield  {author} {\bibinfo {author} {\bibfnamefont {B.~U.}\ \bibnamefont
  {Musch}}, \bibinfo {author} {\bibfnamefont {P.}~\bibnamefont {Hagler}},
  \bibinfo {author} {\bibfnamefont {M.}~\bibnamefont {Engelhardt}}, \bibinfo
  {author} {\bibfnamefont {J.~W.}\ \bibnamefont {Negele}}, \ and\ \bibinfo
  {author} {\bibfnamefont {A.}~\bibnamefont {Schafer}},\ }\bibfield  {title}
  {\enquote {\bibinfo {title} {{Sivers and Boer-Mulders observables from
  lattice QCD}},}\ }\href {\doibase 10.1103/PhysRevD.85.094510} {\bibfield
  {journal} {\bibinfo  {journal} {Phys. Rev. D}\ }\textbf {\bibinfo {volume}
  {85}},\ \bibinfo {pages} {094510} (\bibinfo {year} {2012})}\BibitemShut
  {NoStop}%
\bibitem [{\citenamefont {Musch}\ \emph {et~al.}(2011)\citenamefont {Musch},
  \citenamefont {Hagler}, \citenamefont {Negele},\ and\ \citenamefont
  {Schafer}}]{Musch:2011}%
  \BibitemOpen
  \bibfield  {author} {\bibinfo {author} {\bibfnamefont {B.~U.}\ \bibnamefont
  {Musch}}, \bibinfo {author} {\bibfnamefont {P.}~\bibnamefont {Hagler}},
  \bibinfo {author} {\bibfnamefont {J.~W.}\ \bibnamefont {Negele}}, \ and\
  \bibinfo {author} {\bibfnamefont {A.}~\bibnamefont {Schafer}},\ }\bibfield
  {title} {\enquote {\bibinfo {title} {{Exploring quark transverse momentum
  distributions with lattice QCD}},}\ }\href {\doibase
  10.1103/PhysRevD.83.094507} {\bibfield  {journal} {\bibinfo  {journal} {Phys.
  Rev. D}\ }\textbf {\bibinfo {volume} {83}},\ \bibinfo {pages} {094507}
  (\bibinfo {year} {2011})}\BibitemShut {NoStop}%
\bibitem [{\citenamefont {Ji}(2013)}]{Ji:lQCD}%
  \BibitemOpen
  \bibfield  {author} {\bibinfo {author} {\bibfnamefont {X.}~\bibnamefont
  {Ji}},\ }\bibfield  {title} {\enquote {\bibinfo {title} {{Parton Physics on a
  Euclidean Lattice}},}\ }\href {\doibase 10.1103/PhysRevLett.110.262002}
  {\bibfield  {journal} {\bibinfo  {journal} {Phys. Rev. Lett.}\ }\textbf
  {\bibinfo {volume} {110}},\ \bibinfo {pages} {262002} (\bibinfo {year}
  {2013})}\BibitemShut {NoStop}%
\bibitem [{\citenamefont {Airapetian}\ \emph {et~al.}(2005)\citenamefont
  {Airapetian} \emph {et~al.}}]{hermes_tmds}%
  \BibitemOpen
  \bibfield  {author} {\bibinfo {author} {\bibfnamefont {A.}~\bibnamefont
  {Airapetian}} \emph {et~al.} (\bibinfo {collaboration} {HERMES
  Collaboration}),\ }\bibfield  {title} {\enquote {\bibinfo {title}
  {{Single-spin asymmetries in semi-inclusive deep-inelastic scattering on a
  transversely polarized hydrogen target}},}\ }\href {\doibase
  10.1103/PhysRevLett.94.012002} {\bibfield  {journal} {\bibinfo  {journal}
  {Phys. Rev. Lett.}\ }\textbf {\bibinfo {volume} {94}},\ \bibinfo {pages}
  {012002} (\bibinfo {year} {2005})}\BibitemShut {NoStop}%
\bibitem [{\citenamefont {Adolph}\ \emph {et~al.}(2015)\citenamefont {Adolph}
  \emph {et~al.}}]{compass_tmds}%
  \BibitemOpen
  \bibfield  {author} {\bibinfo {author} {\bibfnamefont {C.}~\bibnamefont
  {Adolph}} \emph {et~al.} (\bibinfo {collaboration} {COMPASS Collaboration}),\
  }\bibfield  {title} {\enquote {\bibinfo {title} {{Collins and Sivers
  asymmetries in muonproduction of pions and kaons off transversely polarised
  protons}},}\ }\href {\doibase 10.1016/j.physletb.2015.03.056} {\bibfield
  {journal} {\bibinfo  {journal} {Phys. Lett. B}\ }\textbf {\bibinfo {volume}
  {744}},\ \bibinfo {pages} {250} (\bibinfo {year} {2015})}\BibitemShut
  {NoStop}%
\bibitem [{\citenamefont {Zhu}\ \emph {et~al.}(2009)\citenamefont {Zhu} \emph
  {et~al.}}]{E866_DY}%
  \BibitemOpen
  \bibfield  {author} {\bibinfo {author} {\bibfnamefont {L.~Y.}\ \bibnamefont
  {Zhu}} \emph {et~al.} (\bibinfo {collaboration} {NuSea Collaboration}),\
  }\bibfield  {title} {\enquote {\bibinfo {title} {{Measurement of Angular
  Distributions of Drell-Yan Dimuons in $p$$+$$p$ Interactions at
  800-GeV/$c$}},}\ }\href {\doibase 10.1103/PhysRevLett.102.182001} {\bibfield
  {journal} {\bibinfo  {journal} {Phys. Rev. Lett.}\ }\textbf {\bibinfo
  {volume} {102}},\ \bibinfo {pages} {182001} (\bibinfo {year}
  {2009})}\BibitemShut {NoStop}%
\bibitem [{\citenamefont {Guanziroli}\ \emph {et~al.}(1988)\citenamefont
  {Guanziroli} \emph {et~al.}}]{NA10_DY}%
  \BibitemOpen
  \bibfield  {author} {\bibinfo {author} {\bibfnamefont {M.}~\bibnamefont
  {Guanziroli}} \emph {et~al.} (\bibinfo {collaboration} {NA10
  Collaboration}),\ }\bibfield  {title} {\enquote {\bibinfo {title} {{Angular
  Distributions of Muon Pairs Produced by Negative Pions on Deuterium and
  Tungsten}},}\ }\href {\doibase 10.1007/BF01549713} {\bibfield  {journal}
  {\bibinfo  {journal} {Z. Phys. C}\ }\textbf {\bibinfo {volume} {37}},\
  \bibinfo {pages} {545} (\bibinfo {year} {1988})}\BibitemShut {NoStop}%
\bibitem [{\citenamefont {Conway}\ \emph {et~al.}(1989)\citenamefont {Conway}
  \emph {et~al.}}]{E615_DY}%
  \BibitemOpen
  \bibfield  {author} {\bibinfo {author} {\bibfnamefont {J.~S.}\ \bibnamefont
  {Conway}} \emph {et~al.},\ }\bibfield  {title} {\enquote {\bibinfo {title}
  {{Experimental Study of Muon Pairs Produced by 252-GeV Pions on Tungsten}},}\
  }\href {\doibase 10.1103/PhysRevD.39.92} {\bibfield  {journal} {\bibinfo
  {journal} {Phys. Rev. D}\ }\textbf {\bibinfo {volume} {39}},\ \bibinfo
  {pages} {92} (\bibinfo {year} {1989})}\BibitemShut {NoStop}%
\bibitem [{\citenamefont {Adamczyk}\ \emph {et~al.}(2016)\citenamefont
  {Adamczyk} \emph {et~al.}}]{STAR_W_An}%
  \BibitemOpen
  \bibfield  {author} {\bibinfo {author} {\bibfnamefont {L.}~\bibnamefont
  {Adamczyk}} \emph {et~al.} (\bibinfo {collaboration} {STAR Collaboration}),\
  }\bibfield  {title} {\enquote {\bibinfo {title} {{Measurement of the
  transverse single-spin asymmetry in $p^\uparrow+p \to W^{\pm}/Z^0$ at
  RHIC}},}\ }\href {\doibase 10.1103/PhysRevLett.116.132301} {\bibfield
  {journal} {\bibinfo  {journal} {Phys. Rev. Lett.}\ }\textbf {\bibinfo
  {volume} {116}},\ \bibinfo {pages} {132301} (\bibinfo {year}
  {2016})}\BibitemShut {NoStop}%
\bibitem [{\citenamefont {Qian}\ \emph {et~al.}(2011)\citenamefont {Qian} \emph
  {et~al.}}]{JLAB_Collins_Sivers}%
  \BibitemOpen
  \bibfield  {author} {\bibinfo {author} {\bibfnamefont {X.}~\bibnamefont
  {Qian}} \emph {et~al.} (\bibinfo {collaboration} {Jefferson Lab Hall A}),\
  }\bibfield  {title} {\enquote {\bibinfo {title} {{Single Spin Asymmetries in
  Charged Pion Production from Semi-Inclusive Deep Inelastic Scattering on a
  Transversely Polarized $^3$He Target}},}\ }\href {\doibase
  10.1103/PhysRevLett.107.072003} {\bibfield  {journal} {\bibinfo  {journal}
  {Phys. Rev. Lett.}\ }\textbf {\bibinfo {volume} {107}},\ \bibinfo {pages}
  {072003} (\bibinfo {year} {2011})}\BibitemShut {NoStop}%
\bibitem [{\citenamefont {Avakian}\ \emph {et~al.}(2010)\citenamefont {Avakian}
  \emph {et~al.}}]{JLAB_clas_asym}%
  \BibitemOpen
  \bibfield  {author} {\bibinfo {author} {\bibfnamefont {H.}~\bibnamefont
  {Avakian}} \emph {et~al.} (\bibinfo {collaboration} {CLAS Collaboration}),\
  }\bibfield  {title} {\enquote {\bibinfo {title} {{Measurement of Single and
  Double Spin Asymmetries in Deep Inelastic Pion Electroproduction with a
  Longitudinally Polarized Target}},}\ }\href {\doibase
  10.1103/PhysRevLett.105.262002} {\bibfield  {journal} {\bibinfo  {journal}
  {Phys. Rev. Lett.}\ }\textbf {\bibinfo {volume} {105}},\ \bibinfo {pages}
  {262002} (\bibinfo {year} {2010})}\BibitemShut {NoStop}%
\bibitem [{\citenamefont {Seidl}\ \emph {et~al.}(2006)\citenamefont {Seidl}
  \emph {et~al.}}]{BELLE_asymm}%
  \BibitemOpen
  \bibfield  {author} {\bibinfo {author} {\bibfnamefont {R.}~\bibnamefont
  {Seidl}} \emph {et~al.} (\bibinfo {collaboration} {BELLE Collaboration}),\
  }\bibfield  {title} {\enquote {\bibinfo {title} {{Measurement of azimuthal
  asymmetries in inclusive production of hadron pairs in $e^+e^-$ annihilation
  at Belle}},}\ }\href {\doibase 10.1103/PhysRevLett.96.232002} {\bibfield
  {journal} {\bibinfo  {journal} {Phys. Rev. Lett.}\ }\textbf {\bibinfo
  {volume} {96}},\ \bibinfo {pages} {232002} (\bibinfo {year}
  {2006})}\BibitemShut {NoStop}%
\bibitem [{\citenamefont {Lees}\ \emph {et~al.}(2014)\citenamefont {Lees} \emph
  {et~al.}}]{BABAR_asymm}%
  \BibitemOpen
  \bibfield  {author} {\bibinfo {author} {\bibfnamefont {J.~P.}\ \bibnamefont
  {Lees}} \emph {et~al.} (\bibinfo {collaboration} {BaBar Collaboration}),\
  }\bibfield  {title} {\enquote {\bibinfo {title} {{Measurement of Collins
  asymmetries in inclusive production of charged pion pairs in e$^+$e$^-$
  annihilation at BABAR}},}\ }\href {\doibase 10.1103/PhysRevD.90.052003}
  {\bibfield  {journal} {\bibinfo  {journal} {Phys. Rev. D}\ }\textbf {\bibinfo
  {volume} {90}},\ \bibinfo {pages} {052003} (\bibinfo {year}
  {2014})}\BibitemShut {NoStop}%
\bibitem [{\citenamefont {Airapetian}\ \emph {et~al.}(2010)\citenamefont
  {Airapetian} \emph {et~al.}}]{HERMES_collins}%
  \BibitemOpen
  \bibfield  {author} {\bibinfo {author} {\bibfnamefont {A.}~\bibnamefont
  {Airapetian}} \emph {et~al.} (\bibinfo {collaboration} {HERMES
  Collaboration}),\ }\bibfield  {title} {\enquote {\bibinfo {title} {{Effects
  of transversity in deep-inelastic scattering by polarized protons}},}\ }\href
  {\doibase 10.1016/j.physletb.2010.08.012} {\bibfield  {journal} {\bibinfo
  {journal} {Phys. Lett. B}\ }\textbf {\bibinfo {volume} {693}},\ \bibinfo
  {pages} {11} (\bibinfo {year} {2010})}\BibitemShut {NoStop}%
\bibitem [{\citenamefont {Adare}\ \emph
  {et~al.}(2014{\natexlab{a}})\citenamefont {Adare} \emph {et~al.}}]{eta_TSSA}%
  \BibitemOpen
  \bibfield  {author} {\bibinfo {author} {\bibfnamefont {A.}~\bibnamefont
  {Adare}} \emph {et~al.} (\bibinfo {collaboration} {PHENIX Collaboration}),\
  }\bibfield  {title} {\enquote {\bibinfo {title} {{Cross section and
  transverse single-spin asymmetry of $\eta$ mesons in $p^{\uparrow}+p$
  collisions at $\sqrt{s}=200$ GeV at forward rapidity}},}\ }\href {\doibase
  10.1103/PhysRevD.90.072008} {\bibfield  {journal} {\bibinfo  {journal} {Phys.
  Rev. D}\ }\textbf {\bibinfo {volume} {90}},\ \bibinfo {pages} {072008}
  (\bibinfo {year} {2014}{\natexlab{a}})}\BibitemShut {NoStop}%
\bibitem [{\citenamefont {Adams}\ \emph {et~al.}(2004)\citenamefont {Adams}
  \emph {et~al.}}]{STAR_pi0_TSSA}%
  \BibitemOpen
  \bibfield  {author} {\bibinfo {author} {\bibfnamefont {J.}~\bibnamefont
  {Adams}} \emph {et~al.} (\bibinfo {collaboration} {STAR Collaboration}),\
  }\bibfield  {title} {\enquote {\bibinfo {title} {{Cross-sections and
  transverse single spin asymmetries in forward neutral pion production from
  proton collisions at $\sqrt{s}=200$ GeV}},}\ }\href {\doibase
  10.1103/PhysRevLett.92.171801} {\bibfield  {journal} {\bibinfo  {journal}
  {Phys. Rev. Lett.}\ }\textbf {\bibinfo {volume} {92}},\ \bibinfo {pages}
  {171801} (\bibinfo {year} {2004})}\BibitemShut {NoStop}%
\bibitem [{\citenamefont {Arsene}\ \emph {et~al.}(2008)\citenamefont {Arsene}
  \emph {et~al.}}]{BRAHMs_TSSA}%
  \BibitemOpen
  \bibfield  {author} {\bibinfo {author} {\bibfnamefont {I.}~\bibnamefont
  {Arsene}} \emph {et~al.} (\bibinfo {collaboration} {BRAHMS Collaboration}),\
  }\bibfield  {title} {\enquote {\bibinfo {title} {{Single Transverse Spin
  Asymmetries of Identified Charged Hadrons in Polarized $p$$+$$p$ Collisions
  at $\sqrt{s}=62.4$ GeV}},}\ }\href {\doibase 10.1103/PhysRevLett.101.042001}
  {\bibfield  {journal} {\bibinfo  {journal} {Phys. Rev. Lett.}\ }\textbf
  {\bibinfo {volume} {101}},\ \bibinfo {pages} {042001} (\bibinfo {year}
  {2008})}\BibitemShut {NoStop}%
\bibitem [{\citenamefont {Belitsky}\ \emph {et~al.}(2003)\citenamefont
  {Belitsky}, \citenamefont {Ji},\ and\ \citenamefont {Yuan}}]{BelitskyYuan}%
  \BibitemOpen
  \bibfield  {author} {\bibinfo {author} {\bibfnamefont {A.~V.}\ \bibnamefont
  {Belitsky}}, \bibinfo {author} {\bibfnamefont {X.}~\bibnamefont {Ji}}, \ and\
  \bibinfo {author} {\bibfnamefont {F.}~\bibnamefont {Yuan}},\ }\bibfield
  {title} {\enquote {\bibinfo {title} {{Final state interactions and gauge
  invariant parton distributions}},}\ }\href {\doibase
  10.1016/S0550-3213(03)00121-4} {\bibfield  {journal} {\bibinfo  {journal}
  {Nucl. Phys. B}\ }\textbf {\bibinfo {volume} {656}},\ \bibinfo {pages} {165}
  (\bibinfo {year} {2003})}\BibitemShut {NoStop}%
\bibitem [{\citenamefont {Brodsky}\ \emph {et~al.}(2002)\citenamefont
  {Brodsky}, \citenamefont {Hwang},\ and\ \citenamefont
  {Schmidt}}]{Brodsky_SIDIS}%
  \BibitemOpen
  \bibfield  {author} {\bibinfo {author} {\bibfnamefont {S.~J.}\ \bibnamefont
  {Brodsky}}, \bibinfo {author} {\bibfnamefont {D.~S.}\ \bibnamefont {Hwang}},
  \ and\ \bibinfo {author} {\bibfnamefont {I.}~\bibnamefont {Schmidt}},\
  }\bibfield  {title} {\enquote {\bibinfo {title} {{Final state interactions
  and single spin asymmetries in semiinclusive deep inelastic scattering}},}\
  }\href {\doibase 10.1016/S0370-2693(02)01320-5} {\bibfield  {journal}
  {\bibinfo  {journal} {Phys. Lett. B}\ }\textbf {\bibinfo {volume} {530}},\
  \bibinfo {pages} {99} (\bibinfo {year} {2002})}\BibitemShut {NoStop}%
\bibitem [{\citenamefont {Collins}(2002)}]{collins_sivers_prediction}%
  \BibitemOpen
  \bibfield  {author} {\bibinfo {author} {\bibfnamefont {J.~C.}\ \bibnamefont
  {Collins}},\ }\bibfield  {title} {\enquote {\bibinfo {title} {{Leading twist
  single transverse-spin asymmetries: Drell-Yan and deep inelastic
  scattering}},}\ }\href {\doibase 10.1016/S0370-2693(02)01819-1} {\bibfield
  {journal} {\bibinfo  {journal} {Phys. Lett. B}\ }\textbf {\bibinfo {volume}
  {536}},\ \bibinfo {pages} {43} (\bibinfo {year} {2002})}\BibitemShut
  {NoStop}%
\bibitem [{\citenamefont {Rogers}\ and\ \citenamefont
  {Mulders}(2010)}]{trogers_factbreaking}%
  \BibitemOpen
  \bibfield  {author} {\bibinfo {author} {\bibfnamefont {T.~C.}\ \bibnamefont
  {Rogers}}\ and\ \bibinfo {author} {\bibfnamefont {P.~J.}\ \bibnamefont
  {Mulders}},\ }\bibfield  {title} {\enquote {\bibinfo {title} {{No Generalized
  TMD-Factorization in Hadro-Production of High Transverse Momentum
  Hadrons}},}\ }\href {\doibase 10.1103/PhysRevD.81.094006} {\bibfield
  {journal} {\bibinfo  {journal} {Phys. Rev. D}\ }\textbf {\bibinfo {volume}
  {81}},\ \bibinfo {pages} {094006} (\bibinfo {year} {2010})}\BibitemShut
  {NoStop}%
\bibitem [{\citenamefont {Bomhof}\ \emph {et~al.}(2006)\citenamefont {Bomhof},
  \citenamefont {Mulders},\ and\ \citenamefont {Pijlman}}]{Mulders:2006}%
  \BibitemOpen
  \bibfield  {author} {\bibinfo {author} {\bibfnamefont {C.~J.}\ \bibnamefont
  {Bomhof}}, \bibinfo {author} {\bibfnamefont {P.~J.}\ \bibnamefont {Mulders}},
  \ and\ \bibinfo {author} {\bibfnamefont {F.}~\bibnamefont {Pijlman}},\
  }\bibfield  {title} {\enquote {\bibinfo {title} {{The Construction of
  gauge-links in arbitrary hard processes}},}\ }\href {\doibase
  10.1140/epjc/s2006-02554-2} {\bibfield  {journal} {\bibinfo  {journal} {Eur.
  Phys. J. C}\ }\textbf {\bibinfo {volume} {47}},\ \bibinfo {pages} {147}
  (\bibinfo {year} {2006})}\BibitemShut {NoStop}%
\bibitem [{\citenamefont {Collins}\ and\ \citenamefont
  {Qiu}(2007)}]{Collins:2007}%
  \BibitemOpen
  \bibfield  {author} {\bibinfo {author} {\bibfnamefont {J.~C.}\ \bibnamefont
  {Collins}}\ and\ \bibinfo {author} {\bibfnamefont {J.~W.}\ \bibnamefont
  {Qiu}},\ }\bibfield  {title} {\enquote {\bibinfo {title} {{$k_{T}$
  factorization is violated in production of high-transverse-momentum particles
  in hadron-hadron collisions}},}\ }\href {\doibase 10.1103/PhysRevD.75.114014}
  {\bibfield  {journal} {\bibinfo  {journal} {Phys. Rev. D}\ }\textbf {\bibinfo
  {volume} {75}},\ \bibinfo {pages} {114014} (\bibinfo {year}
  {2007})}\BibitemShut {NoStop}%
\bibitem [{\citenamefont {Collins}({\natexlab{a}})}]{Collins:2007preprint}%
  \BibitemOpen
  \bibfield  {author} {\bibinfo {author} {\bibfnamefont {J.~C.}\ \bibnamefont
  {Collins}},\ }\href@noop {} {\enquote {\bibinfo {title} {{2-soft-gluon
  exchange and factorization breaking}},}\ } ({\natexlab{a}}),\ \bibinfo {note}
  {arXiv:0708.4410}\BibitemShut {NoStop}%
\bibitem [{\citenamefont {Abe}\ \emph {et~al.}(1994)\citenamefont {Abe} \emph
  {et~al.}}]{CDF_cc}%
  \BibitemOpen
  \bibfield  {author} {\bibinfo {author} {\bibfnamefont {F.}~\bibnamefont
  {Abe}} \emph {et~al.} (\bibinfo {collaboration} {CDF Collaboration}),\
  }\bibfield  {title} {\enquote {\bibinfo {title} {{Evidence for color
  coherence in $p\bar{p}$ collisions at $\sqrt{s}=1.8$ TeV}},}\ }\href
  {\doibase 10.1103/PhysRevD.50.5562} {\bibfield  {journal} {\bibinfo
  {journal} {Phys. Rev. D}\ }\textbf {\bibinfo {volume} {50}},\ \bibinfo
  {pages} {5562} (\bibinfo {year} {1994})}\BibitemShut {NoStop}%
\bibitem [{\citenamefont {Abbott}\ \emph {et~al.}(1997)\citenamefont {Abbott}
  \emph {et~al.}}]{D0_cc}%
  \BibitemOpen
  \bibfield  {author} {\bibinfo {author} {\bibfnamefont {B.}~\bibnamefont
  {Abbott}} \emph {et~al.} (\bibinfo {collaboration} {D0 Collaboration}),\
  }\bibfield  {title} {\enquote {\bibinfo {title} {{Color coherent radiation in
  multijet events from $p\bar{p}$ collisions at $\sqrt{s}=1.8$ TeV}},}\ }\href
  {\doibase 10.1016/S0370-2693(97)01190-8} {\bibfield  {journal} {\bibinfo
  {journal} {Phys. Lett. B}\ }\textbf {\bibinfo {volume} {414}},\ \bibinfo
  {pages} {419} (\bibinfo {year} {1997})}\BibitemShut {NoStop}%
\bibitem [{\citenamefont {Chatrchyan}\ \emph {et~al.}(2014)\citenamefont
  {Chatrchyan} \emph {et~al.}}]{CMS_cc}%
  \BibitemOpen
  \bibfield  {author} {\bibinfo {author} {\bibfnamefont {S.}~\bibnamefont
  {Chatrchyan}} \emph {et~al.} (\bibinfo {collaboration} {CMS Collaboration}),\
  }\bibfield  {title} {\enquote {\bibinfo {title} {{Probing color coherence
  effects in $pp$ collisions at $\sqrt{s}=7$~TeV}},}\ }\href {\doibase
  10.1140/epjc/s10052-014-2901-8} {\bibfield  {journal} {\bibinfo  {journal}
  {Eur. Phys. J. C}\ }\textbf {\bibinfo {volume} {74}},\ \bibinfo {pages}
  {2901} (\bibinfo {year} {2014})}\BibitemShut {NoStop}%
\bibitem [{\citenamefont {Collins}({\natexlab{b}})}]{collins_CSfact_preprint}%
  \BibitemOpen
  \bibfield  {author} {\bibinfo {author} {\bibfnamefont {J.~C.}\ \bibnamefont
  {Collins}},\ }\href@noop {} {\enquote {\bibinfo {title} {{CSS Equation, etc,
  Follow from Structure of TMD Factorization}},}\ } ({\natexlab{b}}),\ \bibinfo
  {note} {arXiv:1212.5974}\BibitemShut {NoStop}%
\bibitem [{\citenamefont {Altarelli}\ and\ \citenamefont
  {Parisi}(1977)}]{DGLAP_equations}%
  \BibitemOpen
  \bibfield  {author} {\bibinfo {author} {\bibfnamefont {G.}~\bibnamefont
  {Altarelli}}\ and\ \bibinfo {author} {\bibfnamefont {G.}~\bibnamefont
  {Parisi}},\ }\bibfield  {title} {\enquote {\bibinfo {title} {{Asymptotic
  Freedom in Parton Language}},}\ }\href {\doibase
  10.1016/0550-3213(77)90384-4} {\bibfield  {journal} {\bibinfo  {journal}
  {Nucl. Phys. B}\ }\textbf {\bibinfo {volume} {126}},\ \bibinfo {pages} {298}
  (\bibinfo {year} {1977})}\BibitemShut {NoStop}%
\bibitem [{\citenamefont {Dokshitzer}(1977)}]{DGLAP_2}%
  \BibitemOpen
  \bibfield  {author} {\bibinfo {author} {\bibfnamefont {Y.~L.}\ \bibnamefont
  {Dokshitzer}},\ }\bibfield  {title} {\enquote {\bibinfo {title} {{Calculation
  of the Structure Functions for Deep Inelastic Scattering and $e^+e^-$
  Annihilation by Perturbation Theory in Quantum Chromodynamics.}}}\
  }\href@noop {} {\bibfield  {journal} {\bibinfo  {journal} {Sov. Phys. JETP}\
  }\textbf {\bibinfo {volume} {46}},\ \bibinfo {pages} {641} (\bibinfo {year}
  {1977})}\BibitemShut {NoStop}%
\bibitem [{\citenamefont {Gribov}\ and\ \citenamefont
  {Lipatov}(1972)}]{DGLAP_3}%
  \BibitemOpen
  \bibfield  {author} {\bibinfo {author} {\bibfnamefont {V.~N.}\ \bibnamefont
  {Gribov}}\ and\ \bibinfo {author} {\bibfnamefont {L.~N.}\ \bibnamefont
  {Lipatov}},\ }\bibfield  {title} {\enquote {\bibinfo {title} {{Deep inelastic
  e p scattering in perturbation theory}},}\ }\href@noop {} {\bibfield
  {journal} {\bibinfo  {journal} {Sov. J. Nucl. Phys.}\ }\textbf {\bibinfo
  {volume} {15}},\ \bibinfo {pages} {438} (\bibinfo {year} {1972})}\BibitemShut
  {NoStop}%
\bibitem [{\citenamefont {Collins}\ and\ \citenamefont
  {Rogers}(2015)}]{collinsrogers_tmdevolutionkernel}%
  \BibitemOpen
  \bibfield  {author} {\bibinfo {author} {\bibfnamefont {J.~C.}\ \bibnamefont
  {Collins}}\ and\ \bibinfo {author} {\bibfnamefont {T.~C.}\ \bibnamefont
  {Rogers}},\ }\bibfield  {title} {\enquote {\bibinfo {title} {{Understanding
  the large-distance behavior of transverse-momentum-dependent parton densities
  and the Collins-Soper evolution kernel}},}\ }\href {\doibase
  10.1103/PhysRevD.91.074020} {\bibfield  {journal} {\bibinfo  {journal} {Phys.
  Rev. D}\ }\textbf {\bibinfo {volume} {91}},\ \bibinfo {pages} {074020}
  (\bibinfo {year} {2015})}\BibitemShut {NoStop}%
\bibitem [{\citenamefont {Landry}\ \emph {et~al.}(2003)\citenamefont {Landry},
  \citenamefont {Brock}, \citenamefont {Nadolsky},\ and\ \citenamefont
  {Yuan}}]{Tevatron_Zboson_resum}%
  \BibitemOpen
  \bibfield  {author} {\bibinfo {author} {\bibfnamefont {F.}~\bibnamefont
  {Landry}}, \bibinfo {author} {\bibfnamefont {R.}~\bibnamefont {Brock}},
  \bibinfo {author} {\bibfnamefont {P.~M.}\ \bibnamefont {Nadolsky}}, \ and\
  \bibinfo {author} {\bibfnamefont {C.~P.}\ \bibnamefont {Yuan}},\ }\bibfield
  {title} {\enquote {\bibinfo {title} {{Tevatron Run-1 $Z$ boson data and
  Collins-Soper-Sterman resummation formalism}},}\ }\href {\doibase
  10.1103/PhysRevD.67.073016} {\bibfield  {journal} {\bibinfo  {journal} {Phys.
  Rev. D}\ }\textbf {\bibinfo {volume} {67}},\ \bibinfo {pages} {073016}
  (\bibinfo {year} {2003})}\BibitemShut {NoStop}%
\bibitem [{\citenamefont {Konychev}\ and\ \citenamefont
  {Nadolsky}(2006)}]{Nadolsky_DYZ_globfit}%
  \BibitemOpen
  \bibfield  {author} {\bibinfo {author} {\bibfnamefont {A.~V.}\ \bibnamefont
  {Konychev}}\ and\ \bibinfo {author} {\bibfnamefont {P.~M.}\ \bibnamefont
  {Nadolsky}},\ }\bibfield  {title} {\enquote {\bibinfo {title} {{Universality
  of the Collins-Soper-Sterman nonperturbative function in gauge boson
  production}},}\ }\href {\doibase 10.1016/j.physletb.2005.12.063} {\bibfield
  {journal} {\bibinfo  {journal} {Phys. Lett. B}\ }\textbf {\bibinfo {volume}
  {633}},\ \bibinfo {pages} {710} (\bibinfo {year} {2006})}\BibitemShut
  {NoStop}%
\bibitem [{\citenamefont {Schweitzer}\ \emph {et~al.}(2010)\citenamefont
  {Schweitzer}, \citenamefont {Teckentrup},\ and\ \citenamefont
  {Metz}}]{Metz_intrin_kt}%
  \BibitemOpen
  \bibfield  {author} {\bibinfo {author} {\bibfnamefont {P.}~\bibnamefont
  {Schweitzer}}, \bibinfo {author} {\bibfnamefont {T.}~\bibnamefont
  {Teckentrup}}, \ and\ \bibinfo {author} {\bibfnamefont {A.}~\bibnamefont
  {Metz}},\ }\bibfield  {title} {\enquote {\bibinfo {title} {{Intrinsic
  transverse parton momenta in deeply inelastic reactions}},}\ }\href {\doibase
  10.1103/PhysRevD.81.094019} {\bibfield  {journal} {\bibinfo  {journal} {Phys.
  Rev. D}\ }\textbf {\bibinfo {volume} {81}},\ \bibinfo {pages} {094019}
  (\bibinfo {year} {2010})}\BibitemShut {NoStop}%
\bibitem [{\citenamefont {Aidala}\ \emph {et~al.}(2014)\citenamefont {Aidala},
  \citenamefont {Field}, \citenamefont {Gamberg},\ and\ \citenamefont
  {Rogers}}]{aidala_rogers_2014}%
  \BibitemOpen
  \bibfield  {author} {\bibinfo {author} {\bibfnamefont {C.~A.}\ \bibnamefont
  {Aidala}}, \bibinfo {author} {\bibfnamefont {B.}~\bibnamefont {Field}},
  \bibinfo {author} {\bibfnamefont {L.~P.}\ \bibnamefont {Gamberg}}, \ and\
  \bibinfo {author} {\bibfnamefont {T.~C.}\ \bibnamefont {Rogers}},\ }\bibfield
   {title} {\enquote {\bibinfo {title} {{Limits on transverse momentum
  dependent evolution from semi-inclusive deep inelastic scattering at moderate
  $Q$}},}\ }\href {\doibase 10.1103/PhysRevD.89.094002} {\bibfield  {journal}
  {\bibinfo  {journal} {Phys. Rev. D}\ }\textbf {\bibinfo {volume} {89}},\
  \bibinfo {pages} {094002} (\bibinfo {year} {2014})}\BibitemShut {NoStop}%
\bibitem [{\citenamefont {Nadolsky}\ \emph {et~al.}(1999)\citenamefont
  {Nadolsky}, \citenamefont {Stump},\ and\ \citenamefont
  {Yuan}}]{SIDIS_evolution}%
  \BibitemOpen
  \bibfield  {author} {\bibinfo {author} {\bibfnamefont {P.}~\bibnamefont
  {Nadolsky}}, \bibinfo {author} {\bibfnamefont {D.~R.}\ \bibnamefont {Stump}},
  \ and\ \bibinfo {author} {\bibfnamefont {C.~P.}\ \bibnamefont {Yuan}},\
  }\bibfield  {title} {\enquote {\bibinfo {title} {{Semiinclusive hadron
  production at HERA: The Effect of QCD gluon resummation}},}\ }\href {\doibase
  10.1103/PhysRevD.64.059903(E), 10.1103/PhysRevD.61.014003} {\bibfield
  {journal} {\bibinfo  {journal} {Phys. Rev. D}\ }\textbf {\bibinfo {volume}
  {61}},\ \bibinfo {pages} {014003} (\bibinfo {year} {1999})},\ \bibinfo {note}
  {[Erratum: Phys. Rev. D {\bf 64},059903(E) (2001)]}\BibitemShut {NoStop}%
\bibitem [{\citenamefont {Adler}\ \emph {et~al.}(2006)\citenamefont {Adler}
  \emph {et~al.}}]{ppg029}%
  \BibitemOpen
  \bibfield  {author} {\bibinfo {author} {\bibfnamefont {S.~S.}\ \bibnamefont
  {Adler}} \emph {et~al.} (\bibinfo {collaboration} {PHENIX Collaboration}),\
  }\bibfield  {title} {\enquote {\bibinfo {title} {{Jet properties from
  dihadron correlations in $p+p$ collisions at $\sqrt{s}=200$ GeV}},}\ }\href
  {\doibase 10.1103/PhysRevD.74.072002} {\bibfield  {journal} {\bibinfo
  {journal} {Phys. Rev. D}\ }\textbf {\bibinfo {volume} {74}},\ \bibinfo
  {pages} {072002} (\bibinfo {year} {2006})}\BibitemShut {NoStop}%
\bibitem [{\citenamefont {Adare}\ \emph
  {et~al.}(2010{\natexlab{a}})\citenamefont {Adare} \emph {et~al.}}]{ppg089}%
  \BibitemOpen
  \bibfield  {author} {\bibinfo {author} {\bibfnamefont {A.}~\bibnamefont
  {Adare}} \emph {et~al.} (\bibinfo {collaboration} {PHENIX Collaboration}),\
  }\bibfield  {title} {\enquote {\bibinfo {title} {{Double-Helicity Dependence
  of Jet Properties from Dihadrons in Longitudinally Polarized $p+p$~Collisions
  at $\sqrt{s}=200$~GeV}},}\ }\href {\doibase 10.1103/PhysRevD.81.012002}
  {\bibfield  {journal} {\bibinfo  {journal} {Phys. Rev. D}\ }\textbf {\bibinfo
  {volume} {81}},\ \bibinfo {pages} {012002} (\bibinfo {year}
  {2010}{\natexlab{a}})}\BibitemShut {NoStop}%
\bibitem [{\citenamefont {Apanasevich}\ \emph {et~al.}(1998)\citenamefont
  {Apanasevich} \emph {et~al.}}]{e706_kt}%
  \BibitemOpen
  \bibfield  {author} {\bibinfo {author} {\bibfnamefont {L.}~\bibnamefont
  {Apanasevich}} \emph {et~al.} (\bibinfo {collaboration} {Fermilab E706}),\
  }\bibfield  {title} {\enquote {\bibinfo {title} {{Evidence for parton $k_T$
  effects in high $p_T$ particle production}},}\ }\href {\doibase
  10.1103/PhysRevLett.81.2642} {\bibfield  {journal} {\bibinfo  {journal}
  {Phys. Rev. Lett.}\ }\textbf {\bibinfo {volume} {81}},\ \bibinfo {pages}
  {2642} (\bibinfo {year} {1998})}\BibitemShut {NoStop}%
\bibitem [{\citenamefont {Angelis}\ \emph {et~al.}(1980)\citenamefont {Angelis}
  \emph {et~al.}}]{CCOR_kt}%
  \BibitemOpen
  \bibfield  {author} {\bibinfo {author} {\bibfnamefont {A.~L.~S.}\
  \bibnamefont {Angelis}} \emph {et~al.} (\bibinfo {collaboration}
  {CERN-Columbia-Oxford-Rockefeller, CCOR}),\ }\bibfield  {title} {\enquote
  {\bibinfo {title} {{A Measurement of the Transverse Momenta of Partons, and
  of Jet Fragmentation as a Function of $\sqrt{s}$ in $pp$ Collisions}},}\
  }\href {\doibase 10.1016/0370-2693(80)90572-9} {\bibfield  {journal}
  {\bibinfo  {journal} {Phys. Lett. B}\ }\textbf {\bibinfo {volume} {97}},\
  \bibinfo {pages} {163} (\bibinfo {year} {1980})}\BibitemShut {NoStop}%
\bibitem [{\citenamefont {Feynman}\ \emph {et~al.}(1978)\citenamefont
  {Feynman}, \citenamefont {Field},\ and\ \citenamefont {Fox}}]{FFF}%
  \BibitemOpen
  \bibfield  {author} {\bibinfo {author} {\bibfnamefont {R.~P.}\ \bibnamefont
  {Feynman}}, \bibinfo {author} {\bibfnamefont {R.~D.}\ \bibnamefont {Field}},
  \ and\ \bibinfo {author} {\bibfnamefont {G.~C.}\ \bibnamefont {Fox}},\
  }\bibfield  {title} {\enquote {\bibinfo {title} {{A Quantum Chromodynamic
  Approach for the Large Transverse Momentum Production of Particles and
  Jets}},}\ }\href {\doibase 10.1103/PhysRevD.18.3320} {\bibfield  {journal}
  {\bibinfo  {journal} {Phys. Rev. D}\ }\textbf {\bibinfo {volume} {18}},\
  \bibinfo {pages} {3320} (\bibinfo {year} {1978})}\BibitemShut {NoStop}%
\bibitem [{\citenamefont {Adare}\ \emph
  {et~al.}(2010{\natexlab{b}})\citenamefont {Adare} \emph {et~al.}}]{ppg095}%
  \BibitemOpen
  \bibfield  {author} {\bibinfo {author} {\bibfnamefont {A.}~\bibnamefont
  {Adare}} \emph {et~al.} (\bibinfo {collaboration} {PHENIX Collaboration}),\
  }\bibfield  {title} {\enquote {\bibinfo {title} {{High $p_T$ direct photon
  and $\pi^0$ triggered azimuthal jet correlations and measurement of $k_T$ for
  isolated direct photons in $p+p$ collisions at $\sqrt{s}=200$ GeV}},}\ }\href
  {\doibase 10.1103/PhysRevD.82.072001} {\bibfield  {journal} {\bibinfo
  {journal} {Phys. Rev. D}\ }\textbf {\bibinfo {volume} {82}},\ \bibinfo
  {pages} {072001} (\bibinfo {year} {2010}{\natexlab{b}})}\BibitemShut
  {NoStop}%
\bibitem [{\citenamefont {Adam}\ \emph {et~al.}(2015)\citenamefont {Adam} \emph
  {et~al.}}]{ALICE_dijet_kt}%
  \BibitemOpen
  \bibfield  {author} {\bibinfo {author} {\bibfnamefont {J.}~\bibnamefont
  {Adam}} \emph {et~al.} (\bibinfo {collaboration} {ALICE Collaboration}),\
  }\bibfield  {title} {\enquote {\bibinfo {title} {{Measurement of dijet $k_T$
  in $p$-Pb collisions at $\sqrt{s}_{NN}=5.02$ TeV}},}\ }\href {\doibase
  10.1016/j.physletb.2015.05.033} {\bibfield  {journal} {\bibinfo  {journal}
  {Phys. Lett. B}\ }\textbf {\bibinfo {volume} {746}},\ \bibinfo {pages} {385}
  (\bibinfo {year} {2015})}\BibitemShut {NoStop}%
\bibitem [{\citenamefont {Adcox}\ \emph
  {et~al.}(2003{\natexlab{a}})\citenamefont {Adcox} \emph
  {et~al.}}]{phenix_central_arms}%
  \BibitemOpen
  \bibfield  {author} {\bibinfo {author} {\bibfnamefont {K.}~\bibnamefont
  {Adcox}} \emph {et~al.} (\bibinfo {collaboration} {PHENIX Collaboration}),\
  }\bibfield  {title} {\enquote {\bibinfo {title} {{PHENIX central arm tracking
  detectors}},}\ }\href@noop {} {\bibfield  {journal} {\bibinfo  {journal}
  {Nucl. Instrum. Methods Phys. Res., Sec. A}\ }\textbf {\bibinfo {volume}
  {499}},\ \bibinfo {pages} {489} (\bibinfo {year}
  {2003}{\natexlab{a}})}\BibitemShut {NoStop}%
\bibitem [{\citenamefont {Aphecetche}\ \emph {et~al.}(2003)\citenamefont
  {Aphecetche} \emph {et~al.}}]{EMCal}%
  \BibitemOpen
  \bibfield  {author} {\bibinfo {author} {\bibfnamefont {L.}~\bibnamefont
  {Aphecetche}} \emph {et~al.} (\bibinfo {collaboration} {PHENIX
  Collaboration}),\ }\bibfield  {title} {\enquote {\bibinfo {title} {{PHENIX
  calorimeter}},}\ }\href {\doibase 10.1016/S0168-9002(02)01954-X} {\bibfield
  {journal} {\bibinfo  {journal} {Nucl. Instrum. Methods Phys. Res., Sec. A}\
  }\textbf {\bibinfo {volume} {499}},\ \bibinfo {pages} {521} (\bibinfo {year}
  {2003})}\BibitemShut {NoStop}%
\bibitem [{\citenamefont {Adare}\ \emph {et~al.}(2012)\citenamefont {Adare}
  \emph {et~al.}}]{directphoton_crosssection}%
  \BibitemOpen
  \bibfield  {author} {\bibinfo {author} {\bibfnamefont {A.}~\bibnamefont
  {Adare}} \emph {et~al.} (\bibinfo {collaboration} {PHENIX Collaboration}),\
  }\bibfield  {title} {\enquote {\bibinfo {title} {{Direct-Photon Production in
  $p+p$ Collisions at $\sqrt{s}=200$ GeV at Midrapidity}},}\ }\href {\doibase
  10.1103/PhysRevD.86.072008} {\bibfield  {journal} {\bibinfo  {journal} {Phys.
  Rev. D}\ }\textbf {\bibinfo {volume} {86}},\ \bibinfo {pages} {072008}
  (\bibinfo {year} {2012})}\BibitemShut {NoStop}%
\bibitem [{\citenamefont {Adare}\ \emph {et~al.}(2016)\citenamefont {Adare}
  \emph {et~al.}}]{ppg186}%
  \BibitemOpen
  \bibfield  {author} {\bibinfo {author} {\bibfnamefont {A.}~\bibnamefont
  {Adare}} \emph {et~al.} (\bibinfo {collaboration} {PHENIX Collaboration}),\
  }\bibfield  {title} {\enquote {\bibinfo {title} {{Inclusive cross section and
  double-helicity asymmetry for $\pi^{0}$ production at midrapidity in
  $p$$+$$p$ collisions at $\sqrt{s}=510$ GeV}},}\ }\href {\doibase
  10.1103/PhysRevD.93.011501} {\bibfield  {journal} {\bibinfo  {journal} {Phys.
  Rev. D}\ }\textbf {\bibinfo {volume} {93}},\ \bibinfo {pages} {011501}
  (\bibinfo {year} {2016})}\BibitemShut {NoStop}%
\bibitem [{\citenamefont {Adare}\ \emph {et~al.}(2011)\citenamefont {Adare}
  \emph {et~al.}}]{ppg107}%
  \BibitemOpen
  \bibfield  {author} {\bibinfo {author} {\bibfnamefont {A.}~\bibnamefont
  {Adare}} \emph {et~al.} (\bibinfo {collaboration} {PHENIX Collaboration}),\
  }\bibfield  {title} {\enquote {\bibinfo {title} {{Cross section and double
  helicity asymmetry for $\eta$ mesons and their comparison to neutral pion
  production in p+p collisions at $\sqrt{s}=200$ GeV}},}\ }\href {\doibase
  10.1103/PhysRevD.83.032001} {\bibfield  {journal} {\bibinfo  {journal} {Phys.
  Rev. D}\ }\textbf {\bibinfo {volume} {83}},\ \bibinfo {pages} {032001}
  (\bibinfo {year} {2011})}\BibitemShut {NoStop}%
\bibitem [{\citenamefont {Adcox}\ \emph
  {et~al.}(2003{\natexlab{b}})\citenamefont {Adcox} \emph {et~al.}}]{DC}%
  \BibitemOpen
  \bibfield  {author} {\bibinfo {author} {\bibfnamefont {K.}~\bibnamefont
  {Adcox}} \emph {et~al.} (\bibinfo {collaboration} {PHENIX Collaboration}),\
  }\bibfield  {title} {\enquote {\bibinfo {title} {{PHENIX central arm tracking
  detectors}},}\ }\href@noop {} {\bibfield  {journal} {\bibinfo  {journal}
  {Nucl. Instrum. Methods Phys. Res., Sec. A}\ }\textbf {\bibinfo {volume}
  {499}},\ \bibinfo {pages} {489} (\bibinfo {year}
  {2003}{\natexlab{b}})}\BibitemShut {NoStop}%
\bibitem [{\citenamefont {Adare}\ \emph {et~al.}(2009)\citenamefont {Adare}
  \emph {et~al.}}]{ppg090}%
  \BibitemOpen
  \bibfield  {author} {\bibinfo {author} {\bibfnamefont {A.}~\bibnamefont
  {Adare}} \emph {et~al.} (\bibinfo {collaboration} {PHENIX Collaboration}),\
  }\bibfield  {title} {\enquote {\bibinfo {title} {{Photon-Hadron Jet
  Correlations in $p+p$ and Au+Au Collisions at $\sqrt{s_{NN}}=200$ GeV}},}\
  }\href@noop {} {\bibfield  {journal} {\bibinfo  {journal} {Phys. Rev. C}\
  }\textbf {\bibinfo {volume} {80}},\ \bibinfo {pages} {024908} (\bibinfo
  {year} {2009})}\BibitemShut {NoStop}%
\bibitem [{\citenamefont {Lai}\ \emph {et~al.}(2010)\citenamefont {Lai},
  \citenamefont {Guzzi}, \citenamefont {Huston}, \citenamefont {Li},
  \citenamefont {Nadolsky}, \citenamefont {Pumplin},\ and\ \citenamefont
  {Yuan}}]{CT10_PDFs}%
  \BibitemOpen
  \bibfield  {author} {\bibinfo {author} {\bibfnamefont {H.~L.}\ \bibnamefont
  {Lai}}, \bibinfo {author} {\bibfnamefont {M.}~\bibnamefont {Guzzi}}, \bibinfo
  {author} {\bibfnamefont {J.}~\bibnamefont {Huston}}, \bibinfo {author}
  {\bibfnamefont {Z.}~\bibnamefont {Li}}, \bibinfo {author} {\bibfnamefont
  {P.~M.}\ \bibnamefont {Nadolsky}}, \bibinfo {author} {\bibfnamefont {Jon}\
  \bibnamefont {Pumplin}}, \ and\ \bibinfo {author} {\bibfnamefont {C.~P.}\
  \bibnamefont {Yuan}},\ }\bibfield  {title} {\enquote {\bibinfo {title} {{New
  parton distributions for collider physics}},}\ }\href {\doibase
  10.1103/PhysRevD.82.074024} {\bibfield  {journal} {\bibinfo  {journal} {Phys.
  Rev. D}\ }\textbf {\bibinfo {volume} {82}},\ \bibinfo {pages} {074024}
  (\bibinfo {year} {2010})}\BibitemShut {NoStop}%
\bibitem [{\citenamefont {de~Florian}\ \emph {et~al.}(2015)\citenamefont
  {de~Florian}, \citenamefont {Sassot}, \citenamefont {Epele}, \citenamefont
  {Hern{\'a}ndez-Pinto},\ and\ \citenamefont {Stratmann}}]{DSS14_FFs}%
  \BibitemOpen
  \bibfield  {author} {\bibinfo {author} {\bibfnamefont {D.}~\bibnamefont
  {de~Florian}}, \bibinfo {author} {\bibfnamefont {R.}~\bibnamefont {Sassot}},
  \bibinfo {author} {\bibfnamefont {M.}~\bibnamefont {Epele}}, \bibinfo
  {author} {\bibfnamefont {R.~J.}\ \bibnamefont {Hern{\'a}ndez-Pinto}}, \ and\
  \bibinfo {author} {\bibfnamefont {M.}~\bibnamefont {Stratmann}},\ }\bibfield
  {title} {\enquote {\bibinfo {title} {{Parton-to-Pion Fragmentation
  Reloaded}},}\ }\href {\doibase 10.1103/PhysRevD.91.014035} {\bibfield
  {journal} {\bibinfo  {journal} {Phys. Rev. D}\ }\textbf {\bibinfo {volume}
  {91}},\ \bibinfo {pages} {014035} (\bibinfo {year} {2015})}\BibitemShut
  {NoStop}%
\bibitem [{\citenamefont {Chatrchyan}\ \emph {et~al.}(2013)\citenamefont
  {Chatrchyan} \emph {et~al.}}]{cms_gammajet}%
  \BibitemOpen
  \bibfield  {author} {\bibinfo {author} {\bibfnamefont {S.}~\bibnamefont
  {Chatrchyan}} \emph {et~al.} (\bibinfo {collaboration} {CMS Collaboration}),\
  }\bibfield  {title} {\enquote {\bibinfo {title} {{Studies of jet quenching
  using isolated-photon+jet correlations in PbPb and $pp$ collisions at
  $\sqrt{s_{NN}}=2.76$ TeV}},}\ }\href {\doibase
  10.1016/j.physletb.2012.11.003} {\bibfield  {journal} {\bibinfo  {journal}
  {Phys. Lett. B}\ }\textbf {\bibinfo {volume} {718}},\ \bibinfo {pages} {773}
  (\bibinfo {year} {2013})}\BibitemShut {NoStop}%
\bibitem [{sup()}]{supp_matt}%
  \BibitemOpen
  \href@noop {} {}\bibinfo {note} {See Supplemental Material at
  [aps-journal-link] for tabulated values of results plotted in Figs. 4, 7, and
  9}\BibitemShut {NoStop}%
\bibitem [{\citenamefont {Clark}\ \emph {et~al.}(1979)\citenamefont {Clark}
  \emph {et~al.}}]{jt_cern}%
  \BibitemOpen
  \bibfield  {author} {\bibinfo {author} {\bibfnamefont {A.~G.}\ \bibnamefont
  {Clark}} \emph {et~al.},\ }\bibfield  {title} {\enquote {\bibinfo {title}
  {{Large Transverse Momentum Jets in High-energy Proton Proton Collisions}},}\
  }\href@noop {} {\bibfield  {journal} {\bibinfo  {journal} {Nucl. Phys. B}\
  }\textbf {\bibinfo {volume} {160}},\ \bibinfo {pages} {397} (\bibinfo {year}
  {1979})}\BibitemShut {NoStop}%
\bibitem [{\citenamefont {Aad}\ \emph {et~al.}(2011)\citenamefont {Aad} \emph
  {et~al.}}]{atlas_fragfunc}%
  \BibitemOpen
  \bibfield  {author} {\bibinfo {author} {\bibfnamefont {G.}~\bibnamefont
  {Aad}} \emph {et~al.} (\bibinfo {collaboration} {ATLAS Collaboration}),\
  }\bibfield  {title} {\enquote {\bibinfo {title} {{Measurement of the jet
  fragmentation function and transverse profile in proton-proton collisions at
  a center-of-mass energy of 7 TeV with the ATLAS detector}},}\ }\href
  {\doibase 10.1140/epjc/s10052-011-1795-y} {\bibfield  {journal} {\bibinfo
  {journal} {Eur. Phys. J. C}\ }\textbf {\bibinfo {volume} {71}},\ \bibinfo
  {pages} {1795} (\bibinfo {year} {2011})}\BibitemShut {NoStop}%
\bibitem [{\citenamefont {Sjostrand}\ \emph {et~al.}()\citenamefont
  {Sjostrand}, \citenamefont {Mrenna},\ and\ \citenamefont {Skands}}]{pythia}%
  \BibitemOpen
  \bibfield  {author} {\bibinfo {author} {\bibfnamefont {T.}~\bibnamefont
  {Sjostrand}}, \bibinfo {author} {\bibfnamefont {S.}~\bibnamefont {Mrenna}}, \
  and\ \bibinfo {author} {\bibfnamefont {P.~Z.}\ \bibnamefont {Skands}},\
  }\href@noop {} {\enquote {\bibinfo {title} {{PYTHIA 6.4 Physics and
  Manual}},}\ }\bibinfo {note} {{J. High Energy Phys. {\bf 05 (2006)}
  026}}\BibitemShut {NoStop}%
\bibitem [{\citenamefont {Pumplin}\ \emph {et~al.}()\citenamefont {Pumplin},
  \citenamefont {Stump}, \citenamefont {Huston}, \citenamefont {Lai},
  \citenamefont {Nadolsky},\ and\ \citenamefont {Tung}}]{CTEQ6L1}%
  \BibitemOpen
  \bibfield  {author} {\bibinfo {author} {\bibfnamefont {J.}~\bibnamefont
  {Pumplin}}, \bibinfo {author} {\bibfnamefont {D.~R.}\ \bibnamefont {Stump}},
  \bibinfo {author} {\bibfnamefont {J.}~\bibnamefont {Huston}}, \bibinfo
  {author} {\bibfnamefont {H.~L.}\ \bibnamefont {Lai}}, \bibinfo {author}
  {\bibfnamefont {P.~M.}\ \bibnamefont {Nadolsky}}, \ and\ \bibinfo {author}
  {\bibfnamefont {W.~K.}\ \bibnamefont {Tung}},\ }\href@noop {} {\enquote
  {\bibinfo {title} {{New generation of parton distributions with uncertainties
  from global QCD analysis}},}\ }\bibinfo {note} {{J. High Energy Phys. {\bf 07
  (2002)} 012}}\BibitemShut {NoStop}%
\bibitem [{\citenamefont {Aurenche}\ \emph {et~al.}(1984)\citenamefont
  {Aurenche}, \citenamefont {Douiri}, \citenamefont {Baier}, \citenamefont
  {Fontannaz},\ and\ \citenamefont {Schiff}}]{NLO_dirphots}%
  \BibitemOpen
  \bibfield  {author} {\bibinfo {author} {\bibfnamefont {P.}~\bibnamefont
  {Aurenche}}, \bibinfo {author} {\bibfnamefont {A.}~\bibnamefont {Douiri}},
  \bibinfo {author} {\bibfnamefont {R.}~\bibnamefont {Baier}}, \bibinfo
  {author} {\bibfnamefont {M.}~\bibnamefont {Fontannaz}}, \ and\ \bibinfo
  {author} {\bibfnamefont {D.}~\bibnamefont {Schiff}},\ }\bibfield  {title}
  {\enquote {\bibinfo {title} {{Prompt Photon Production at Large $p_T$ in QCD
  Beyond the Leading Order}},}\ }\href {\doibase 10.1016/0370-2693(84)91053-0}
  {\bibfield  {journal} {\bibinfo  {journal} {Phys. Lett. B}\ }\textbf
  {\bibinfo {volume} {140}},\ \bibinfo {pages} {87} (\bibinfo {year}
  {1984})}\BibitemShut {NoStop}%
\bibitem [{\citenamefont {Adamczyk}\ \emph {et~al.}(2014)\citenamefont
  {Adamczyk} \emph {et~al.}}]{STARcorr:2014PRL}%
  \BibitemOpen
  \bibfield  {author} {\bibinfo {author} {\bibfnamefont {L.}~\bibnamefont
  {Adamczyk}} \emph {et~al.} (\bibinfo {collaboration} {STAR Collaboration}),\
  }\bibfield  {title} {\enquote {\bibinfo {title} {{Jet-Hadron Correlations in
  $\sqrt{s_{NN}}=200$ GeV $p+p$ and Central Au+Au Collisions}},}\ }\href
  {\doibase 10.1103/PhysRevLett.112.122301} {\bibfield  {journal} {\bibinfo
  {journal} {Phys. Rev. Lett.}\ }\textbf {\bibinfo {volume} {112}},\ \bibinfo
  {pages} {122301} (\bibinfo {year} {2014})}\BibitemShut {NoStop}%
\bibitem [{\citenamefont {Aybat}\ and\ \citenamefont
  {Rogers}(2011)}]{AybatRogers:2011}%
  \BibitemOpen
  \bibfield  {author} {\bibinfo {author} {\bibfnamefont {S.~M.}\ \bibnamefont
  {Aybat}}\ and\ \bibinfo {author} {\bibfnamefont {T.~C.}\ \bibnamefont
  {Rogers}},\ }\bibfield  {title} {\enquote {\bibinfo {title} {{TMD Parton
  Distribution and Fragmentation Functions with QCD Evolution}},}\ }\href
  {\doibase 10.1103/PhysRevD.83.114042} {\bibfield  {journal} {\bibinfo
  {journal} {Phys. Rev. D}\ }\textbf {\bibinfo {volume} {83}},\ \bibinfo
  {pages} {114042} (\bibinfo {year} {2011})}\BibitemShut {NoStop}%
\bibitem [{\citenamefont {Aaltonen}\ \emph {et~al.}(2009)\citenamefont
  {Aaltonen} \emph {et~al.}}]{CDF_kt_MLLA}%
  \BibitemOpen
  \bibfield  {author} {\bibinfo {author} {\bibfnamefont {T.}~\bibnamefont
  {Aaltonen}} \emph {et~al.} (\bibinfo {collaboration} {CDF Collaboration}),\
  }\bibfield  {title} {\enquote {\bibinfo {title} {{Measurement of the $k_T$
  Distribution of Particles in Jets Produced in $p\bar{p}$ Collisions at
  $\sqrt{s}=1.96$~TeV}},}\ }\href {\doibase 10.1103/PhysRevLett.102.232002}
  {\bibfield  {journal} {\bibinfo  {journal} {Phys. Rev. Lett.}\ }\textbf
  {\bibinfo {volume} {102}},\ \bibinfo {pages} {232002} (\bibinfo {year}
  {2009})}\BibitemShut {NoStop}%
\bibitem [{\citenamefont {Aidala}\ \emph {et~al.}(2013)\citenamefont {Aidala},
  \citenamefont {Bass}, \citenamefont {Hasch},\ and\ \citenamefont
  {Mallot}}]{physrevmod}%
  \BibitemOpen
  \bibfield  {author} {\bibinfo {author} {\bibfnamefont {C.~A.}\ \bibnamefont
  {Aidala}}, \bibinfo {author} {\bibfnamefont {S.~D.}\ \bibnamefont {Bass}},
  \bibinfo {author} {\bibfnamefont {D.}~\bibnamefont {Hasch}}, \ and\ \bibinfo
  {author} {\bibfnamefont {G.~K.}\ \bibnamefont {Mallot}},\ }\bibfield  {title}
  {\enquote {\bibinfo {title} {{The Spin Structure of the Nucleon}},}\ }\href
  {\doibase 10.1103/RevModPhys.85.655} {\bibfield  {journal} {\bibinfo
  {journal} {Rev. Mod. Phys.}\ }\textbf {\bibinfo {volume} {85}},\ \bibinfo
  {pages} {655} (\bibinfo {year} {2013})}\BibitemShut {NoStop}%
\bibitem [{\citenamefont {Adare}\ \emph
  {et~al.}(2014{\natexlab{b}})\citenamefont {Adare} \emph {et~al.}}]{ppg135}%
  \BibitemOpen
  \bibfield  {author} {\bibinfo {author} {\bibfnamefont {A.}~\bibnamefont
  {Adare}} \emph {et~al.} (\bibinfo {collaboration} {PHENIX Collaboration}),\
  }\bibfield  {title} {\enquote {\bibinfo {title} {{Measurement of
  transverse-single-spin asymmetries for midrapidity and forward-rapidity
  production of hadrons in polarized p+p collisions at $\sqrt{s}=200$ and 62.4
  GeV}},}\ }\href {\doibase 10.1103/PhysRevD.90.012006} {\bibfield  {journal}
  {\bibinfo  {journal} {Phys. Rev. D}\ }\textbf {\bibinfo {volume} {90}},\
  \bibinfo {pages} {012006} (\bibinfo {year} {2014}{\natexlab{b}})}\BibitemShut
  {NoStop}%
\bibitem [{\citenamefont {Rogers}(2013)}]{rogers:2013:extra_asymm}%
  \BibitemOpen
  \bibfield  {author} {\bibinfo {author} {\bibfnamefont {T.~C.}\ \bibnamefont
  {Rogers}},\ }\bibfield  {title} {\enquote {\bibinfo {title} {{Extra spin
  asymmetries from the breakdown of transverse-momentum-dependent factorization
  in hadron-hadron collisions}},}\ }\href {\doibase 10.1103/PhysRevD.88.014002}
  {\bibfield  {journal} {\bibinfo  {journal} {Phys. Rev. D}\ }\textbf {\bibinfo
  {volume} {88}},\ \bibinfo {pages} {014002} (\bibinfo {year}
  {2013})}\BibitemShut {NoStop}%
\bibitem [{\citenamefont {Skands}(2010)}]{perugia_tune}%
  \BibitemOpen
  \bibfield  {author} {\bibinfo {author} {\bibfnamefont {P.~Z.}\ \bibnamefont
  {Skands}},\ }\bibfield  {title} {\enquote {\bibinfo {title} {{Tuning Monte
  Carlo Generators: The Perugia Tunes}},}\ }\href {\doibase
  10.1103/PhysRevD.82.074018} {\bibfield  {journal} {\bibinfo  {journal} {Phys.
  Rev. D}\ }\textbf {\bibinfo {volume} {82}},\ \bibinfo {pages} {074018}
  (\bibinfo {year} {2010})}\BibitemShut {NoStop}%
\bibitem [{\citenamefont {Aaltonen}\ \emph {et~al.}(2012)\citenamefont
  {Aaltonen} \emph {et~al.}}]{CDFzbosons}%
  \BibitemOpen
  \bibfield  {author} {\bibinfo {author} {\bibfnamefont {T.}~\bibnamefont
  {Aaltonen}} \emph {et~al.} (\bibinfo {collaboration} {CDF Collaboration}),\
  }\bibfield  {title} {\enquote {\bibinfo {title} {{Transverse momentum cross
  section of $e^+e^-$ pairs in the $Z$-boson region from $p\bar{p}$ collisions
  at $\sqrt{s}=1.96$ TeV}},}\ }\href {\doibase 10.1103/PhysRevD.86.052010}
  {\bibfield  {journal} {\bibinfo  {journal} {Phys. Rev. D}\ }\textbf {\bibinfo
  {volume} {86}},\ \bibinfo {pages} {052010} (\bibinfo {year}
  {2012})}\BibitemShut {NoStop}%
\end{thebibliography}

%
 
\end{document}